\documentclass[traditabstract]{aa}
\usepackage[utf8]{inputenc}
\usepackage{threeparttable,pifont}
\usepackage{natbib}
\usepackage{color}
\usepackage{multicol}
\usepackage{amsmath,amssymb,url}
\usepackage{graphicx}
\usepackage{xcolor}
\usepackage{txfonts,longtable,dcolumn,verbatim}
\bibpunct{(}{)}{;}{a}{}{,}
\usepackage{array} % fixes the corners
\usepackage[normalem]{ulem}
\usepackage{lscape}

\usepackage[squaren,textstyle]{SIunits}

\usepackage{hyperref}
\hypersetup{
    unicode=true,           % non-Latin characters in Acrobat's bookmarks
    pdftoolbar=true,        % show Acrobat's toolbar?
    pdfmenubar=true,        % show Acrobat's menu?
    pdffitwindow=true,      % page fit to window when opened
    pdftitle={Physical models of Jupiter trojan asteroids},
    pdfauthor={Hanus et al.},
    pdfsubject={Planetary Science},
    pdfkeywords={},         % list of keywords
    pdfnewwindow=true,      % links in new window
    colorlinks=true,        % false: boxed links; true: colored links
    linkcolor=gray,         % color of internal links
    citecolor=blue,         % color of links to bibliography
    filecolor=gray,         % color of file links
    urlcolor=gray           % color of external links
}

\usepackage{xspace}

\newcommand{\adam}{\texttt{ADAM}\xspace}

\pdfoutput=1

\begin{document}

\title{Shape models and spin states of Jupiter Trojans}
\subtitle{Testing the streaming instability formation scenario
    \thanks{Tables B.1-5 are  available  at  the  CDS  via  anonymous   ftp  to \url{http://cdsarc.u-strasbg.fr/} or via \url{http://cdsarc.u-strasbg.fr/viz-bin/qcat?J/A+A/xxx/Axxx}}}

\titlerunning{Physical models of Trojans}

\author{  
    J.~Hanu{\v s}\inst{\ref{auuk}}          \and % 3D modeling, physical properties of JTs
    D.~Vokrouhlick\'y\inst{\ref{auuk}}      \and % dynamics
    D.~Nesvorn\'y\inst{\ref{swri}}          \and % dynamics
    J.~\v Durech\inst{\ref{auuk}}           \and % modeling 
    R.~Stephens\inst{\ref{stephens}}        \and
   % O.~Pejcha \inst{\ref{utf}}              \and% ASAS-SN data 
    V.~Benishek \inst{\ref{benishek}}       \and
    J.~Oey \inst{\ref{oey}}                 \and
    P.~Pokorn\'y \inst{\ref{pok1},\ref{pok2},\ref{pok3}}           
} 

\institute{
      %---- AUUK Hanus, Vokrouhlicky, Durech
     Charles University, Faculty of Mathematics and Physics, Institute of Astronomy, V Hole\v sovi\v ck\'ach 2, CZ-18000, Prague 8, Czech Republic\label{auuk}\\
     \email{josef.hanus@mff.cuni.cz}
     \and %SWRI
     Department of Space Studies, Southwest Research Institute, 1050 Walnut St., Suite 300, Boulder, CO 80302, USA\label{swri}
     \and
     Center for Solar System Studies, 9302 Pittsburgh Ave. Suite 200, Rancho Cucamonga, CA 91730, USA\label{stephens}
     %\and
     % Charles University, Faculty of Mathematics and Physics, Institute of Theoretical Physics, V~Hole{\v s}ovi{\v c}k{\'a}ch 2, 18000 Prague, Czech Republic\label{utf}
     \and
     Belgrade Astronomical Observatory, Volgina 7, 11060 Belgrade 38, Serbia\label{benishek}
     \and
     Blue Mountains Observatory, Leura, Australia\label{oey}
     \and
     Department of Physics, The Catholic University of America, Washington, DC 20064, USA\label{pok1}
     \and
     Astrophysics Science Division, NASA Goddard Space Flight Center, Greenbelt, MD 20771, USA\label{pok2}
     \and
     Center for Research and Exploration in Space Science and Technology, NASA/GSFC, Greenbelt, MD 20771, USA\label{pok3}
}

   \date{Received x-x-2022 / Accepted x-x-2022}
% \abstract{}{}{}{}{} 
% 5 {} token are mandatory
 
\abstract
{The leading theory for the origin of Jupiter Trojans (JTs) assumes that JTs were captured to their orbits near the Lagrangian points of Jupiter during the early reconfiguration of the giant planets. The natural source region for the majority of JTs would then be the population of planetesimals born in a massive trans-Neptunian disk. If true, JTs represent the most accessible stable population of small Solar System bodies that formed in the outer regions of the Solar System. For this work, we compiled photometric datasets for about 1000 JTs and applied the convex inversion technique in order to assess their shapes and spin states. We obtained full solutions for $79$ JTs, and partial solutions for an additional $31$ JTs. We found that the observed distribution of the pole obliquities of JTs is broadly consistent with expectations from the streaming instability, which is the leading mechanism for the formation of planetesimals in the trans-Neptunian disk. The observed JTs' pole distribution has a slightly smaller prograde vs. retrograde asymmetry (excess of obliquities $>130^\circ$) than what is expected from the existing streaming instability simulations. However, this discrepancy can be plausibly reconciled by the effects of the post-formation collisional activity. Our numerical simulations of the post-capture spin evolution indicate that the JTs' pole distribution is not significantly affected by dynamical processes such as the eccentricity excitation in resonances, close encounters with planets, or the effects of nongravitational forces. However, a few JTs exhibit large latitude variations of the rotation pole and may even temporarily transition between prograde- and retrograde-rotating categories.
} 

\keywords
{Minor planets, asteroids: Jupiter Trojans -- Surveys --  Methods: observational --   Methods: data analysis}

\maketitle

%%%%%%%%%%%%%%% INTRO %%%%%%%%%%%%%%%%%%%%%%%%%%%%%%%%%%%%%%%%%%%%%%%%%%%%%%%%%%%%%
\section{Introduction}\label{sec:introduction}

Jupiter Trojans (JTs) are minor bodies co-orbiting with Jupiter in the proximity of its Lagrangian points L$_4$ and L$_5$. Bodies librating about the leading L$_4$ point are commonly referred to as the Greek camp (or clan or group), while those near the trailing L$_5$ point are referred to as the Trojan camp. As for the currently known population of Greeks and Trojans, we adopt in this work a list of JTs as identified by the orbit classification flags in the MPC Orbit (MPCORB) 
database\footnote{\texttt{MPCORB.DAT} file available at \url{https://www.minorplanetcenter.net/iau/MPCORB.html}}.

The origin of JTs remains an open problem. The currently leading theory assumes that JTs were captured to their orbits near the Lagrangian points during the early reconfiguration of the giant planets \citep{Morbidelli2005, Nesvorny2013}. The natural source region for the majority of JTs would then be the population of planetesimals born in a massive trans-Neptunian disk. As a result, JTs should share the physical properties of the currently observed trans-Neptunian objects (TNOs), as well as the comets and irregular satellites of giant planets. Other theories avoid involving the planetary reconfiguration event and postulate that JTs formed at their current location together with Jupiter \citep[see reviews in][]{mar2002,emery2015}, or were born in the Jupiter co-orbital zone and accompanied its early inward migration \citep{pirani2019}. In this paper, we adopt the capture model for JTs as a baseline hypothesis, because several other pieces of evidence support the view that giant planets underwent a violent instability at some early moment of the Solar System evolution \citep[see, e.g.,][]{n2018}. In any case, studying the physical properties of JTs is an obvious way how to test various theories of the JTs' origin and eventually decide which theory is more in line with the observing evidence. In fact, this is also one of the main goals of our work. 

Previous physical studies have already revealed interesting properties of JTs. For instance, analysis of visible and near-infrared spectra suggested that a color bimodality exists in the JTs' population \citep[e.g.,][]{Emery2011,wb2016}: the majority of bodies belong to the so-called red (D-type) and less red (P-type) groups. This color bimodality hints at the relationship between the spectral properties of JTs and of comets and TNOs. On the other hand, C-type bodies, which represent about 10\% of the JT population and -- for instance -- include the largest collisional JT family Eurybates, are spectrally more similar to a primitive outer asteroidal belt or Cybele group objects \citep{Fornasier2007, DeLuise2010}. 

In closer relation to our work, we note that a significant amount of data was collected about JTs' rotation rate since the 1980s \citep[see reviews in][]{bar2002,emery2015}. Hints from these early studies have been recently confirmed and extended by analysis of space-born observations from Kepler and Transiting Exoplanet Survey Satellite (TESS), and dedicated survey programs using large ground-based instruments \citep[e.g.,][]{Szabo2016, Ryan2017, Kalup2021, Chang2021}. These studies find (i) an excess (possibly even separate population) of slowly rotating objects, and (ii) a presence of a size-dependent lower limit of the rotation period, such that smaller JTs may reach shorter periods to near $4$~h, while larger JTs tend to have this limit near $5$~h. The latter  nicely supports a rubble-pile structure of JTs with a characteristic bulk density of $\simeq 0.9$ g~cm$^{-3}$, and the emerging role of the Yarkovsky-O'Keefe-Radzievski-Paddack (YORP) effect for small JTs \citep[e.g.,][]{vetal15}. The population of unusually slowly rotating JTs is interesting in the context of our work since it has been linked to unbound components from tidally evolved binary planetesimals captured among Trojans \citep{netal2020}.

While a significant amount of data about the rotation rates of JTs have been collected and analyzed, much less is presently known about the complete characterization of their spin state (i.e., rotation rate and pole direction).
Here we aim to fill this missing piece of information by deriving rotation state properties and convex shapes for many JTs. Particularly, we are interested in the direction of the rotation axis with respect to the orbital plane -- the pole obliquity. This parameter reflects the object's dynamical history and could help constrain various theories aiming to explain the population's origin.

Our paper is organized as follows. We gathered available optical photometric data (Sec.~\ref{sec:data}) from various sources and analyzed them using a convex inversion method \citep{Kaasalainen2001b, Kaasalainen2001a} following, with some minor adjustments, the scheme of \citet{Hanus2021} (Sec.~\ref{sec:inversion}). We present derived physical properties in Sec.~\ref{sec:modeling}. Adopting the capture model for JTs, we also performed numerical simulations to (i)~access the original spin distribution of the planetesimal population that is assumed to be the source population of JTs (Sec.~\ref{sec:streeming}), and (ii)~estimate the post-capture spin evolution of JTs (Sec.~\ref{sec:dynamics}). Finally, we discuss our findings in Sec.~\ref{sec:discussion}, and conclude our work in Sec.~\ref{sec:conclusions}.

%%%%%%%%%%%%%%% DATA %%%%%%%%%%%%%%%%%%%%%%%%%%%%%%%%%%%%%%%%%%%
\section{Data}\label{sec:data}

We gathered optical disk-integrated photometry from various sources. First, data for asteroids with published shape models are usually available in the Database of Asteroid Models from Inversion Techniques\footnote{\url{https://astro.troja.mff.cuni.cz/projects/damit/}} \citep[DAMIT;][]{Durech2010}. Additional dense-in-time light curves were downloaded from the ALCDEF\footnote{\url{https://minplanobs.org/alcdef/}} database or were obtained from individual observers. Moreover, we also make use of sparse-in-time photometric data from various surveys. These include Catalina Sky Survey \citep[CSS;][]{Larson2003}, the US Naval Observatory in Flagstaff (USNO-Flagstaff), the Asteroid Terrestrial-impact Last Alert System \citep[ATLAS;][]{Tonry2018}, the All-Sky Automated Survey for Supernovae \citep[ASAS-SN;][]{Shappee2014, Kochanek2017, Hanus2021}, Gaia Data Release 3
\citep[Gaia DR3;][]{Tanga2023, Babusiaux2023}, the Zwicky Transient Facility \citep[ZTF;][]{bellm2019}, Kepler K2 \citep{Szabo2016, Kalup2021}, Palomar Transient Factory \citep[PTF;][]{Waszczak2015}, and TESS \citep{Pal2020}.

Individual photometric measurements from CSS, USNO-Flagstaff, and ZTF are available via AstDyS-2 database\footnote{\url{https://newton.spacedys.com/astdys/}}, from where we downloaded them and processed them following the approach of \citet{Hanus2011}. Data from ASAS-SN, Gaia DR3, K2, TESS, and PTF were released together with the corresponding publications. We downloaded the data from the repositories and processed them similarly to the other sparse-in-time data. We already used ATLAS and ASAS-SN data in previous studies \citep{Durech2020, Hanus2021}, to which we refer for additional information about the data processing. 

Altogether, we obtained optical datasets for 1,009 JTs. In all cases, sparse-in-time datasets are included, usually from multiple sources. Dense-in-time light curves are available for a subsample of 164 JTs.

\section{Light curve inversion}\label{sec:inversion}

\begin{figure*}
    \centering
  \includegraphics[width=\textwidth]{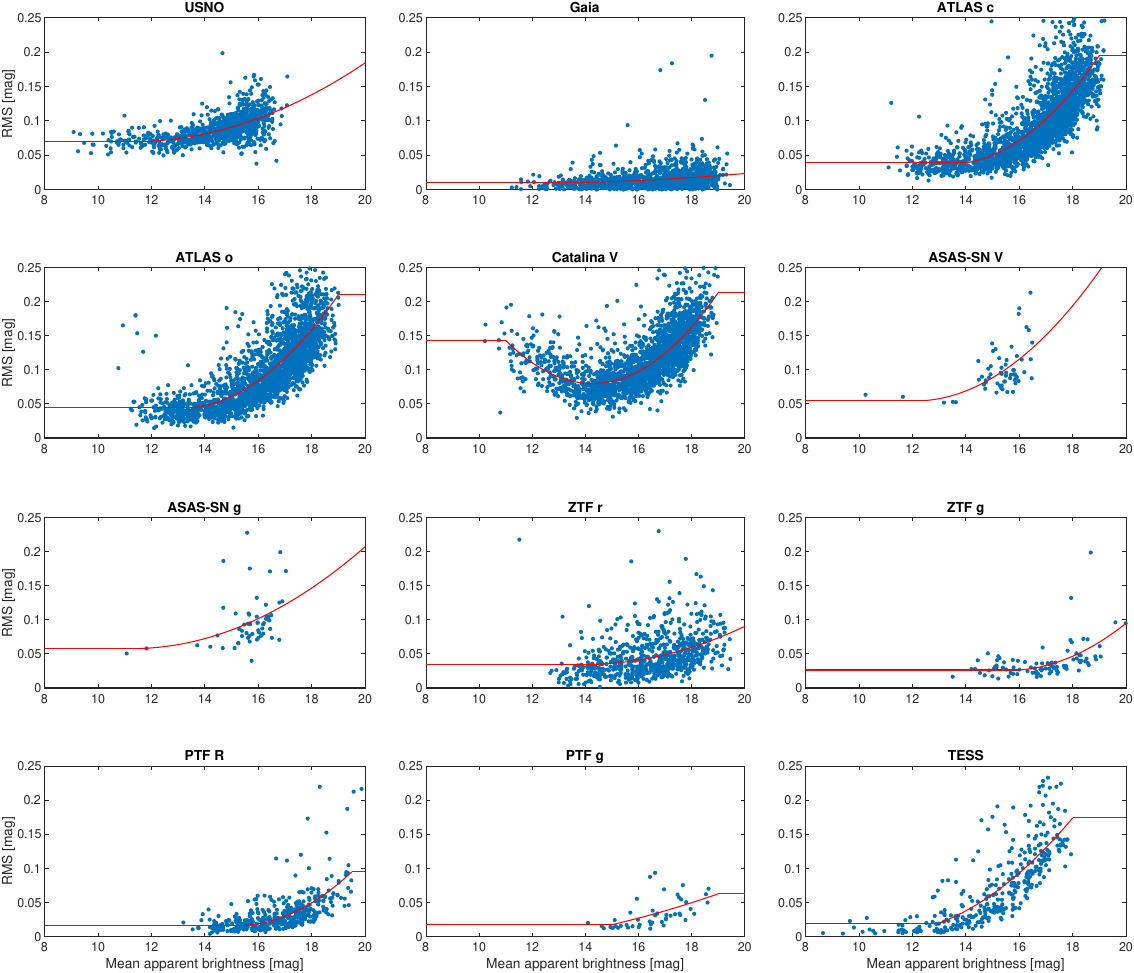}
 \caption{\label{fig:rms}RMS values as a function of mean brightness for all available sparse datasets. The red curve represents a fit to the data defined by the quadratic approximation (and constant parts) in Eq.~(\ref{eq:rms}).}
\end{figure*}

We analyzed the optical photometry data using the convex inversion method \citep{Kaasalainen2001b, Kaasalainen2001a}. This commonly used gradient-based inversion method represents the shape model as a convex polyhedron; convexity is a necessary assumption for the shape solution to be stable and unique as only disk-integrated data are utilized \citep{Kaasalainen2006}. 
The method converges to a unique shape solution for fixed values of the rotation state parameters -- sidereal rotation period and spin axis orientation.\footnote{In all cases reported here the JTs rotate in the energetically lowest state, namely about the shortest principal axis of the inertia tensor. No obvious signs of significant tumbling are in fact expected for most objects in our sample of large JTs, because the corresponding damping timescale is typically smaller than a gigayear \citep[e.g.,][]{pravec2014}. The only exception might be JT 88229 as it has a rotation period of $\sim 1,500$~h and corresponding nutation damping timescale $\gg 4$~Gyr.}
Including these (usually unknown) parameters into the optimization, the $\chi^2$ behavior over the parameter space becomes complicated by having a myriad of local minima. The most common, albeit computationally cumbersome, way how to find the global minimum in the parameter space is to analyze all the relevant local minima. In practice, this represents undergoing a CPU time-consuming procedure of running convex inversion for all relevant combinations of rotation state parameters and constructing the corresponding $\chi^2$ maps. 

We followed the shape modeling procedure described in \citet{Hanus2021} with one major difference -- data weighting. By definition, the best model is defined as having the lowest $\chi^2$ value. In order to compute $\chi^2$, one should know the uncertainty of individual observations, especially when multisource datasets are utilized. However, these uncertainties are generally not known or are not reliable for the majority of data we have. Moreover, the data we utilized have very different uncertainty -- from precise Gaia photometry and high-quality dense light curves, to usually noisy sparse data from sky surveys. 

To assign realistic weights to individual datasets, we estimated their characteristic uncertainty by computing RMS residuals for existing shape models in the DAMIT database. Our procedure was as follows. First, we selected data for asteroids with models in DAMIT for each sparse dataset. We used the period and pole parameters as initial values for the optimization and applied the light curve inversion method. For each model, we computed the final RMS and assumed that this number is a proxy for the uncertainty of the measurements. In Fig.~\ref{fig:rms}, we plot RMS values as a function of mean apparent brightness for all datasets. There is a general trend of increasing RMS with increasing magnitude, which naturally arises from the fact that fainter asteroids have larger photometric uncertainties. An exception is Catalina Sky Survey, for which brighter asteroids have also larger errors, likely due to saturation. For each dataset, we fit a quadratic function for the RMS vs. magnitude $m$ dependence with constant parts outside the central interval:
\begin{equation}\label{eq:rms}
    {\rm RMS} = A\, m^2 + B\, m + C\;,
\end{equation}
where $A$, $B$, and $C$ are free parameters and $m$ is the brightness in magnitudes. This function (red curve in Fig.~\ref{fig:rms}) was then used to compute formal errors of all datasets for all asteroids. The magnitude was computed as a mean apparent brightness over all data.

Although this approach has several caveats -- DAMIT models are not fully realistic, the brightness changes significantly for one asteroid due to observing geometry, telescope photometric performance can change during its operation, etc. -- it serves well for our purpose to assign operationally acceptable relative weights to different datasets.

The relative weights $w_i$ of all available datasets for each asteroid were then computed by

\begin{equation}
    w_i = \frac{1}{{\rm RMS}_i^2}\; ,
\end{equation}
where RMS$_i$ is the formal error for each dataset.

Finally, we normalized $w_i$ such that 
\begin{equation}
   \sum_{i=1}^{N}w_i=N\; , 
\end{equation}
where $N$ is the number of light curves in the dataset. Each sparse dataset is considered as a single light curve.  

%%%%%%%%%%% RESULTS %%%%%%%%%%%%%%%%%
\section{Spin and shape modeling of JTs}\label{sec:modeling}

\begin{figure}%[ht!]
\begin{center}
  \resizebox{1.0\hsize}{!}{\includegraphics{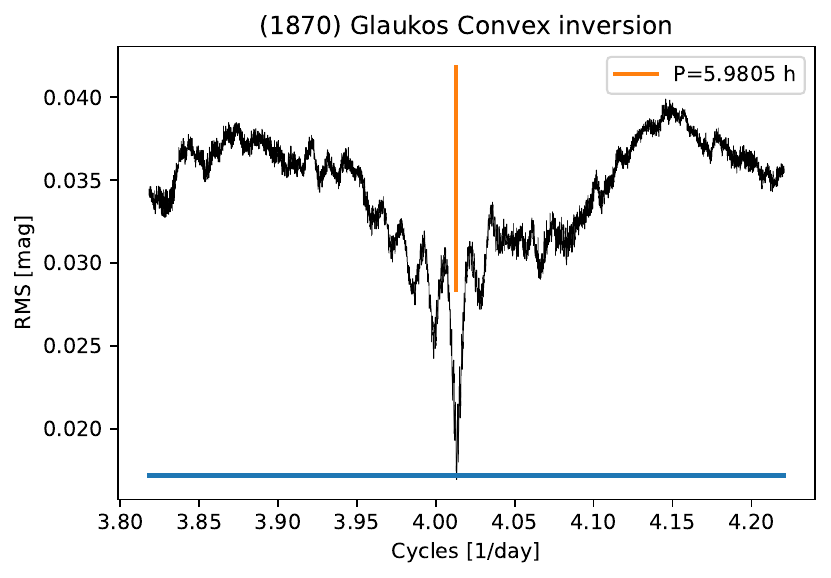}}
\end{center}
 \caption{\label{fig:periodogram}Periodogram in the rotation frequency domain for L$_5$ JT (1870)~Glaukos. The black line connects the minima over the trial runs sampling all local RMS values at a fixed rotation period but walking through all other parameters of the model. The blue horizontal line indicates the RMS threshold as defined by Eq.~(\ref{eq:chi2limit}), while the orange vertical line represents the best-fitting sidereal rotation period.}
\end{figure}

In total, we applied the convex inversion to 881 JTs with a sufficient amount of photometric measurements. Following the standard modeling approach \citep[e.g.,][]{Hanus2021}, we constructed the periodogram for each asteroid (see an example in Fig.~\ref{fig:periodogram}). To reduce the CPU time requirements, whenever available, we searched the period only on a short interval near the reported period in the LightCurve DataBase \citep[LCDB,][]{Warner2009}. We considered the reliability of the reported values (through the validity flag) by selecting larger period intervals for less reliable periods. For poor period estimates or unknown periods, we searched for the period on an interval of 2--5,000 hours. 

We considered the period as unique if the best-fitting (i.e., global) solution with $\chi^2_{\mathrm{min}}$ is the only one within the threshold of
\begin{equation}\label{eq:chi2limit}
\chi^2_{\mathrm{tr}} = \left(1+0.66\sqrt{\frac{2}{\nu}}\right)\,\chi^2_{\mathrm{min}}\; ,
\end{equation}
where $\nu$ is the number of degrees of freedom (number of observations minus the number of free parameters). We consider three parameters for the description of the rotation state, three parameters for the scattering law, and 7$^2$ parameters for the shape model (55 free parameters in total).

Next, we searched the pole orientation by scanning tens of different values isotropically distributed on a sphere. We then set the same condition for the unique solution as for the period search (Eq.~\ref{eq:chi2limit}). Due to the symmetry of the light curve inversion problem \citep{Kaasalainen2006}, two pole solutions usually have similar fits -- they have a similar shape model and the pole-latitude, but their pole-longitude is $\simeq 180^\circ$ different. 

\subsection{Partial solutions and new rotation period estimates}\label{sec:partials}

For 31 JTs, we obtained more than two pole solutions within the $\chi^2$ threshold set by Eq.~(\ref{eq:chi2limit}). Although we are not able to report the unique spin state and shape solution in these cases, the rotation period is derived unambiguously. Moreover, for the majority of these so-called partial solutions, the pole-ecliptic latitude is well constrained (i.e., is similar within the multiple pole solutions), and therefore, it is possible to decide whether the asteroid is a prograde or a retrograde rotator (see Table~B.1). Interestingly, Table~B.1 contains 11 JTs with no prior value of the rotation period in the LCDB database (e.g., 6997, 15398, 32370). In the remaining 20 cases, our sidereal rotation periods are consistent with the LCDB rotation periods. Our values usually have smaller uncertainty than the synodic periods from the LCDB, often reaching fractions of a second.

\subsection{New and improved shape and rotation state solutions}\label{sec:models}

For 79 JTs, we derived their unique spin state and shape solutions (Table~B.2). We visually checked the periodograms and the fit to the light curve data and rejected all suspicious solutions. We also computed the principal moments of inertia \citep{Dobrovolskis1996} for each shape model solution and dismissed those that were nonphysical. Finally, we also compared derived rotation periods with those from the LCDB database and individually investigated all inconsistent cases. We identified six such cases, but none of them is significant, because their LCDB reliability flags indicate that the synodic rotation period estimates could be inaccurate, and thus still consistent with our new, not too different, values. 

As there is an overlap between our solutions and their availability in DAMIT, we also verified their consistency. However, these models are usually not fully independent as some of the optical data are mutual. Despite that, the DAMIT solutions represent a useful reliability check of the methodology. The overlap with the DAMIT database is for 13 JTs. Additionally, we considered the spin state solution for (884)~Priamus of \citet{Stephens2017} here, although it has not yet been included in DAMIT. We list all the previous spin state solutions in Table~B.2.

Six solutions differ by more than 30 degrees in the pole direction, while all the rotation periods are very similar. The inconsistent solutions are for (659)~Nestor, (911)~Agamemnon, (3391)~Sinon, (4489)~Dracius, (4709)~Ennomos, and (15663)~Periphas. The previous solutions, all published in \citet{Durech2019}, are based on combined Gaia DR2 and Lowell data. As especially the Lowell data have poor photometric accuracy, it is not surprising that the revised models could be significantly different considering the spin vector direction. Interestingly, the solutions usually agree in ecliptic pole-latitude but differ in pole-longitude. Our solutions are based on substantially larger photometric datasets including, in some cases, dense light curves, therefore we prefer those. Moreover, in some cases, we either rejected one of the two previously reported pole solutions or have the mirror solution (i.e., two poles) instead of a single pole. The remaining 65 shape model solutions are new -- we list the physical properties of these JTs in Table~B.2.

We provide optical data of asteroids with new and revised shape model solutions that are not included in DAMIT in Table~B.5. New and revised solutions will be available in the DAMIT database (optical light curves, light curve fits, rotation state parameters, and shape models).

\subsection{Assessing the model uncertainties by bootstrapping the photometric datasets}\label{sec:bootstrap}

In order to assess the pole and shape uncertainties quantitatively, we performed additional modeling. For asteroids with shape and spin state solutions from Sec.~\ref{sec:models}, we bootstrapped their photometric datasets and applied the convex inversion. We created the bootstrapped datasets for each asteroid by the following procedure: We randomly selected $N$ dense light curves from the original dense dataset of $N$ light curves. Therefore, the bootstrapped datasets could contain individual light curves multiple times, while some have to be inevitably omitted. The sparse datasets were bootstrapped internally: We randomly selected the same number of measurements as in the original sparse dataset from each sparse data source (i.e., ASAS-SN, ATLAS, etc.). We computed the weights following the procedure described in Sec.~\ref{sec:inversion}.

In the shape modeling, we used the nominal spin state solutions as initial inputs and found the bootstrapped solutions near these local minima. We repeated the modeling with ten different bootstrapped datasets. We summarize the mean values and standard deviations of pole directions within the bootstrapped solutions in Table~B.4. This gives us a rough estimate of the uncertainty -- usually below 5 degrees, and quite rarely $>$10 degrees. Moreover, we also constructed the cumulative obliquity distributions for each bootstrapped dataset to assess its stability later in Sec.~\ref{sec:phys}.

\subsection{Stellar occultations}

For eight JTs listed in Tables~B.2 and B.3, there were successfully observed stellar occultations with a sufficient number of chords that enabled us to scale the shape models and in some cases also to reject one of the two possible pole solutions. The stellar occultations were assessed through the Occult software\footnote{\url{http://www.lunar-occultations.com/iota/occult4.htm}}. A thorough description of the occultation data can be found in \citet{Herald2020}. For comparing the occultations with our shape models, we used the same approach as \citet{Durech2011}. We computed the orientation of the shape model at the time of the occultation, projected it to the fundamental plane, and fit the asteroid's size to get the best agreement between the model's silhouette and observed chords. This way, we scaled models of (588)~Achilles (Fig.~\ref{fig:occ_588}), (884)~Priamus (Fig.~\ref{fig:occ_884}), (911)~Agamemnon (Fig.~\ref{fig:occ_911}), (1437)~Diomedes (Fig.~\ref{fig:occ_1437}), (1867)~Deiphobus (Fig.~\ref{fig:occ_1867}), (2207)~Antenor (Fig.~\ref{fig:occ_2207}), (4709)~Ennomos (Fig.~\ref{fig:occ_4709}), and (31344)~Agathon (Fig.~\ref{fig:occ_31344}). If one of the two possible pole solutions gave a significantly better agreement with the occultation, we rejected the second pole -- this concerns JTs Priamus, Diomedes, and Agathon. We indicate these cases in Table~B.2.

For Deiphobus the agreement between the occultation and the shape projection was suboptimal. Therefore, we also reconstructed the shape model by a different approach -- by the All-Data Asteroid Modelling (\adam{}) inversion technique \citep{Viikinkoski2015, Viikinkoski2016}. \adam{} allows using the stellar occultation for the shape reconstruction contrary to the convex inversion where the shape is scaled in size only. The \adam{} shape model of Deiphobus agrees better with the occultation and provides a more realistic size estimate. The alternative \adam{} model is listed in Table~B.2.

\subsection{Spin states of JTs}\label{sec:phys}

\begin{figure*}%[ht!]
\begin{center}
  \resizebox{1.0\hsize}{!}{\includegraphics{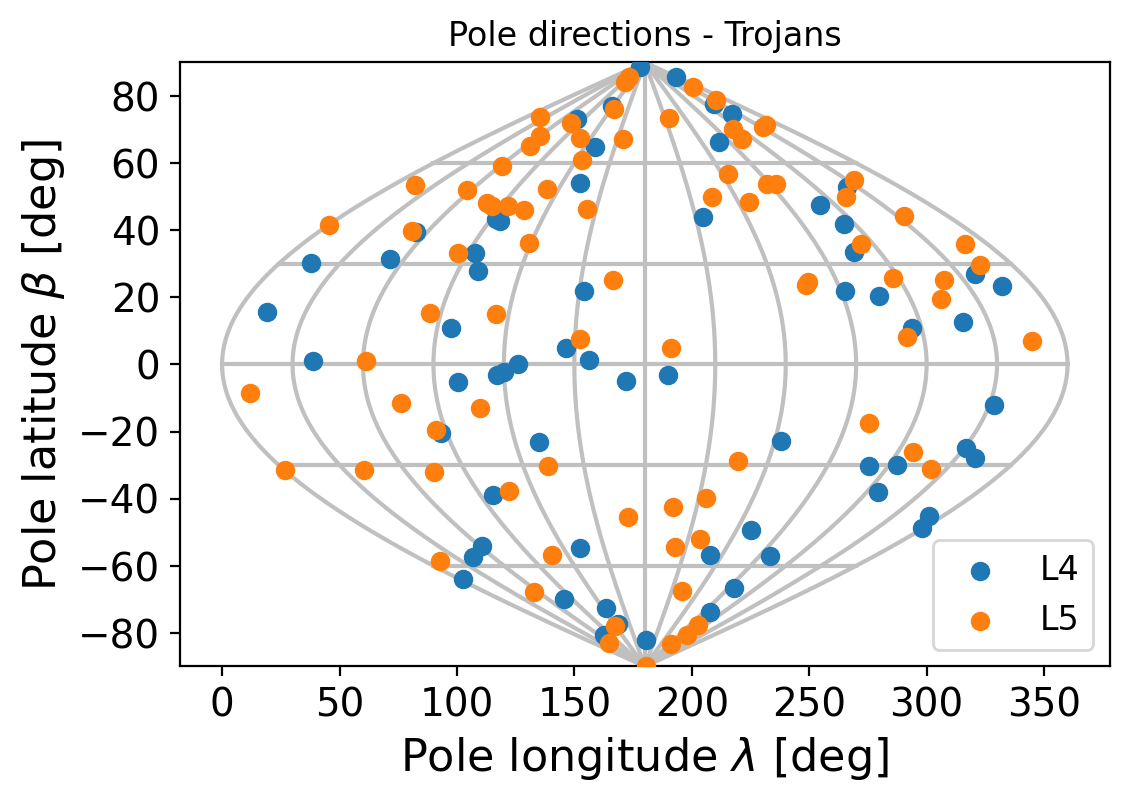}\includegraphics{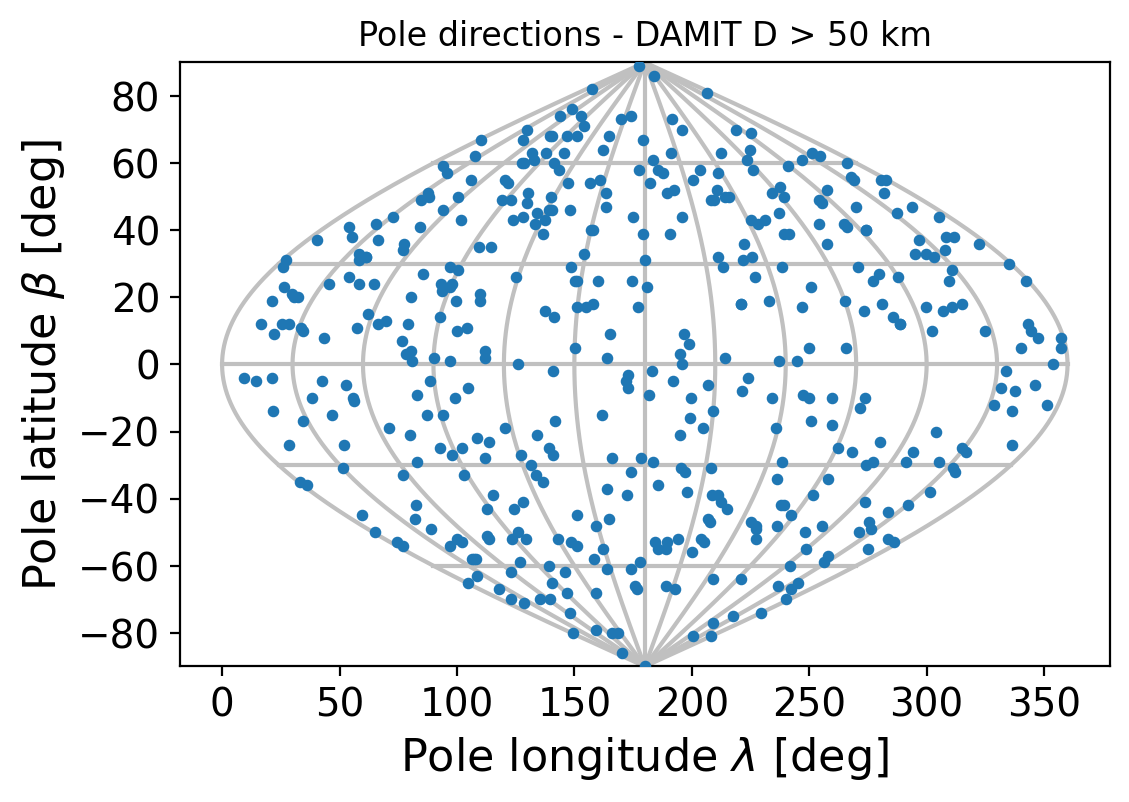}}\\
\end{center}
 \caption{\label{fig:poles}Spin vector distribution of JTs (left panel) and of main belt asteroids with sizes $>$50~km for comparison (right panel). Spin vectors of MBAs are taken from the DAMIT database. If two possible pole solutions exist, typically separated by $\simeq 180^\circ$ in longitude, we plot both. L4 and L5 represent the two camps of JTs.}
\end{figure*}

Our analysis is based on full shape and spin state solutions of 90 JTs -- 79 from Sec.~\ref{sec:models} and 11 solutions adopted from DAMIT (Table~B.3), and partial models of 31 JTs (Sec.~\ref{sec:partials}). We identified 49 members of the L$_4$ cloud and 72 of the L$_5$ cloud. In what follows, we focus on the analysis of the rotation pole distribution, as this is a new result in this paper. We omit to comment on the rotation rate distribution. This is because previous works, such as
\citet{Szabo2016, Ryan2017, Kalup2021, Chang2021} covered this issue with even more data.

In the left panel of Fig.~\ref{fig:poles}, we show the pole directions for JTs. The pole distribution is consistent with being grossly isotropic, similar to the pole distribution among the large ($D>50$~km) main-belt asteroids (for models in DAMIT) that is shown in the right panel of Fig.~\ref{fig:poles} for comparison. Both populations consist of bodies that are large enough not to be significantly affected by the YORP effect, so the pole directions should primarily reflect the initial distribution in the source region and possibly modification due to the collision history (if important enough). The nonuniform distribution of poles caused by the YORP effect is evident only for main-belt asteroids with $D<30$~km \citep{Hanus2011}. Unfortunately, we do not have spin state solutions for such small ($D<30$~km) JTs. So far, the first tentative evidence of the YORP effect acting on JTs $<10-20$~km was reported by \citet{Chang2021} -- smaller JTs have a lower limit of the rotation period near $4$~h than the previously published result of $5$~h found for larger JTs. There is no obvious difference between the pole orientations within the L$_4$ and L$_5$ clouds.

Deciding the sense of rotation with respect to the orbital plane requires the knowledge of the spin obliquity $\varepsilon$. We computed the obliquity for cases with the full spin solutions. We consider the sense of rotation to be inconclusive if the obliquity is between $80^\circ$ and $100^\circ$, while obliquity $<80^{\circ}$ indicates a prograde rotator and obliquity $>100^{\circ}$ a retrograde rotator in our simple approach. We find there are more prograde rotators than retrograde ones in our sample, namely 49 versus 33. This hints at a factor of $\simeq 1.5$ between a number of prograde and retrograde populations, or $\simeq 60$\% abundance of prograde rotators in the overall JT population. We shall operationally work with this result while admitting that the hypothesis of a similar number of prograde and retrograde rotators among JTs may be valid at $\simeq 10$\% level due to the small size of the sample. For instance, if the sample is doubled, and the ratio of the prograde vs retrograde solutions remains $\simeq 1.5$, the room for the isotropy would shrink to only about $1.5$\% probability. This is a large motivation for determining more spin models among JTs. By splitting the sample at obliquity equal to $90^\circ$, we have 54 prograde vs 36 retrograde JTs, thus a similar abundance of prograde rotators. Due to a larger sample, the significance of the isotropy of the spin poles of JTs decreases to the $\simeq 7$\% level. The excess in prograde rotators is also present considering the two clouds: there are 22 prograde and 18 retrograde rotators within the L$_4$ cloud (55\% vs. 45\%), and 32 prograde and 17 retrograde rotators within the L$_5$ cloud (65\% vs. 35\%).

\begin{figure*}%[ht!]
\begin{center}
 \resizebox{1.0\hsize}{!}{\includegraphics{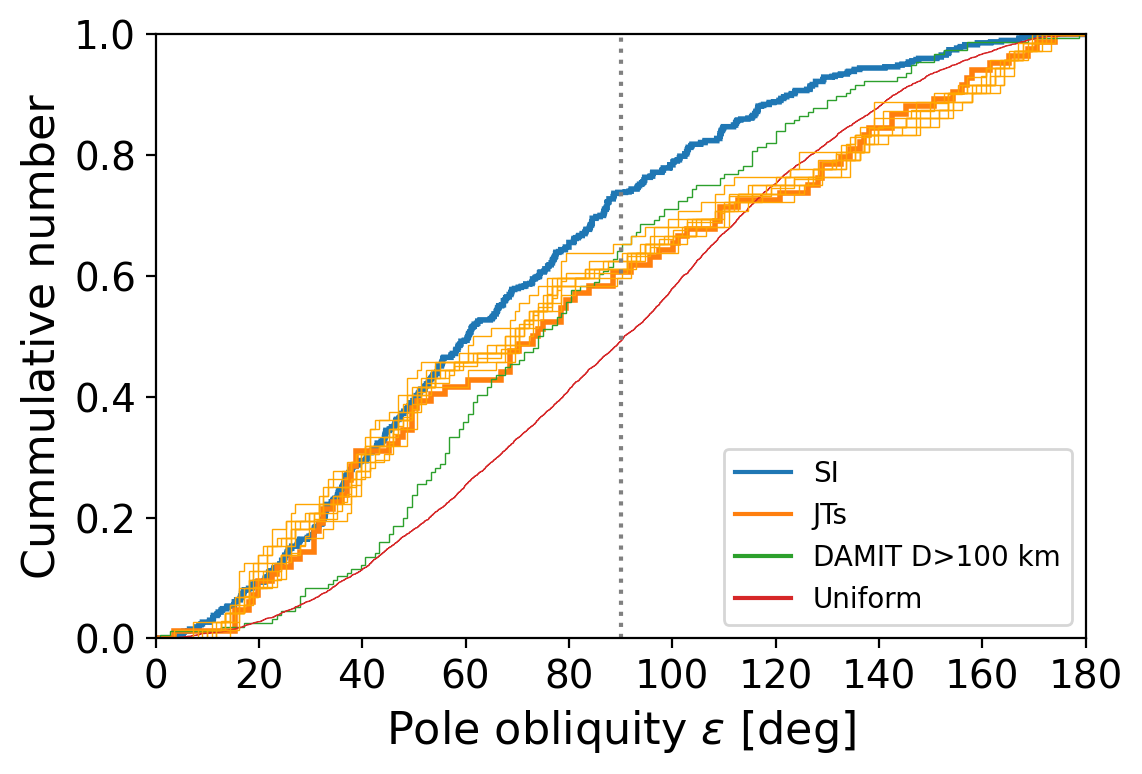}\includegraphics{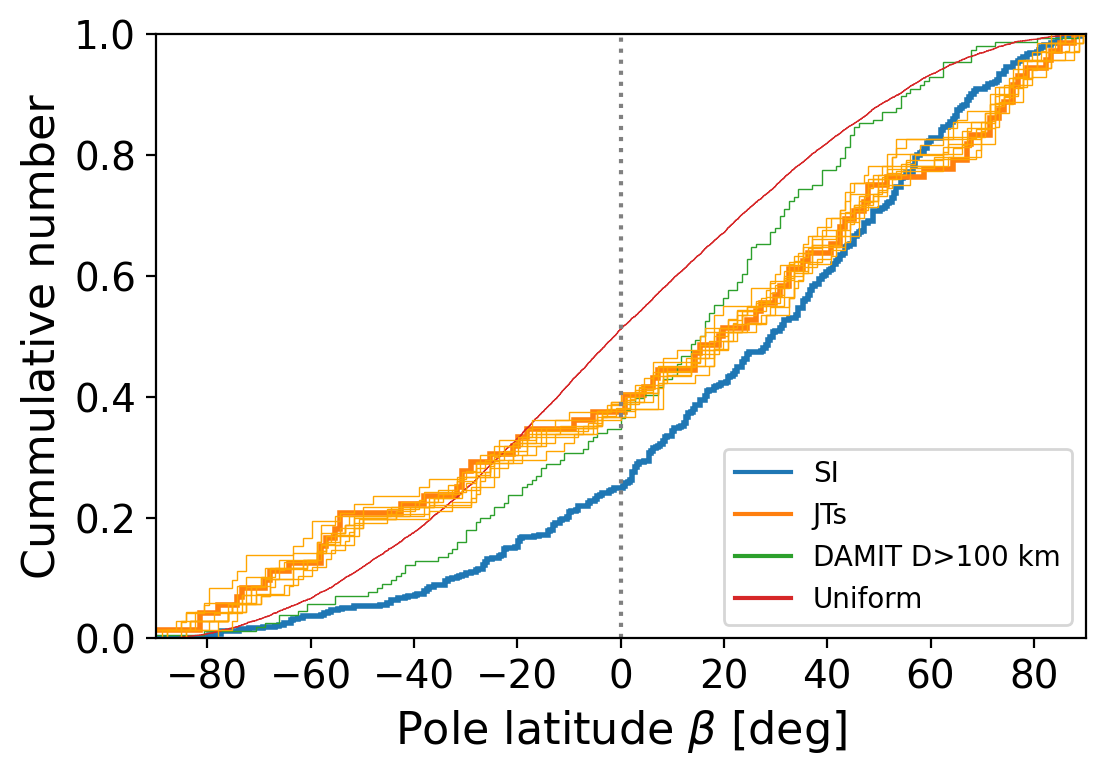}}\\
\end{center}
\caption{\label{fig:spins}Spin obliquities and pole ecliptic latitudes within various populations. Left panel: Cumulative distribution of obliquities for JTs (orange) and of planetesimals obtained in our simulations of the streaming instability (SI) model (blue). The thick orange line corresponds to the nominal spin state solutions, while the thin orange lines represent the spin states based on the bootstrapped photometric datasets. The spins in SI are predominantly prograde with $\simeq 75$\% having obliquity $\varepsilon<90^\circ$. We also plot the cumulative distributions for randomly oriented spins (red), and large MBAs (green, spin states adopted from DAMIT). Right panel: Same but now for the ecliptic latitude of the rotational pole instead of the spin obliquity.}
\end{figure*}

In Fig.~\ref{fig:spins}, we show the cumulative obliquity distribution of JTs, which provides more complete information than just the simple ratio between the number of prograde and retrograde rotators. Additionally, we also included in Fig.~\ref{fig:spins} the obliquity distributions for randomly oriented spins, and for the population of MBAs larger than $100$~km, which spins were adopted from DAMIT. Clearly, none of the populations have similar obliquity distributions and none are consistent with randomly distributed spins. In addition, the sample of all MBAs with known spins (not shown in the figures) is dominated by objects in the size range of 10--30 km, for which the YORP effect is the main driver shaping the spin vector distribution -- there is a lack of spin vectors with obliquities $\sim90^\circ$ and about the same number of prograde and retrograde rotators \citep{Hanus2011}. 
The obliquities of MBAs larger than 100~km are not affected by YORP and their initial spin distribution was modified by collisional and orbital evolution. We observe a significant excess of prograde rotators ($\sim60\%$, Fig.~\ref{fig:spins}). JTs have a similar excess of prograde rotators of $\sim60\%$ but contain a larger fraction of objects with obliquities $0^\circ$ to $50^\circ$. The cumulative obliquity distributions based on bootstrapped datasets are qualitatively similar to the nominal distribution, therefore, our conclusions are not strongly dependent on the pole uncertainties.

In what follows (Secs.~\ref{sec:streeming} and \ref{sec:dynamics}) we aim at understanding if the slight asymmetry toward the prograde sense of rotation among JTs and the overall obliquity distribution is consistent with the expectations within the JTs capture model (Sec.~\ref{sec:introduction}). At this moment, we do not extend our analysis to alternative models of the JTs' origin.

\subsection{Shape models}\label{sec:shapes}

\begin{figure*}%[ht!]
\begin{center}
  \resizebox{1.0\hsize}{!}{\includegraphics{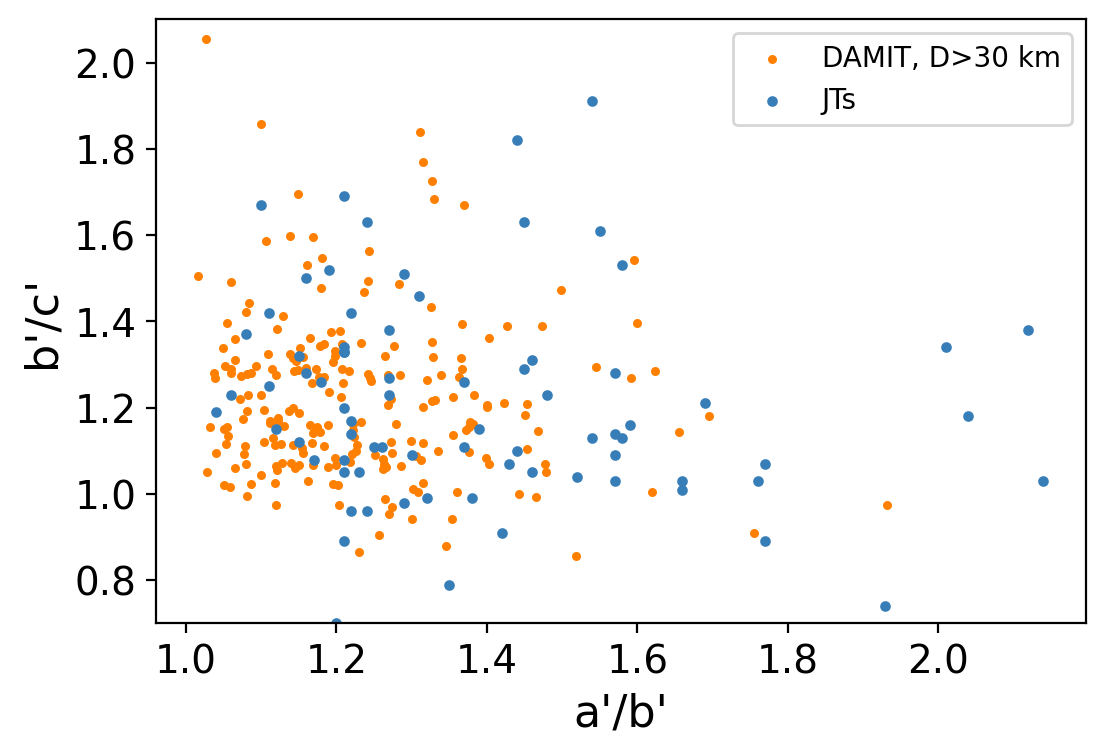}\includegraphics{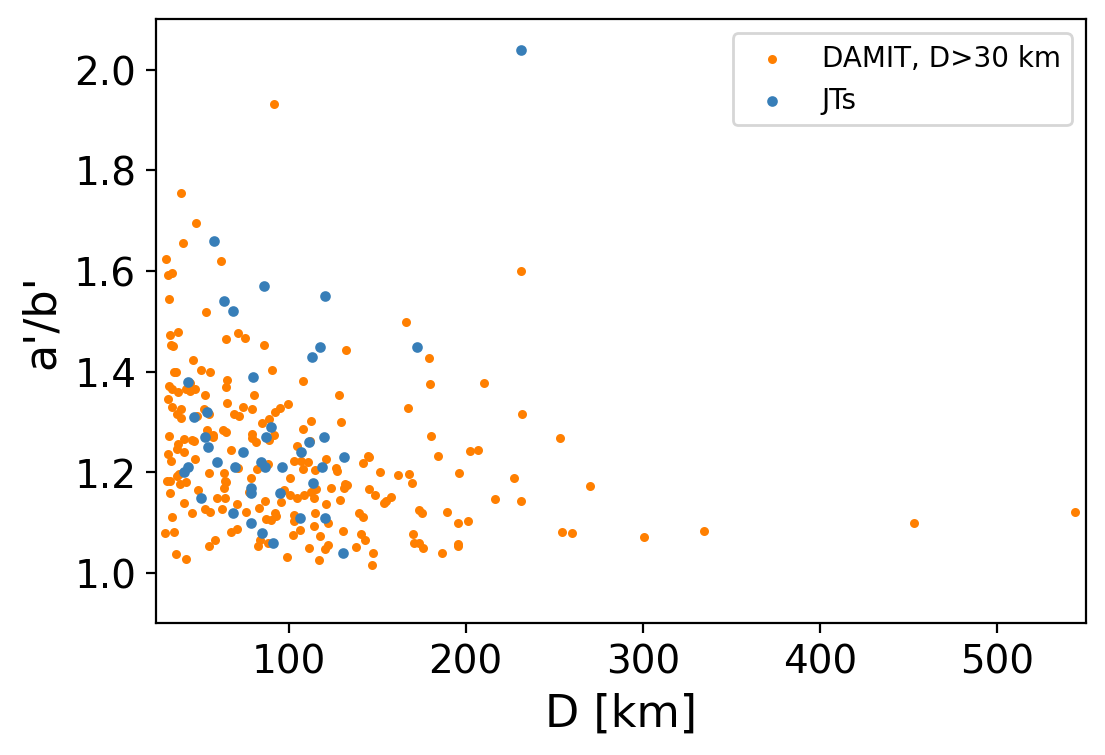}}\\
\end{center}
\caption{\label{fig:ratios}Shape properties within JTs and large MBAs. Left panel: The $a'/b'$ vs. $b'/c'$ axis ratios of JTs, and of large MBAs ($D>30$~km). Right panel: The $a'/b'$ ratio as a function of diameter $D$ for the same populations.}
\end{figure*}

We also analyze the shape models of JTs and make a comparison with the population of MBAs. The convex shape models can be best characterized by the ratios of dimensions ($a'$, $b'$, $c'$) along the main axes. We compute the relative dimensions (shape models are not scaled in size) by the "overall dimensions" method of \citet{Torppa2008} and normalize them such that $c'$=1. We note that the $c'$ dimension is usually the least accurate. In Fig.~\ref{fig:ratios}, we plot the $a'/b'$ vs. $b'/c'$ axis ratios of JTs, and of large MBAs ($D>30$~km in this case), and the dependence of $a'/b'$ ratio on the size $D$. We used the shape models of MBAs available in DAMIT and included only those having at least ten optical light curves, thus the most realistic solutions. According to the two-sample Kolmogorov-Smirnov test, the $a'/b'$ distributions are different ($p$-value=10$^{-6}$), while the $b'/c'$ distributions are similar ($p$-value=0.14). We note that the two-sample K-S test does not depend on the binning of the data.

The bootstrapping described in Sec.~\ref{sec:bootstrap} simultaneously provides shape models that we used for assessing the uncertainties of the ratios of dimensions along the main axes. For each asteroid, we obtained ten bootstrapped shape models for each pole solution. Out of 79 asteroids, only seven have relative uncertainties in $a'$ larger than 20\% and 22 larger than 10\%. Similarly, only four asteroids have relative uncertainties in $b'$ larger than 20\% and 24 larger than 10\%. This means that the JT population has a rather stable distribution of $a'/b'$ and $b'/c'$ axis ratios considering the uncertainties in the dimensions, and thus the statistical tests above should be statistically meaningful.

\subsection{Assessing the systematic uncertainties by using assumed shapes}\label{sec:systematics}

As we aim above to compare the obliquity and axial ratio distributions in a quantitative manner, we also need to understand the possible effects of systematic uncertainties in the modeling assumptions on these distributions. In Sec.~\ref{sec:bootstrap}, we test the sensitivity of the shape models and spin states on the photometric datasets (by bootstrapping the optical light curves), but under the same model assumptions. Therefore, these uncertainties represent only a lower bound of the total uncertainties. We admit that fully accessing the total uncertainties in the shape models and rotation state parameters requires a robust synthetic study, where various sources of systematic errors should be investigated (noise in data, observing geometries, shapes, spin-state distributions, amount of data, modeling assumptions, inversion method, etc.). This is largely beyond the scope of this work and we leave it for future efforts. However, some selection effects were already quantified in our previous study through a small-scale synthetic study \citep{Hanus2011} and the mathematical stability of the convex inversion method for well-formed data sets is well manifested in the literature \citep[e.g.,][]{Kaasalainen2001a, Kaasalainen2001b}.

Here, we decided to assess whether the original shape itself affects the pole-latitude distribution and how well the axis ratios are preserved. Therefore, we performed the following exercise. We assumed that each asteroid in our final sample of solutions has an assumed fixed-shape model -- we selected the nonconvex shape models of asteroids (433)~Eros (PDS, 433 Eros Plate Model MSI 1708) and (3)~Juno \citep{Viikinkoski2015}. For each asteroid, we used its derived spin state (Table~B.2) and the assumed "fake" shape model to generate the new set of optical light curves corresponding to the original epochs. We added noise to each light curve according to its original rms. Then, we derived the spin state and the shape model based on the generated light curve dataset. Finally, we constructed in Fig.~\ref{fig:ratios_systematics} the cumulative obliquity distributions and the axis ratios for each "fake-shape" dataset to assess their stability.

The shapes of Eros and Juno represent substantially different bodies. While Eros is highly elongated and largely concave, Juno is rather round without large concavities. Thus, our exercise covers two end members of the asteroid population.  However, we note that JTs should have shapes qualitatively more similar to Juno than Eros. In both cases, the cumulative distribution of obliquities is similar to that of the JTs from Sec.~\ref{sec:phys}. This means that the spin vectors were preserved almost independently on the assumed shape model. Only, in the case of Juno, fewer unique solutions were derived, likely because the original models were more elongated, thus with larger imprints in the light curves. This smaller Juno sample still has a similar cumulative obliquity distribution to that of JTs, thus the missing solutions do not have preferential obliquities. The axis ratios of Juno and Eros synthetic JT populations are well preserved, and most of the solutions cluster within 10\% in $a'/b'$ and $b'/c'$ (Fig.~\ref{fig:ratios_systematics}). Such uncertainties are comparable to those assessed by the bootstrapping of the optical data in Sec.~\ref{sec:bootstrap}. We note that the dimensions of the original concave shape of Eros lie outside the Eros JTs cluster having higher $b'/c'$ and borderline $a'/b'$. The main reasons are the peculiar shape of Eros and its qualitative difference from the convex approximation. The main takeaway from this analysis is that (i)~the shape dimensions are relatively well preserved, and (ii)~ there is no indication of overestimated $a'/b'$ ratios (the population of JTs has larger $a'/b'$ ratios than the larger MBAs ($D>30$~km), probably not due to the above-analyzed systematic uncertainty).

%%%%%%%%%%%%%%%%%%%%%%%%%%%%%%%%%%%%%%%%%%%% Streaming instability %%%%%%%%%%%%%%%%%%%%%%%%%%%%%%%%%%%%%%%%%%%%%%%%%%%%%%%%%%%%
\section{Spin vectors of planetesimals in the streaming instability model}\label{sec:streeming}

In this Section, we return to the issue of the possible asymmetry between a prograde and retrograde sense of JTs rotation. We first determine what obliquity distribution is expected among the planetesimals born in the trans-Neptunian disk prior to their capture into the JT population. We adopt the currently leading model of planetesimal formation mechanism, namely the streaming instability (SI) in a cold proto-planetary disk.

Streaming instability is a mechanism to aerodynamically concentrate small particles in a proto-planetary  disk \citep{yg2005}. Here we analyze the results of 3D simulations of the streaming instability reported in \citet{netal2019,netal2021}. The simulations accounted for the hydrodynamic flow of gas, aerodynamic forces on particles, backreaction of particles on the gas flow, and particle self-gravity. They were performed with the ATHENA code \citep{bs2010}. The obliquity distribution of particle clumps in the ATHENA run A12 was reported in \citet{netal2019}. It shows an 80\% preference for prograde rotation, which matches the observed distribution of binaries in the Kuiper belt. We performed additional integrations for selected clumps, where we followed their gravitational collapse into individual planetesimals, to understand the implications for the spin states of JTs.

Gravitational collapse of particle clouds was followed with a modified version of the $N$-body cosmological code PKDGRAV \citep{s2001}, described in \citet{retal2000}. PKDGRAV is a scalable, parallel tree code that is the fastest code available to us for the proposed simulations. A unique feature of the code is the ability to rapidly detect and realistically treat collisions between particles. The ATHENA simulations were used to set up the initial conditions for PKDGRAV. In total, we performed simulations for ten representative clumps from A12 \citep[see][]{netal2021}. Five simulations were completed in each case, where we used slightly modified initial conditions to understand the statistical variability of the results.

All planetesimals with diameters $D>25$ km that formed in the PKDGRAV simulations were identified, 103 bodies in total. For each of them, we computed the orientation of the spin vector relative to the reference frame. Since the reference frame of PKDGRAV simulations is set such that the $X$--$Y$ plane is perpendicular to the initial angular momentum of the cloud, the colatitude of the spin vector gives us the obliquity distribution of planetesimals relative to the clump's initial angular momentum vector. Finally, we convolved the two distributions to generate predictions for the obliquity distribution of planetesimals with respect to the Solar System plane. Figure~\ref{fig:spins} shows the result.

There is a notable $\simeq 75$\% preference for prograde spin, which is similar to predictions of the streaming instability for the orbital inclinations of Kuiper belt binaries \citep[compare with Fig.~3 in][]{netal2019}. Here the preference is slightly weaker, 75\% vs. 80\%, probably because individual objects experienced stochastic growth during gravitational collapse, smearing the initially stronger preference for prograde rotation (binaries preserved it better). Indeed, we see in the PKDGRAV simulation that the spin vectors of individual objects can significantly change during the collapse. For comparison, from the analysis of light curves in this work, we find that JTs show a weaker, $\simeq 60$\% preference for prograde rotation (Sec.~\ref{sec:phys}). There are several options to reconcile these results. For example, as JTs evolved collisionally prior to their capture \citep[e.g.,][]{netal2018}, the initial distribution of obliquities could have been partially randomized. 
We do not perform any collisional simulations in this paper. The results of collisional simulations depend on many unknown parameters, such as the overall excitation of orbits in the planetesimal disk, momentum transfer in subcatastrophic collisions, or the lifetime of the disk prior to the onset of planetary instability. We thus defer quantitative modeling of these processes to future work.

Apart from the general preference for prograde spins, our numerical 
simulations of the streaming instability also predict the distribution of planetesimal obliquities that can be
compared to the observed obliquity distribution of JTs (Fig.~\ref{fig:spins}).
%from Sec.~\ref{sec:phys}. 
There are some significant differences between the two distributions.
The observed obliquity distribution is, in general, wider than the 
simulated one. It closely follows the model distribution for $\varepsilon<50^\circ$ and starts diverging from it for $\varepsilon>50^\circ$. We present a more quantitative comparison in Sec.~\ref{sec:discussion}. 

For completeness, we also analyzed the distribution of the ecliptic latitude $\beta$ of the pole for the model and the different observed populations in the right panel of Fig.~\ref{fig:spins}. In the model case, the transformation is trivial: because of the initial cold planetesimal disk, the obliquity is
simply $90^\circ-\beta$. However, the transformation is not so simple
for MBAs and JTs. This is because the orbital inclinations
extend to large values up to $\simeq 40^\circ$. For JTs, the inclinations
were excited prior to and during their capture by gravitational perturbations
from planets (and possibly other massive objects in the planetesimal
disk). It is not exactly known whether, or to what degree, the spins
of the captured JTs followed this excitation. So considering both the
obliquity and the ecliptic latitude distributions we cover both extreme
possibilities: (i) a complete binding of the spins to the orbits during
the excitation process, or (ii) spins not being affected by processes
that excited the orbital planes. While understanding this issue in detail
may again require a complex model, data shown in Fig.~\ref{fig:spins}
luckily manifest that conclusions are the same, independently of whether obliquity
or the pole latitude is used. In order to bring quantitative evidence, we tested the corresponding obliquity and ecliptic latitude samples (i.e., for JTs, and DAMIT $D>100$~km) by the standard two-sample Kolmogorov-Smirnov test. In both cases, the samples are consistent with being drawn from the same distribution. For instance, the K-S test for the two JTs samples (i.e., obliquities vs. ecliptic latitudes) gives $p$-value=0.09.

%%%%%%%%%%%%%%%%%%%%%%%%%%%%%%%%%%%%%%%%%%%%%%%%% Post-capture evolution %%%%%%%%%%%%%%%%%%%%%%%%%%%%%%%%%%%%%%%%%%%%%%%%%%%%%%%%%%%%%%%%%%%
\section{Post-capture spin evolution}\label{sec:dynamics}
%%% By DV
%\section{Post-capture spin evolution}\label{sec:dynamics}
In this section, we consider what happens to the Trojan spin state during the long
period of time after they have been captured in the coorbital state with Jupiter.
We consider dynamical, rather than collisional effects in this Section. This
is because we assume that the collisional activity among Trojan clouds after the capture
was dwarfed by a much more intense period before capture.
Given the commonly accepted point of view, the capture happened
during the early evolution of the Solar System \citep[possibly $4.4-4.5$~Gyr ago or
so; e.g.,][]{n2018}, when giant planets underwent a chaotic reconfiguration
terminated by Jupiter's inward jump to near its current orbit \citep[e.g.,][]{Nesvorny2013}.
Planetesimals roaming in the planet-crossing zone that happened to be suitably
situated with respect to the Jupiter orbit at that moment formed the early
Trojan population. The implication of this model is the marginal stability of
a fraction of the Trojan population such that over eons that followed about
$23$\% of them leaked away \citep[e.g.,][]{holt2020}, leaving even some of the
currently observed Trojans on unstable orbits ready to escape from the population in the
close future (Sect.~\ref{unst}). Even some of the Trojans residing currently
stable orbits (that will stay in the population for another gigayear), might have
underwent orbital evolution since their origin due to subtle chaoticity in
their orbital phase space. As a result, it is not our aim to describe in detail the past
gigayear-lasting evolution of spin state even for our limited sample of 90 Trojans,
for which we know it today, as this is a task that is too difficult.

Instead, we probe the shorter-term evolution of the Trojan spin state and consider it
an operational proxy of what happens on longer timescales. In particular, we
propagate, using the model described in Sect.~\ref{mode}, the spin states of
our sample of Trojans for $50$~Myr forward in time. Note this is information equivalent
to the propagation backward in time. Given the main arguments in this paper
(Sec.~\ref{sec:phys}), we are principally interested
to know whether there is a substantial evolutionary flow of Trojan spins between
the prograde- and retrograde-rotating groups. Finding there is little of such
interchange on the monitored $50$~Myr time interval, we dare to extrapolate the
conclusion to gigayear-long evolution.

\subsection{Dynamical model}\label{mode}
Assume the body rotates about the axis of the largest inertia tensor. If no 
torques are applied, the rotational angular momentum ${\bf L}$ is conserved. This 
implies that both (i) the angular rotation rate $\omega$, and (ii) the unit vector 
${\bf s}$ of the rotation pole in the inertial frame are also constant. However,
if we aim to know what happens to the Trojan rotation state over a long period of time,
we need to include relevant torques of which the most important are due to the
gravitational effects of the Sun and planets \citep[the radiation-related YORP torques
would have been important only for timespan exceeding $\simeq 5$~Gyr for $D<40$~km 
Trojans, e.g.,][]{vetal15}. Each of the perturbing bodies may be represented by 
a point source of mass $M$ and its gravitational field expanded in the local
frame of the Trojan to the quadrupole level. Assuming the origin of the local frame
coincides with the center-of-mass of the Trojan, the rotation-averaged gravitational
torques may be expressed in simple terms as \citep[e.g.,][ Chap.~4]{c1966,bfv2003}
\begin{equation}
 {\bf T} = - \frac{3GM}{r^5}\left[C-\frac{1}{2}\left(A+B\right)\right]\left({\bf s}
  \cdot {\bf r}\right)\left({\bf s}\times {\bf r}\right)\; , \label{tor}
\end{equation}
where $G$ is the gravitational constant, ${\bf r}$ position vector of the Trojan
with respect to the source ($r=|{\bf r}|$), and $(A\leq B\leq C)$ are the principal moments
of the inertia tensor. Because ${\bf L}=C\omega\, {\bf s}$ in the principal axis
rotation state, the Euler equation $d{\bf L}/dt={\bf T}$ implies that $\omega$ is
conserved in this approximation. However the pole direction ${\bf s}$ evolves
according to
\begin{equation}
 \frac{d{\bf s}}{dt} = - \frac{3GM}{r^5}\frac{\Delta}{\omega}\left({\bf s}\cdot
  {\bf r}\right) \left({\bf s}\times {\bf r}\right)\; , \label{eu1}
\end{equation}
where $\Delta=\left[C-\frac{1}{2}\left(A+B\right)\right]/C$ is the dynamical flattening.
For nearly spherical objects $\Delta$ is very small, and it reaches maximum values
near to $0.5$ for highly irregular-shaped bodies. Contributions from different massive
bodies in the Solar System linearly superpose on the righthand side of Eq.~(\ref{eu1}).

In our most complete simulations, we included the Sun and Jupiter as sources of the
gravitational torques. In this approach, we numerically integrated Eq.~(\ref{eu1}) using
a simple Euler scheme and a short timestep of $3$~days. The value of the rotation frequency
$\omega$ was obtained from our solution of the rotation period, and the dynamical
flattening $\Delta$ from the nominal shape of the body and assumption of a homogeneous
density distribution \citep[see, e.g.,][]{Dobrovolskis1996}. We embedded this unit
into a general-purpose orbit-propagation package {\tt swift\_rmvs4}, which provided
the position vectors ${\bf r}$ for each of the sources (and used the same timestep of
$3$~days for the orbit integration). All planets, and the Sun, were propagated using 
{\tt swift\_rmvs4}\footnote{\url{http://www.boulder.swri.edu/~hal/swift.html}}, together with our sample of Trojans for which rotation states
were inferred in Sect.~\ref{sec:modeling}. The planetary and Trojan heliocentric position vectors
were integrated in a global reference frame defined by the invariable plane of the
Solar System. For that reason, we also transformed Trojan pole directions ${\bf s}$
into the same system at the beginning of the integration. The corresponding tilt is, however,
very small, as the ecliptic plane has an inclination of only $\simeq 1.5^\circ$
to the invariable plane. Therefore the pole latitudes differed in the two systems
at maximum by this value. We performed the simulations forward in time over
$50$~Myr interval, over which most of the orbits were stable (see though Sec.~\ref{unst}
for more comments).

In our first set of integrations, we included the gravitational torque due to Jupiter's
gravity on the righthand side
of Eq.~(\ref{eu1}) for the sake of completeness and also for its special status with
respect to the Trojan clouds. However, we found that its effect is minimal over the
timescale of $50$~Myr we considered (a result that likely would not change even
over a longer timescale). This is because the Solar torque is much
more significant owing to its much larger mass (in spite of the fact that some Trojans with large
libration amplitude in our sample may approach Jupiter at a distance 
that is slightly more than half of the Solar distance). We thus find that adequate
results can be obtained when considering the Solar gravitational torque only on the
righthand side of Eq.~(\ref{eu1}). This model has further important advantages.

First, we observe that the secular changes of ${\bf s}$ have periods comparable
with that of the precession of the Trojan orbital plane rather than a much shorter revolution
period about the Sun. It is thus possible, and convenient, to further average the
torque (\ref{tor}) over the revolution cycle about the Sun. Assuming osculating
elliptical orbit, this may be performed analytically using
\begin{equation}
 \left\langle \frac{{\bf r}{\bf r}}{r^5} \right\rangle = \frac{1}{2b^3} \left({\bf E} 
  - {\bf n}{\bf n}\right) \; , \label{aver}
\end{equation}
where $b=a\sqrt{1-e^2}$ is the semiminor axis of the elliptic orbit ($a$ and $e$ being
the respective semimajor axis and eccentricity), ${\bf E}$ is the unitary $3\times 3$
matrix and ${\bf n}$ is the unit vector normal to the osculating orbital plane in
the direction of the orbital angular momentum. Therefore, ${\bf n}^T=(\sin I\sin\Omega,-
\sin I\cos\Omega,\cos I)$, where $I$ is the orbit inclination to the reference plane and
$\Omega$ is the longitude of the node. As a result, Eq.~(\ref{eu1}) now takes the form
\citep[e.g.,][]{c1966}
\begin{equation}
 \frac{d{\bf s}}{dt} = - \alpha \left({\bf s}\cdot {\bf n}\right) \left({\bf s}\times {\bf n}
  \right)\; ,\label{eu2}
\end{equation}
where
\begin{equation}
 \alpha = \frac{3}{2} \frac{GM}{b^3}\frac{\Delta}{\omega}\; \label{pc}
\end{equation}
is the precession constant. The fundamental difference between (\ref{eu1}) and (\ref{eu2})
is that none of the parameters on the righthand side of the latter changes on the
timescale of Trojan's orbital revolution about the Sun (neglecting very small short-period
changes in the semiminor axis $b$ in the definition of the precession constant $\alpha$).
Therefore, Eq.~(\ref{eu2}) may be numerically integrated with a much larger timestep,
obtaining results quite more efficiently. In our simulations, we used a $50$~yr timestep,
instead of the $3$~day timestep needed for integration of Eq.~(\ref{eu1}). Additionally,
\citet{bnv2005} developed an efficient symplectic scheme for the propagation of (\ref{eu2}),
which conveniently conserves its first integrals. We implemented their LP2 scheme. The
knowledge of the orbital evolution is still needed, as the normal vector ${\bf n}$
changes its direction in the inertial space due to the orbit precession and inclination
variations (that is, both $I$ and $\Omega$ are time-dependent due to planetary perturbations).
In order to fully describe both, we still use the orbital integration by
{\tt swift\_rmvs4} package and include the Trojan spin integrator of (\ref{eu2}) as a
separate subunit of the code.

Second, a great advantage of using the second model described by Eq.~(\ref{eu2})
consists of the fact that its
solutions have been thoroughly studied since \citet{c1966}. This past and well-known
analysis provides a key insight into why some of the Trojan pole solutions exhibit only
very small variations of the pole latitude, while in other cases large variations of
the latitude are possible. Importantly, while many perturbations, reflecting planetary
orbits and their own evolution, participate in the time evolution of Trojan orbital
${\bf n}$ vector, not all are equally prominent. In fact, the evolution of ${\bf n}$ is
nearly always dominated by a contribution from a single, proper term. In this case,
the inclination $I$ is constant and the longitude of the node exhibits a steady precession
with the proper frequency $s$ (thus $\Omega=s\,t + \Omega_0$). In this simplified
model, called the Colombo top \citep[e.g.,][]{c1966,hm1987}, the variety of solutions
of Eq.~(\ref{eu2}) depends on three parameters: (i) the two frequencies $\alpha$ and
$s$, and (ii) the orbit inclination $I$. The complexity stems from the possibility
that $\alpha$ and $s$ enter in a resonance. As a result, the flow of ${\bf s}$ on
a unit sphere has two distinct regimes according to the ratio $\kappa=|\alpha/s|$
(our notation adopted here, recalling that $\alpha$ is positive and $s$ negative,
may differ from the signature in other references). When,
$\kappa<\kappa_\star$, where $\kappa_\star=\left(\sin^{2/3} I + \cos^{2/3} I\right)^{3/2}$,
${\bf s}$ exhibits a simple circulation about two fixed points called Cassini state~2
and 3. In the limit of very small $\kappa$ value, these stationary points have obliquity
of $I$ and $90^\circ-I$. The spin vector ${\bf s}$ thus circulates about the north or south
poles of the invariable frame in the inertial system, and the rotation pole
latitude has only small variations. A more complex regime onsets when $\kappa>
\kappa_\star$ (note that $\kappa_\star$ ranges between the value $1$ at very small
inclination and reaches a maximum of $2$ when $I=45^\circ$). In this situation, the flow
of ${\bf s}$ navigates between four fixed points, called Cassini state~1 to 4. Cassini
states~2 and 4 are stable and unstable stationary points of a resonant zone. The
appearance of this zone triggers large obliquity oscillations, which are then reflected
by large oscillations of the rotation pole latitude. Transitions between the
prograde and retrograde rotation regimes are in principle possible. Mathematical
details of the Colombo top problem may be found in a number of publications
\citep[e.g.,][]{c1966,hm1987,hap2020}.

We need to decide on Which of the two regimes is to be expected more typical for Trojan spin evolution. 
For small eccentricities, we may adopt $b\simeq a\simeq 5.2$~au in the Trojan region,
and thus $\alpha\simeq 9.4\,\Delta\,(P/{\rm 6\, hr})$ arcsec yr$^{-1}$ (see Eq.~\ref{pc}).
Additionally, the population of large Trojans we are interested in has rather regular
shapes, typically with $\Delta\simeq 0.1-0.2$. As a result, the precession constant values
are typically $\sim$arcsec yr$^{-1}$ unless the rotation period $P$ is long
(several tens of hours or so). In the same time, the proper frequencies $s$ have values
typically in the range $-5$ to $-30$ arcsec yr$^{-1}$, and only exceptionally have
a value smaller \citep[e.g.,][]{m1993}. As a result, the situation we should encounter most
often among Trojans is the limit of very small $\kappa$, implying thus only small variations
of the rotation pole latitude. Exceptions are to be expected when the rotation period
$P$ is very long, and/or proper frequency $s$ is anomalously small. 
%%%%%%%%%%%%%%%%%%%%%%%%%%%%%%%%%%%%%%%%%%%%%%%%%%%%%%%%%%%%%%%%%%%%%%%%%%%%%%%%%%%%
\begin{table*}[tpb]
\caption{\label{table_L4}
 Limits of oscillations of the rotation pole latitude for L$_4$ Trojans. The first two
 columns provide the denomination of the Trojan. The next three columns apply to the
 first pole solution (P1), followed by three columns for the second pole solution (P2);
 if the pole solution is unique, the P2 columns are empty. We note that $b_0$ is the nominal pole
 latitude of the current epoch in the invariable frame of the Solar System. Furthermore, $b_{\rm min}$
 and $b_{\rm max}$ are the minimum and maximum latitude values in the next $50$~Myr from
 our secular model. The last column gives the proper frequency $s$ (in arcsec per year)
 of the orbital plane precession from {\tt AstDyS}.}
 \vspace*{3mm} \centering
\begin{tabular}{rlrrrrrrr}
\hline \hline
%\rule{0pt}{2ex}
\multicolumn{2}{c}{Body} & \multicolumn{1}{c}{$b_0$} & \multicolumn{1}{c}{$b_{\rm min}$} & \multicolumn{1}{c}{$b_{\rm max}$} & \multicolumn{1}{c}{$b_0$} & \multicolumn{1}{c}{$b_{\rm min}$} & \multicolumn{1}{c}{$b_{\rm max}$} & \multicolumn{1}{c}{$s$} \\
\hline
  588 & Achilles    & --6.42 & --6.50 & --5.71 &         &        &        &--11.03 \\ 
  624 & Hektor      &--23.91 &--25.18 &--20.68 &         &        &        &--12.89 \\ 
  659 & Nestor      &--73.09 &--76.37 &--71.91 &         &        &        &--16.91 \\ 
  911 & Agamemnon   &  34.57 &  26.09 &  38.12 &         &        &        & --8.58 \\ 
 1143 & Odysseus    &--53.05 &--53.68 &--52.42 &--50.16  &--50.84 &--49.61 &--10.91 \\ 
 1404 & Ajax        &  42.16 &  28.54 &  50.36 &         &        &        &--14.77 \\ 
 1437 & Diomedes    &   3.81 &   3.16 &   7.96 &  13.42  &  12.24 &  21.84 &--20.14 \\ 
 1868 & Thersites   &  43.78 &  35.91 &  44.62 &         &        &        &--20.37 \\ 
 2797 & Teucer      &--30.30 &--30.73 &--26.28 & --2.74  & --3.49 & --2.32 &--17.00 \\ 
 2920 & Automedon   & --4.89 & --5.30 & --3.76 &         &        &        &--12.71 \\ 
 3391 & Sinon       &  88.03 &  72.82 &  89.99 &         &        &        &--14.35 \\ 
 3564 & Talthybius  &--58.20 &--61.09 &--46.17 &--69.12  &--88.24 &--66.07 & --9.64 \\ 
 3709 & Polypoites  &   0.11 & --0.55 &   1.69 &  24.84  &  23.57 &  37.32 & --7.15 \\ 
 4063 & Euforbo     &  75.52 &  60.07 &  80.88 &  40.77  &  33.83 &  41.70 & --8.49 \\ 
 4068 & Menestheus  &--55.89 &--57.40 &--48.21 &--23.90  &--28.27 &--23.15 &--13.46 \\ 
 4086 & Podalirius  &  17.27 &  14.42 &  19.58 &  42.39  &  38.34 &  58.97 &--15.15 \\ 
 4489 &--           &  28.10 &  25.58 &  33.41 &  11.03  &   9.46 &  18.26 &--13.66 \\ 
 4543 & Phoinix     &--55.43 &--60.03 &--51.80 &--43.87  &--49.13 &--41.47 &--22.51 \\ 
 4834 & Thoas       &  33.56 &  26.18 &  89.99 &         &        &        & --4.58 \\ 
 4836 & Medon       &--57.41 &--60.43 &--49.96 &--23.83  &--28.57 &--22.07 &--10.47 \\ 
 5027 & Androgeos   &  32.68 &  23.01 &  48.21 &  33.57  &  21.71 &  48.52 & --7.44 \\ 
 5209 &--           &  76.50 &  72.56 &  80.86 &  65.61  &  57.94 &  66.54 &--20.19 \\ 
 5244 & Amphilochos &  78.30 &  77.97 &  85.94 &  64.05  &  52.04 &  64.46 &--11.39 \\ 
 5283 & Pyrrhus     &--10.89 &--11.08 & --9.50 &  20.89  &  20.39 &  25.52 & --8.35 \\ 
 5285 & Krethon     &--74.04 &--84.64 &--67.58 &         &        &        &--14.46 \\ 
 5436 & Eumelos     &  21.57 &  18.56 &  32.45 &  11.31  &   9.32 &  14.10 & --7.89 \\ 
 9694 & Lycomedes   &  53.30 &  52.23 &  54.91 &  53.90  &  52.55 &  55.03 &--29.94 \\ 
11429 & Demodokus   & --3.26 & --6.10 & --2.79 &--37.48  &--39.65 &--24.50 &--16.34 \\ 
13060 &--           &--26.76 &--89.26 &--22.21 &         &        &        & --9.76 \\ 
13229 & Echion      &--62.36 &--63.27 &--62.00 &--83.73  &--84.19 &--82.38 &--18.72 \\ 
13372 &--           &  42.88 &  33.27 &  69.40 &  47.46  &  32.09 &  76.28 & --9.53 \\ 
14268 &--           &--38.70 &--39.75 &--37.68 &--29.54  &--30.41 &--28.79 &--14.90 \\ 
15436 &--           &   2.57 &   2.25 &   4.30 &         &        &        & --4.50 \\ 
15527 &--           &  86.94 &  80.06 &  90.00 &  73.63  &  61.47 &  74.52 &--12.34 \\ 
15663 & Periphas    &--19.85 &--62.46 &--17.68 &         &        &        & --1.61 \\ 
18062 &--           &  20.36 &  14.95 &  21.24 &         &        &        & --8.86 \\ 
18263 & Anchialos   &--79.62 &--84.32 &--79.22 &--47.11  &--50.96 &--46.40 &--21.13 \\ 
23135 & Pheidas     &  31.78 &  31.36 &  34.87 & --4.76  & --5.01 & --4.29 &--15.81 \\ 
24485 &--           &--77.78 &--79.86 &--69.16 &--66.96  &--76.40 &--65.37 &--11.69 \\ 
25911 &--           &  28.21 &  27.82 &  35.38 &         &        &        &--12.14 \\
\hline
\end{tabular}
\end{table*}
%%%%%%%%%%%%%%%%%%%%%%%%%%%%%%%%%%%%%%%%%%%%%%%%%%%%%%%%%%%%%%%%%%%%%%%%%%%%%%%%%%%%
%%%%%%%%%%%%%%%%%%%%%%%%%%%%%%%%%%%%%%%%%%%%%%%%%%%%%%%%%%%%%%%%%%%%%%%%%%%%%%%%%%%%
\begin{table*}[tpb]
\caption{\label{table_L5}
 Same as Table~\ref{table_L4} but now for L5 Trojans. Only 5144 Achates, whose
 orbit was exhibiting stable chaos over a period of 50~Myr, was unsuitable for the
 accurate determination of the proper frequency $s$.}
 \vspace*{3mm} \centering
\begin{tabular}{rlrrrrrrr}
\hline \hline
%\rule{0pt}{2ex}
\multicolumn{2}{c}{Body} & \multicolumn{1}{c}{$b_0$}  & \multicolumn{1}{c}{$b_{\rm min}$} & \multicolumn{1}{c}{$b_{\rm max}$} & \multicolumn{1}{c}{$b_0$}  & \multicolumn{1}{c}{$b_{\rm min}$} & \multicolumn{1}{c}{$b_{\rm max}$} & \multicolumn{1}{c}{$s$} \\
 \hline
   884 & Priamus    &--29.99 &--31.75 & --29.23 &--46.89 &--47.80 &--44.73 &--11.82 \\  
  1173 & Anchises   &--55.87 &--56.02 & --52.66 &        &        &        &--25.04 \\  
  1208 & Troilus    &--30.09 &--59.40 &    6.51 &--10.53 &--53.78 &   9.49 & --0.41 \\  
  1867 & Deiphobus  &  80.07 &--14.15 &   85.11 &        &        &        & --6.23 \\  
  1870 & Glaukos    &  14.80 &  13.89 &   17.64 &  26.17 &  26.09 &  28.45 & --9.48 \\  
  1873 & Agenor     &  37.38 &  34.43 &   55.38 &   6.49 &   4.18 &  11.69 &--10.38 \\  
  2207 & Antenor    &--18.81 &--18.89 & --18.26 &        &        &        &--15.69 \\  
  2241 & Alcathous  &  67.54 &  64.34 &   69.01 &  48.59 &  46.41 &  49.43 &--11.63 \\  
  2363 & Cebriones  &--37.76 &--38.51 &  --8.07 &        &        &        & --4.06 \\  
  2893 & Peiroos    &  72.73 &  63.41 &   75.17 &        &        &        &--12.90 \\  
  2895 & Memnon     &   8.25 &   8.20 &   27.04 &        &        &        & --3.63 \\  
  3240 & Laocoon    &  72.77 &  71.91 &   77.27 &        &        &        &--12.91 \\  
  3317 & Paris      &  35.98 &  24.44 &   50.76 &  30.91 &  21.69 &  40.63 & --3.33 \\  
  3451 & Mentor     &  15.90 &  11.98 &   16.57 &        &        &        &--16.29 \\  
  4348 & Poulydamas &  47.69 &  41.30 &   49.88 &        &        &        &--11.47 \\  
  4707 & Khryses    &  75.52 &  72.13 &   78.08 &  70.24 &  68.02 &  71.86 &--12.87 \\  
  4709 & Ennomos    &  74.02 &  58.28 &   75.12 &        &        &        & --6.77 \\  
  4715 &--          &  56.15 &  48.80 &   61.35 &        &        &        &--13.85 \\  
  4722 & Agelaos    &   8.25 &   6.27 &   10.00 &  23.96 &  23.62 &  28.16 &--11.28 \\  
  4792 & Lykaon     &  45.33 &  25.13 &   46.37 &  66.11 &  58.90 &  69.29 &--12.01 \\  
  4828 & Misenus    &  84.11 &  69.86 &   86.81 &  53.35 &  45.87 &  53.84 &--11.05 \\  
  5130 & Ilioneus   &--25.46 &--28.87 & --24.97 &--25.46 &--28.87 &--24.97 &--16.14 \\   
  5144 & Achates    &--30.96 &--31.49 & --30.50 &        &        &        &        \\    
 11089 &--          &--66.62 &--66.84 & --64.86 &        &        &        &--22.82 \\   
 15502 &--          &--68.83 &--69.24 & --59.08 &--30.14 &--35.93 &--29.96 &--14.58 \\  
 16428 &--          &  66.17 &  53.91 &   68.32 &  55.81 &  32.93 &  56.46 &--11.42 \\  
 16560 & Daitor     &  60.32 &  56.41 &   61.07 &  72.00 &  62.34 &  73.58 &--25.72 \\  
 17314 & Aisakos    &--80.47 &--80.89 & --77.67 &--56.88 &--60.36 &--56.55 &--13.10 \\  
 17365 &--          &  69.13 &  56.88 &   83.25 &        &        &        &--11.25 \\  
 17414 &--          &--77.48 &--89.98 & --71.12 &        &        &        &--12.20 \\  
 23549 & Epicles    &--88.24 &--89.97 & --73.36 &        &        &        &--12.67 \\  
 23694 &--          & --6.84 & --7.29 &  --2.48 &   3.23 &   1.20 &  53.73 &--18.67 \\  
 24471 &--          &  75.38 &  15.98 &   89.92 &        &        &        &--12.03 \\  
 30705 & Idaios     &  47.35 &  31.72 &   54.99 &  55.18 &  44.82 &  65.94 &--10.34 \\  
 31342 &--          &  45.92 &  31.22 &   81.78 &  26.05 &  14.40 &  43.35 & --2.99 \\  
 31344 & Agathon    &  52.05 &  51.68 &   69.57 &  36.05 &  28.27 &  36.53 &--28.28 \\  
 31819 &--          &  48.93 &  48.73 &   86.32 &  22.85 &  13.75 &  22.95 &--20.12 \\  
 32339 &--          &  33.68 &  25.50 &   33.71 &        &        &        &--11.90 \\  
 32615 &--          &  43.10 &  41.24 &   43.87 &  85.93 &  81.89 &  89.05 &--26.03 \\  
 32811 & Apisaon    &  20.19 &  13.40 &   20.91 &  60.28 &  56.90 &  85.48 &--15.43 \\  
 34746 & Thoon      &  23.88 &  21.56 &   63.91 &   2.24 & --0.15 &  11.17 & --7.67 \\  
 34835 &--          &  50.56 &  50.19 &   51.85 &  45.44 &  44.06 &  47.52 &--30.12 \\  
 51364 &--          &  67.23 &  42.39 &   81.49 &  53.31 &  39.11 &  61.85 &--19.92 \\  
 51984 &--          &--83.34 &--89.81 & --64.69 &--78.23 &--87.11 &--77.89 &--11.35 \\  
 55474 &--          &  71.90 &  42.45 &   75.05 &  41.22 &  14.79 &  43.43 &--12.85 \\  
 58931 &--          &--81.71 &--82.66 & --72.87 &--53.47 &--62.09 &--52.65 &--15.64 \\  
 63923 &--          &--57.13 &--58.10 & --54.52 &--41.41 &--43.66 &--40.78 &--16.94 \\  
 76867 &--          &--43.98 &--61.86 & --30.57 &--31.12 &--50.57 &--23.29 & --3.05 \\  
 99943 &--          &--17.70 &--21.01 & --13.45 &--12.91 &--13.90 & --9.58 &--25.24 \\  
124729 &--          &  47.44 &  29.98 &   48.34 &  53.32 &  52.92 &  79.85 &--12.23 \\   
\hline
\end{tabular}
\end{table*}
%%%%%%%%%%%%%%%%%%%%%%%%%%%%%%%%%%%%%%%%%%%%%%%%%%%%%%%%%%%%%%%%%%%%%%%%%%%%%%%%%%%%

Results from our simulations confirm these conclusions. Tables~\ref{table_L4} and
\ref{table_L5} provide results from simulations where we propagated all our pole
solutions from Sec.~\ref{sec:modeling} for $50$~Myr interval to the future. We used
the most complete
torque model, including both the Sun and the Jupiter effects in Eq.~(\ref{eu1}),
although the results from the secular model using Eq.~(\ref{eu2}) are nearly identical.
We were mainly interested in the behavior of the rotation pole latitude, as this parameter
helps us classify Trojans into prograde- and retrograde-rotating groups. The columns
$b_0$ give the initial latitude in the invariable-frame reference system. For each
body we propagate both pole solutions P1 and P2; only in cases, for which the pole solution
is unique, we have just the P1 data. The columns denoted $b_{\rm min}$ and $b_{\rm max}$
give minimum and maximum latitude values attained over the monitored $50$~Myr time
interval. In most of the cases, the latitude variations are very small, reflecting
a large separation between $\alpha$ and $s$ frequencies. The rotation of these bodies is
well represented by the $\kappa\ll \kappa_\star$ solutions of the Colombo top. Importantly,
this majority of Trojans thus remain safely within their group, either prograde or
retrograde rotators, and do  not confuse our conclusions by possible transitions
between them. As anticipated, in a few cases we observe a larger range of variations in the
pole latitude. Here, $\kappa\simeq \kappa_\star$ or even $\kappa>\kappa_\star$ for either
of the two reasons mentioned above.
The pole solution may still remain in its category, but in exceptional cases, it may
even temporarily transition between prograde- and retrograde-rotating categories (in
Sec.~\ref{col}, we illustrate three of such cases in some detail). The overall balance of 
this flow is excepted to be directed toward retrograde states, helping slightly to solve 
the difference between the observed JT spin states and the modeled original spins 
(Sec.~\ref{sec:streeming}). This is because the Cassini resonance is predominantly 
located in the prograde rotation zone.
% FIG 1 %%%%%%%%%%%%%%%%%%%%%%%%%%%%%%%%%%%%%%%%%%%%%%%%%%%%%%%%%%%%%%%%%%%%%%%%%%%%%%%%%%%%%%%%%%%%%%%
\begin{figure*}[tb]
 \includegraphics[width=0.48\textwidth]{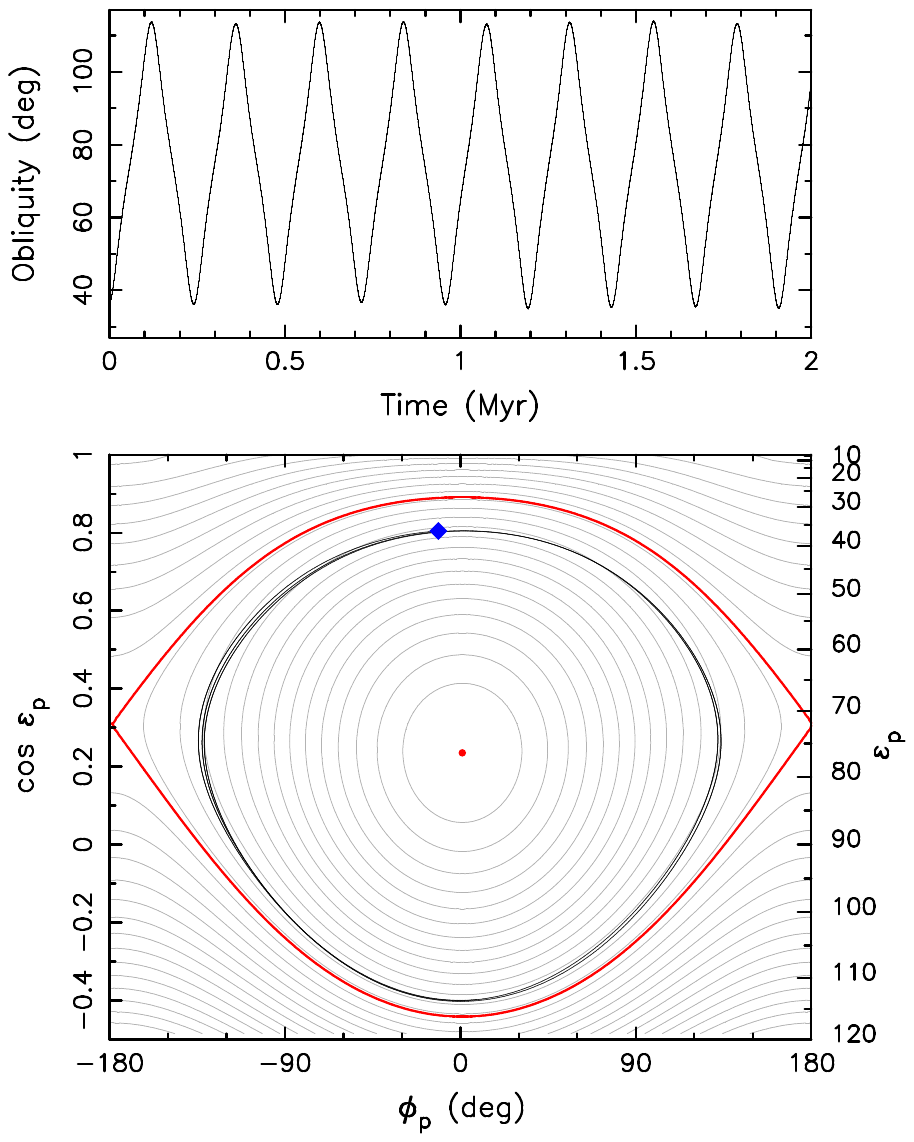}\includegraphics[width=0.48\textwidth]{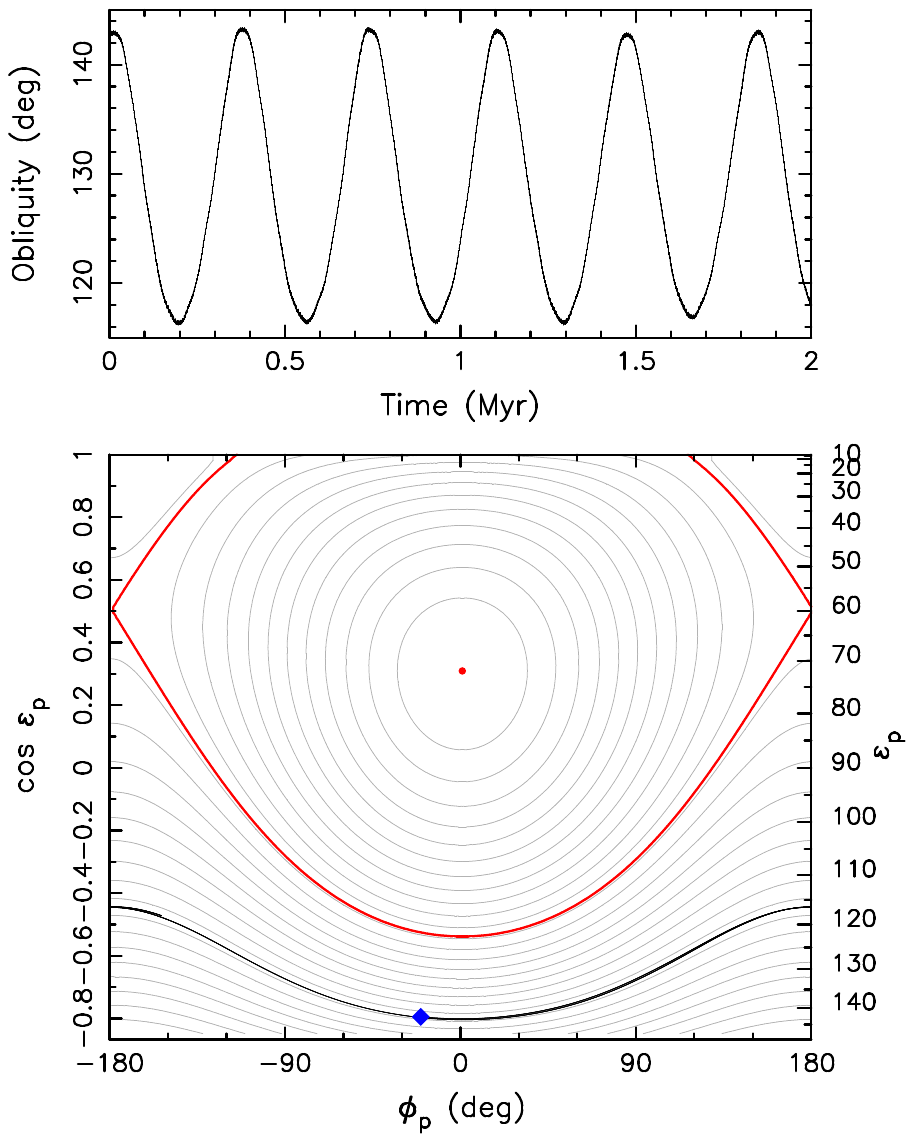}
 \caption{Results from a short-term integration of the rotational pole direction
   ${\bf s}$ for the L$_5$ Trojan (1867) Deiphobus (left panels) and for the L$_4$ Trojan (15663) Periphas (right panels). The top panel shows the osculating obliquity
   as a function of time. The bottom panel provides a projection of the ${\bf s}$ evolution over
   the $2$~Myr timespan onto the phase space defined by the proper-frame variables: (i) obliquity
   $\varepsilon_{\rm p}$ on the ordinate, and (ii) longitude $\phi_{\rm p}$ at the
   abscissa. The numerically integrated evolution is shown by a black line with the
   initial conditions shown by the blue diamond. The gray lines are idealized flow-lines
   from the simple Colombo top model. Red line is the separatrix of the resonant
   zone, and red points are locations of the Cassini state~2 (the Cassini state~4
   is at the junction of the two separatrix branches with $\phi_{\rm p}=\pm 180^\circ$).}
 \label{fig1}
\end{figure*}
%%%%%%%%%%%%%%%%%%%%%%%%%%%%%%%%%%%%%%%%%%%%%%%%%%%%%%%%%%%%%%%%%%%%%%%%%%%%%%%%%%%%%%%%%%%%%%%%%%%%%%%
% FIG 2 %%%%%%%%%%%%%%%%%%%%%%%%%%%%%%%%%%%%%%%%%%%%%%%%%%%%%%%%%%%%%%%%%%%%%%%%%%%%%%%%%%%%%%%%%%%%%%%
%\begin{figure}[tb]
% \includegraphics[width=0.48\textwidth]{figs/15663.pdf}
% \caption{Results from a short-term integration of the rotational pole direction
%   ${\bf s}$ for the L$_4$ Trojan (15663) Periphas. The top panel shows the osculating obliquity
%   as a function of time. The bottom panel provides a projection of the ${\bf s}$ evolution over
%   the $2$~Myr timespan onto the phase space defined by the proper-frame variables: (i) obliquity
%   $\varepsilon_{\rm p}$ on the ordinate, and (ii) longitude $\phi_{\rm p}$ at the
%   abscissa. The numerically integrated evolution is shown by a black line with the
%   initial conditions shown by the blue diamond. The gray lines are idealized flow-lines
%   from the simple Colombo top model. Red line is the separatrix of the resonant
%   zone, and red points are locations of the Cassini state~2 (the Cassini state~4
%   is at the junction of the two separatrix branches with $\phi_{\rm p}=\pm 180^\circ$).}
% \label{fig2}
%\end{figure}
%%%%%%%%%%%%%%%%%%%%%%%%%%%%%%%%%%%%%%%%%%%%%%%%%%%%%%%%%%%%%%%%%%%%%%%%%%%%%%%%%%%%%%%%%%%%%%%%%%%%%%%
% FIG 3 %%%%%%%%%%%%%%%%%%%%%%%%%%%%%%%%%%%%%%%%%%%%%%%%%%%%%%%%%%%%%%%%%%%%%%%%%%%%%%%%%%%%%%%%%%%%%%%
\begin{figure}[tb]
 \includegraphics[width=0.48\textwidth]{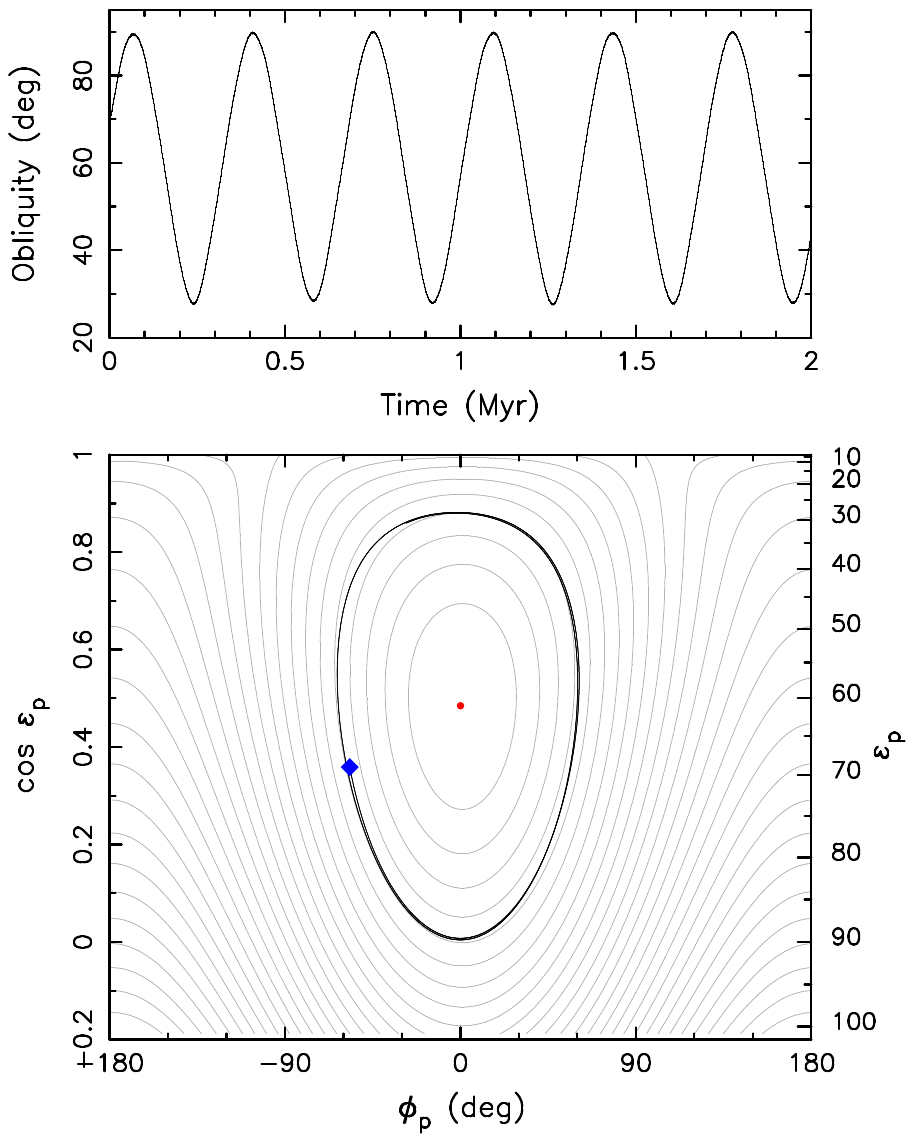}
 \caption{Results from a short-term integration of the rotational pole direction
   ${\bf s}$ for the L$_4$ Trojan (4834) Thoas. The top panel shows the osculating obliquity
   as a function of time. The bottom panel provides a projection of the ${\bf s}$ evolution over
   the $2$~Myr timespan onto the phase space defined by the proper-frame variables: (i) obliquity
   $\varepsilon_{\rm p}$ on the ordinate, and (ii) longitude $\phi_{\rm p}$ at the
   abscissa. The numerically integrated evolution is shown by a black line with the
   initial conditions shown by the blue diamond. The gray lines are idealized flow-lines
   from the simple Colombo top model. Red point marks the location of the Cassini state~2.}
 \label{fig3}
\end{figure}
%%%%%%%%%%%%%%%%%%%%%%%%%%%%%%%%%%%%%%%%%%%%%%%%%%%%%%%%%%%%%%%%%%%%%%%%%%%%%%%%%%%%%%%%%%%%%%%%%%%%%%%

\subsection{Trojans with large variations of pole latitude}\label{col}
In this section, we illustrate reasons for large latitude variations of
the rotation pole for three exceptional Trojans. We start with the pole solution of
(1867)~Deiphobus, the L$_5$ cloud member, for which we predict the largest effect
in this group (see Table~\ref{table_L5}). In this case, the orbit has a proper inclination
$I\simeq 28.3^\circ$ and proper frequency $s\simeq -6.23$ arcsec yr$^{-1}$.
Our estimated $\Delta\simeq 0.246$ from the shape model and rotation period
$P\simeq 59.18$~h provide $\alpha\simeq 21.86$ arcsec yr$^{-1}$. As a
result, we have $\kappa\simeq 3.509$ and $\kappa_\star\simeq 1.886$. Thus
$\kappa>\kappa_\star$, principally due to the very slow rotation of
Deiphobus and the resonant situation with four Cassini states apply. The
obliquity $\varepsilon_2$ of the resonant center, the Cassini state~2, is among
the solutions of the equation of $\kappa\sin 2\varepsilon_2 =
2\sin(\varepsilon_2-I)$ \citep[e.g.,][]{c1966}. Interestingly, this quartic
equation may be solved analytically \citep[e.g.,][]{hap2020}, providing
$\varepsilon_2 \simeq 77.2^\circ$, quite larger than the proper inclination
value. Because of the large proper inclination $I$ and proximity of $\kappa$
to $\kappa_\star$ the resonant zone occupies a large portion of the
phase space: it spans from $\simeq 28.9^\circ$ to $\simeq 117.1^\circ$ in
obliquity along the $\phi_{\rm p}=0^\circ$ section. Finally,
the present-date obliquity value of our unique pole for Deiphobus is
$\simeq 37^\circ$, not too close to the Cassini state~2. This implies that
while circulating about the Cassini state~2, Deiphobus pole obliquity
will exhibit large oscillations and these will be reflected in large
variations of the pole latitude in the inertial frame.

Data shown on Fig.~\ref{fig1} confirm these conclusions. While the
estimates mentioned above used the simplified Colombo top model, the solution
shown here is obtained by full-fledged numerical simulations. Its closeness
to what the Colombo model predicts justifies its validity. In particular,
the top panel shows osculating obliquity as a function of time over $2$~Myr
initial segment of our simulation. The large-amplitude oscillation with a
period of $\simeq 243$~kyr is the projected circulation about the Cassini
state~2. To see the effect more fully, we also transformed the ${\bf s}$
evolution into the variables defined in the proper-term orbital frame
(that is, the frame having the reference plane inclined by the proper orbital
inclination value and precessing in the inertial frame with exactly $s$
frequency): (i) the proper obliquity $\varepsilon_{\rm p}$, namely the angle
between ${\bf s}$ and the z-axis of the proper frame, and (ii) the proper
longitude $\phi_{\rm p}$, namely the angle between the  ${\bf s}$ projected
onto the reference proper plane and the x-axis rotated $90^\circ$ away from
the proper node. The Cassini state~2 is shown by the red dot at $\phi_{\rm p}
=0^\circ$, and the gray curves are exact solutions of the idealized
Colombo top model \citep[e.g.,][]{hap2020}. The true ${\bf s}$ evolution in
these variables follow very closely the neighboring gray lines justifying
the approximate validity of the Colombo model. The takeaway message is that the large
obliquity oscillations, between $\simeq 35^\circ$ and $\simeq 113^\circ$, are
indeed triggered by oscillations about the Cassini state~2, which itself is
tilted to a large $\varepsilon_2$ 
value.\footnote{The confinement of the P1 pole solution for Deiphobus in the
 $s$-type Cassini resonance is reminiscent of the spin state of (433)~Eros
 \citep{eros2005}.}

The resonant zone about the Cassini state 2 is predominantly located in
the prograde rotation zone, but its presence may considerably affect solutions
in the retrograde rotation zone. An example of this behavior is the L$_4$
member (15663) Periphas (see Table~\ref{table_L4}). In this case, our
solution yields $\Delta\simeq 0.233$ and the rotation period $P\simeq 9.92$~h,
not suspiciously large. Together with other orbital parameters, they provide
a rather small value of the precession constant $\alpha\simeq 3.54$ arcsec yr$^{-1}$.
However, the really small value of the orbital precession frequency (partly
due to large inclination value $I\simeq 33.94^\circ$), $s\simeq -1.61$ arcsec
yr$^{-1}$, makes this case anomalous. We thus obtain $\kappa\simeq 2.20$ and
$\kappa_\star\simeq 1.95$. The proximity of $\kappa$ to $\kappa_\star$ implies the
resonant zone about the Cassini state~2, here at $\varepsilon_2 \simeq 72.6^\circ$,
is large. This is demonstrated at the bottom panel of Fig.~\ref{fig1}. However,
the rotation of Periphas is retrograde, with the current obliquity value of
$\simeq 142^\circ$, and it does not interact with the Cassini resonance directly.
The evolutionary track of the Periphas' spin evolution is shown on Fig.~\ref{fig1}
confirms it circulates about the Cassini state~3. Still, the Cassini resonance
reaching up to obliquity of $\varepsilon$ $\simeq 123^\circ$ pushes even the retrograde solutions
to perform large obliquity variation, nearly $30^\circ$ as shown in the upper panel
of Fig.~\ref{fig1}.

In order to demonstrate that the existence of the Cassini resonant zone is not
a necessary condition for large latitude oscillations we consider (4834) Thoas,
another L$_4$ Trojan (Table~\ref{table_L4}). In this case, the
proper inclination has a value $I\simeq 27.05^\circ$ and the proper frequency
$s\simeq -4.58$ arcsec yr$^{-1}$. The rotation period is not suspiciously
long, $P\simeq 18.19$~h, and thus the precession constant is rather low $\alpha\simeq
6.20$ arcsec yr$^{-1}$. As a result $\kappa<\kappa_\star$ in this case, but not
much smaller due to the moderately small value of the proper frequency $s$ and moderately
long rotation period ($\kappa\simeq 1.35$,
while $\kappa_\star\simeq 1.87$). There are only two Cassini states, but due to
the proximity of $\kappa$ to $\kappa_\star$ the Cassini state~2 is forced to already
a large obliquity value of $\varepsilon_2\simeq 61.6^\circ$. Figure~\ref{fig3}
illustrates the whole situation and also shows Thoas' spin evolution over
the next $2$~Myr interval starting with our unique pole solution. The bottom panel
shows a projection of the spin evolution into the phase-space variables of the Colombo
model associated with the proper variables. The spin axis of Thoas performs a slow
circulation about the Cassini state~2 shifted to an anomalously large value of the
obliquity. This evolution then triggers a large amplitude of Thoas' obliquity
oscillations, also reflected in large latitude excursions.

The takeaway message from this section is that all cases for which the polar
latitude was found to oscillate in a large interval of values in our numerical
simulations (Tables~\ref{table_L4} and \ref{table_L5}) may be understood using
the Colombo top model. They correspond to
the situations when $\kappa$ is not much smaller than $\kappa_\star$, therefore
when the Cassini resonance exists of is about to emerge. Due to the typically
large proper inclination of Trojan orbits, the Cassini resonance zone occupies
a large portion of the phase space and drives large obliquity oscillations.

\subsection{Unstable Trojans}\label{unst}
For completeness, we also comment on orbital stability for Trojans
in our sample with resolved rotation poles. In most cases, their orbits were fairly
stable over the interval of $50$~Myr used for monitoring the variations of
the pole latitude. But in five cases we noticed orbital instability onset
even during this period of time. Here we report only those deemed ``macroscopically
unstable'', in which orbital eccentricity or inclination exhibited large jumps,
and we omit cases exhibiting traces of only ``stable chaos'' on the 50~Myr
timescale.

Analysis of the Trojan cloud stability has a long history, and it is
not our goal to provide a comprehensive review here. The observed population
analysis (as opposed to purely mathematical studies) was provided by \citet{m1993},
who computed synthetic proper elements
of a set of 174 Trojans based on a megayear-long numerical integration. Some 13\%
of these orbits showed various traces of instability evidenced by the Lyapunov
timescales between $60$ and $600$~kyr. This sample of chaotic orbits was later
studied in more detail using numerical integrations spanning a longer timescale
of $50$ to $100$~Myr \citep[e.g.,][]{pilat1999,dt2000,tsi2000}. These studies
helped to describe macroscopic orbital instability and identified the associated
dynamical reasons. Even longer, the gigayear timescale orbital stability
of Trojans was studied by \citet{lev1997}, as afforded by development in
symplectic numerical tools. These authors estimated that some 12\% of real
Trojans escape over the Solar System age from their population. This fraction was
recently even increased to about 23\% by \citet{holt2020}, perhaps by using
data for many more, especially small Trojans. Overall, the nearly steady
leakage of even large Trojans may be understood by realizing their formation
mechanism that allows them to fully fill their phase space up to limits of
instability \citep[e.g.,][]{Nesvorny2013}. Finding that few of our studied 90
Trojans reside on noticeably unstable orbits is therefore not surprising.
% FIG 4 %%%%%%%%%%%%%%%%%%%%%%%%%%%%%%%%%%%%%%%%%%%%%%%%%%%%%%%%%%%%%%%%%%%%%%%%%%%%%%%%%%%%%%%%%%%%%%%
\begin{figure*}[t!]
 \begin{tabular}{cc}
  \hspace{-1cm}
  \includegraphics[width=0.48\textwidth]{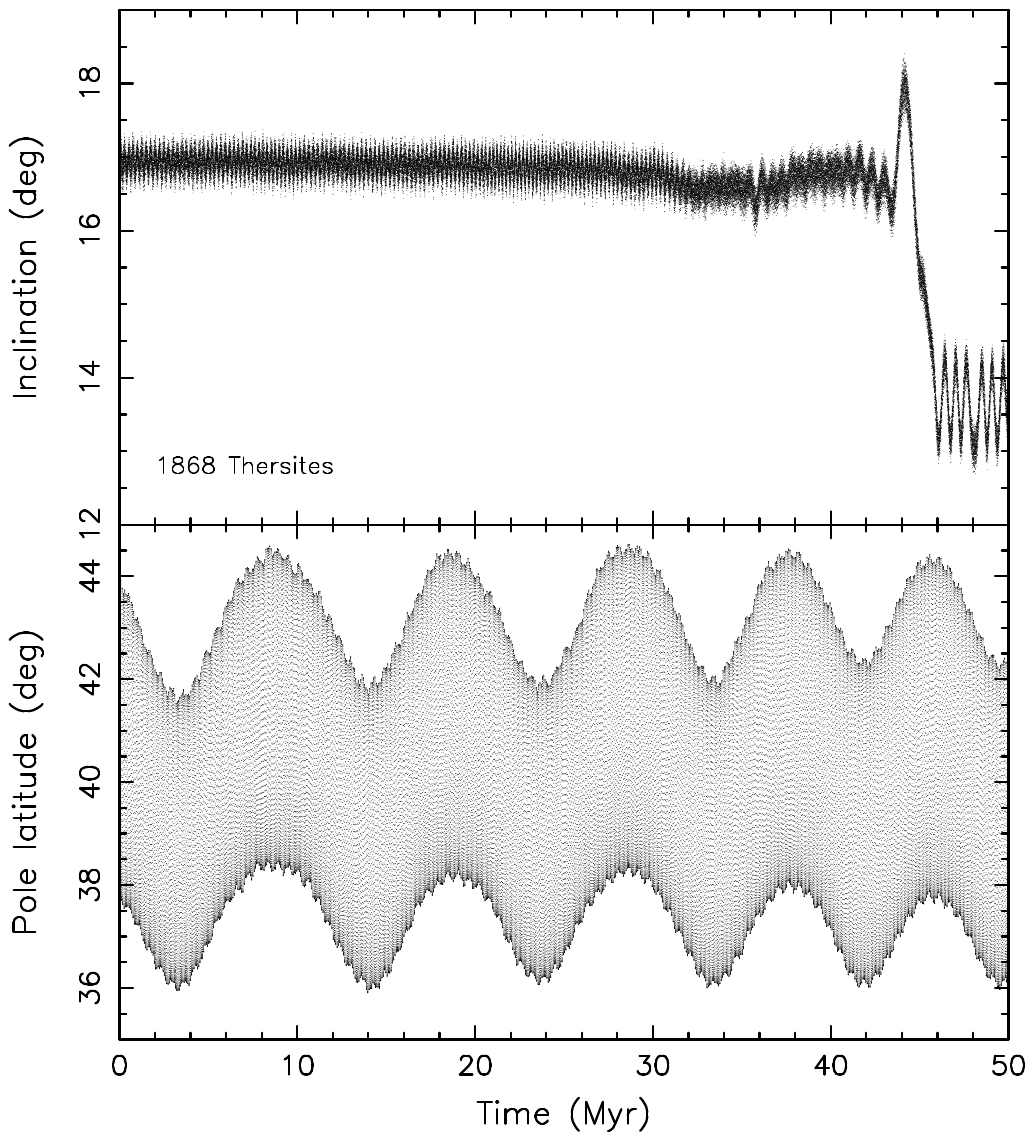} &
  \includegraphics[width=0.48\textwidth]{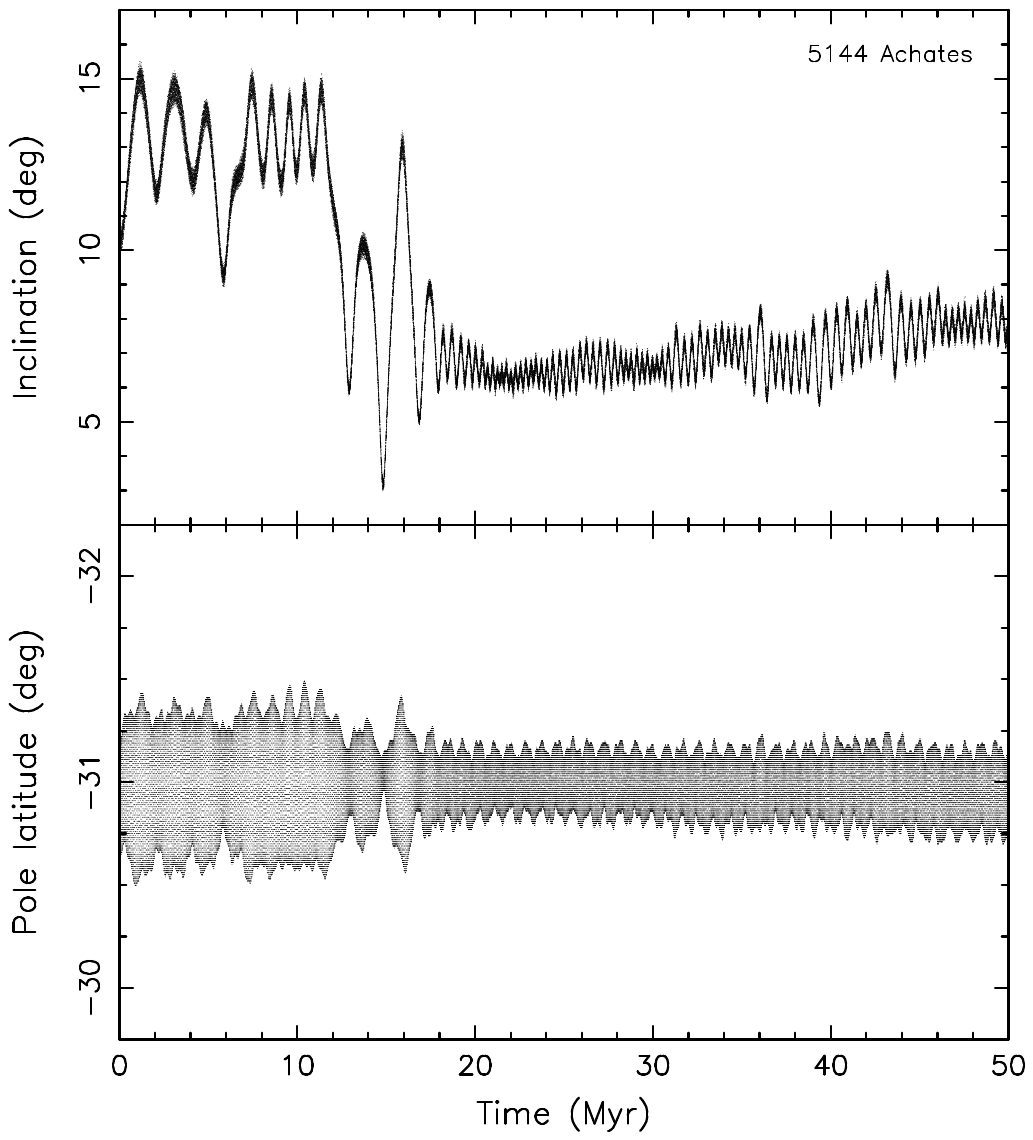} \\
 \end{tabular}
 \caption{Two orbitally unstable Trojans: (i) L$_4$ object (1868)~Thersites
  in the left panels, and (ii) L$_5$ object (5144)~Achates in the right panels. The top
  panels show time dependence of the osculating orbital inclination with respect to
  the invariable plane. Bottom panels show the evolution of the pole P1 latitude in
  the same reference system.}
 \label{fig4}
\end{figure*}
%%%%%%%%%%%%%%%%%%%%%%%%%%%%%%%%%%%%%%%%%%%%%%%%%%%%%%%%%%%%%%%%%%%%%%%%%%%%%%%%%%%%%%%%%%%%%%%%%%%%%%%

Figure~\ref{fig4} shows results from our simulations of two macroscopically
unstable Trojans in our sample: (i) (1868)~Thersites (left), and
(ii) (5144)~Achates (right).%
\footnote{Results for the remaining three Trojans on unstable orbits in our
 sample, namely (1173) Anchises, (32615) 2001~QU277, and (51364) 2000~SU333
 all in the L$_5$ camp, are very similar to that of Achates.}
Especially the large eccentricity
Achates' orbit indicates macroscopic chaos witnessed here by
irregular evolution of the inclination \cite[top panel on Fig.~\ref{fig4}, see
also][]{dt2000} due to intermittent interaction with several secular
resonances (of which $\nu_{16}$ is the most prominent). In fact,
we numerically propagated the nominal orbit of (5144)~Achates
together with 49 very close orbital clones (all starting from the 
uncertainty interval of its currently determined orbit) for
1~Gyr forward in time. We found that a median timescale before
ejection from the Trojan region was $\simeq 220$~Myr, in good
agreement with the data in Fig.~1 of \citet{lev1997}, and
comparable to the case of yet another unstable Trojan (1173)~Anchises
\citep[see][]{hor2012}. Similarly, (1868)~Thersites is a well
studied unstable orbit \citep[left panel; see][]{tsi2000}. In our
simulation, the inclinations instability onsets at $\simeq 44$~Myr.
In both cases, however, the orbital instability has little
effect on the rotation pole latitude (bottom panels on Fig.~\ref{fig4}
where we used the P1 solution for both Trojans). This is because
both objects have sufficiently short rotation periods, such that
always $\kappa\ll \kappa_\star$ during the integrated time interval.
As a result, the unstable nature of the Thersites and Achates orbits
does not seem to confuse our conclusions as far as the rotation pole
categories (prograde or retrograde) are concerned.
 %%% To speed up compilation when working on the manuscript

\section{Discussion}\label{sec:discussion}

The axis ratios of JTs with shape models are formally compatible in $b'/c'$ and incompatible in $a'/b'$ with the population of larger ($D>30$~km) MBAs. JTs contain less near-spherical and more elongated bodies than the sample of MBAs. There are, however, several caveats concerning the JTs. Namely, 
(i)~the number of objects is rather small compared to MBAs (90  vs. 218),
(ii)~the axis ratios have large uncertainties in cases where the shape models are based on limited photometric datasets, and, in general, concerning the $c$ dimension,
(iii)~the sample is biased toward objects with larger amplitudes, thus we likely miss (oblate) spheroids. Therefore, it is difficult to conclude whether the shapes are indeed similar within the two populations. The open question here is whether oblate spheroids dominate in the JT population, which might indicate the lesser importance of collisions since their formation. We do not see many in our sample, which is a common property with MBAs. However, we already learned an important lesson about the bias in the main belt concerning oblate spheroids. Only recently, the shape models or large asteroids of (10)~Hygiea, (31)~Euphrosyne, (324)~Bamberga, or (702)~Interamnia were derived thanks to disk-resolved images from the high-resolution SPHERE instrument mounted on the 8 meter-class Very Large Telescope \citep{Vernazza2020, Yang2020a, Vernazza2021, Hanus2020a}. All these asteroids have shapes consistent with an oblate spheroid and have not been derived before using optical photometry only (thus no previous convex shape models). It is likely that the difficulty in deriving the shapes of such bodies applies to JTs as well.

In our study, we focus on testing whether the observed physical properties of JTs are consistent with those predicted based on the currently most acceptable formation scenario -- capture to their current orbits near the Lagrangian points of Jupiter during the early reconfiguration of the giant planets from the planetesimals born in the massive trans-Neptunian disk. The numerical simulations of the streaming instability, the leading mechanism for the planetesimals formation, predict the spin obliquity distribution and the ratio between the number of prograde and retrograde rotators. Both predictions are testable by observed properties. The streaming instability predicts a somewhat larger fraction of prograde rotators than we observe (75\% vs. 60\%). However, this could be attributed to, for example, (i)~partial randomization due to collisional evolution prior to the capture, or (ii)~due to post-capture collisional and orbital evolution. We modeled the latter mechanism in Sec.~\ref{sec:dynamics} and showed that it can be responsible for some minor depletion of prograde rotators. 

In Sec.~\ref{sec:streeming}, we consider only planetesimals with $D>25$~km. This limit corresponds to the absolute magnitude of $H\sim12$ mag (assuming a geometric visible albedo of $p_{\mathrm{V}}=0.05$). Thus, we limit our JT sample to objects with $H\leq12$ mag for the analysis of the pole obliquities. This limit also conveniently excludes objects whose obliquities could be affected by YORP thermal forces. Although YORP is weaker for JTs than for MBAs given the difference in the distance from the Sun, the size at which it is important is roughly the same for the two populations. This is due to the JTs having significantly longer collisional lifetimes than similarly sized MBAs. Considering JTs smaller than $\sim25$~km in the spin vector analysis could affect our interpretations as those spin vectors could be significantly evolved by the YORP.

In Fig.~\ref{fig:spins}, we compare the obliquity distributions of several populations, including JTs, and of the prediction based on streaming instability (SI). The obliquity distribution of JTs seems to be visually the closest match to the SI distribution, however, the K-S test does not support the hypothesis that the two samples are drawn from the same distribution ($p$-value=0.015). Indeed, JTs lack members with obliquities in the interval 50$^\circ$--70$^\circ$ and contain more retrograde bodies. As discussed earlier, several dynamical processes could affect the distribution. In addition, the observing bias should be present as well. The shape and spin solutions are derived by the convex inversion method and it was shown by \citet{Hanus2011} that objects with obliquities near 0$^\circ$ and 180$^\circ$ are derived more successfully than bodies with obliquities near 90$^\circ$, namely even by 30--40\%. This is because the former bodies lack the pole-on observing geometry, which provides only low-amplitude brightness changes often comparable to the noise scatter. By correcting the observed obliquity distribution on this bias, we would obtain more objects with mid-value obliquities and thus a better agreement with the SI distribution. However, the excess of retrograde rotators would still exist.
Interestingly, the obliquities of JTs and large main-belt asteroids are the most similar (although inconsistent) with the K-S test $p$-value=0.02. The main difference is for small obliquities ($\varepsilon<$50$^\circ$). 

Although the observed sample of JT obliquities is still rather small, the comparison with the streaming instability predicted obliquities suggests that this formation scenario could be consistent with the observed properties of JTs. The main issue is the overabundance of retrograde rotators with $\varepsilon>130^\circ$.

\section{Conclusions}\label{sec:conclusions}

Deriving physical properties of JTs is challenging due to various obstacles -- the observing geometry is changing much slower than for MBAs and is limited to phase angles (i.e., angle Sun-asteroid-Earth) up to just a few degrees; their large distance to the Earth together with their low albedo makes them rather faint objects accessible to at least meter-class telescopes. Therefore, their photometry is scarce and often affected by large uncertainties. Despite that, the currently available photometric datasets can be successfully used for the physical characterization of several tens of JTs.

By analyzing optical datasets for $\sim$1000 JTs, we obtained spin state and shape solutions in $79$ cases (Sec.~\ref{sec:modeling}). We found that the observed distribution of the pole obliquities/latitudes of JTs is broadly consistent with expectations from the streaming instability (Sec.~\ref{sec:streeming}), which is currently considered the leading mechanism for the formation of planetesimals in the trans-Neptunian disk. Observed JTs latitude/obliquity distribution has a slightly smaller prograde/retrograde asymmetry (excess of obliquities $>130^\circ$) than that expected from the existing streaming instability simulations. However, this discrepancy can be plausibly reconciled by the effects of the post-formation collisional activity. Our numerical simulations of the post-capture spin evolution in Sec.~\ref{sec:dynamics} indicate that the JTs' pole distribution is not significantly affected by dynamical processes such as the eccentricity excitation in resonances, close encounters with planets, or the effects of non-gravitational forces. However, a few JTs exhibit large latitude variations of the rotation pole and may even temporarily transition between prograde- and retrograde-rotating categories.

As surveys with better accuracy, faster cadence, and higher limiting magnitudes are substituting or supplementing the already established surveys, better data become available. This will lead to an increase in the number of spin state solutions within the JTs, which should increase the significance of our results and possibly reveal hidden dependencies within various physical properties of JTs. Ideally, the larger statistical sample should further constrain theoretical models aiming at explaining the origin of the JT population.

JTs are also commonly targeted by the stellar occultation hunters -- the event predictions are now rather accurate leading to many positive detections, which further improves the ephemeris, thus future stellar occultation predictions. Convex shape models, together with the stellar occultation profiles can provide direct measurements of the dimensions, volume, and eventually bulk density if an accurate mass estimate is available (e.g., from system multiplicity). Occultation measurements also provide reliable estimates of axis ratios, which will help to assess whether the shapes of JTs are similar to the shapes of MBAs. 

The remote-like studies of physical properties of JTs such as ours represent invaluable support for the insitu exploration of several JTs by the Lucy mission \citep{Levison2021}. Only a combination of both approaches can lead to the most complete understanding of the JT population and, if properly modeled, also the trans-Neptunian population.

\begin{acknowledgements}
The work of JH and JD has been supported by the Czech Science Foundation through grant 22-17783S and by the Erasmus+ program of the European Union under grant number 2020-1-CZ01-KA203-078200. The work of DV has been supported by the Czech Science Foundation through grant 21-11058S. The observational work by V.~Benishek at the Sopot Astronomical Observatory in Serbia was kindly supported by the Shoemaker NEO Grants from the Planetary Society for 2017 and 2021. The work of PP has been supported through a NASA Solar System Workings award number 80NSSC21K0153.
This research has used the Minor Planet Physical Properties Catalogue (MP3C) of the Observatoire de la Côte d’Azur and the IMCCE's Miriade VO tool. 
We thank Brian A. Skiff from the Lowell Observatory for providing the data for asteroid (5209) 1989 CW1. 
This work has made use of data from the Asteroid Terrestrial-impact Last Alert System (ATLAS) project. ATLAS is primarily funded to search for near-Earth asteroids through NASA grants NN12AR55G, 80NSSC18K0284, and 80NSSC18K1575; byproducts of the NEO search include images and catalogs from the survey area.  The ATLAS science products have been made possible through the contributions of the University of Hawaii Institute for Astronomy, the Queen's University Belfast, the Space Telescope Science Institute, the South African Astronomical Observatory (SAAO), and the Millennium Institute of Astrophysics (MAS), Chile.

We would like to thank our reviewer for taking the time and effort necessary to review the manuscript. We sincerely appreciate all valuable comments and suggestions, which helped us to improve the quality of the manuscript.
\end{acknowledgements}

\bibliography{mybib,literatura}
\bibliographystyle{aa}

%\newpage
\onecolumn
\begin{appendix}
\section{Additional figures}\label{appendix:occ}%[t]

\begin{figure*}[!htb]
\begin{center}
  \resizebox{1.0\hsize}{!}{\includegraphics{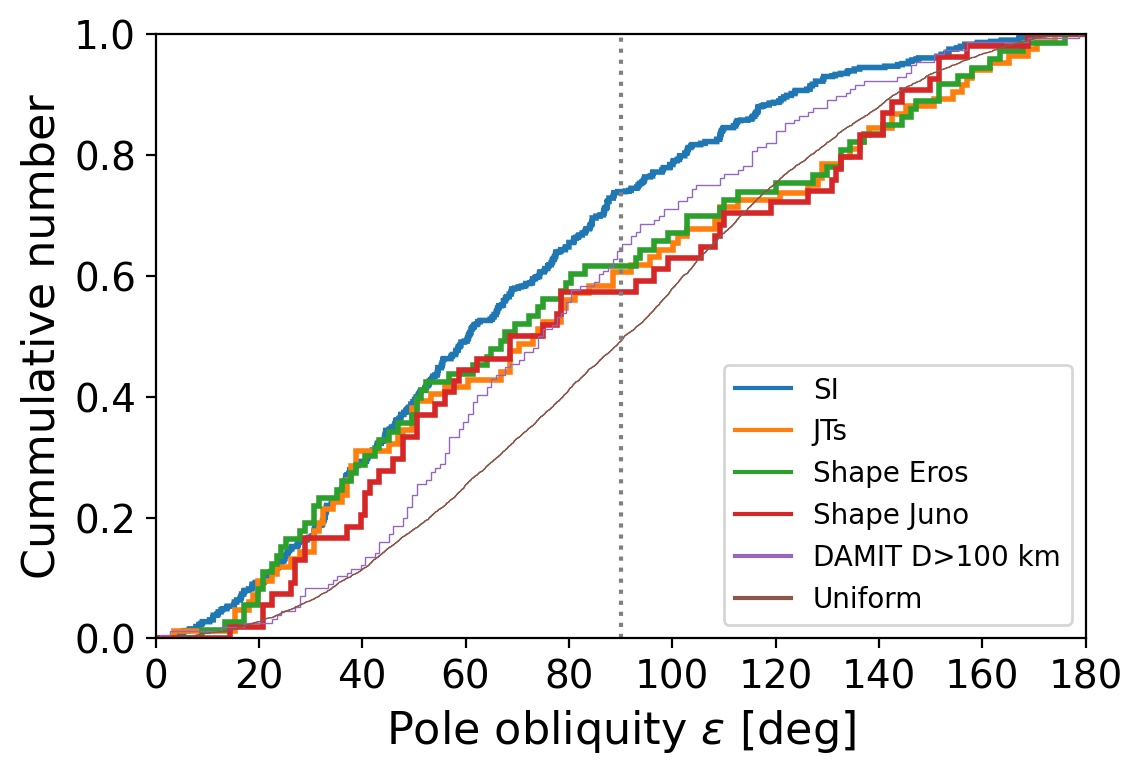}\includegraphics{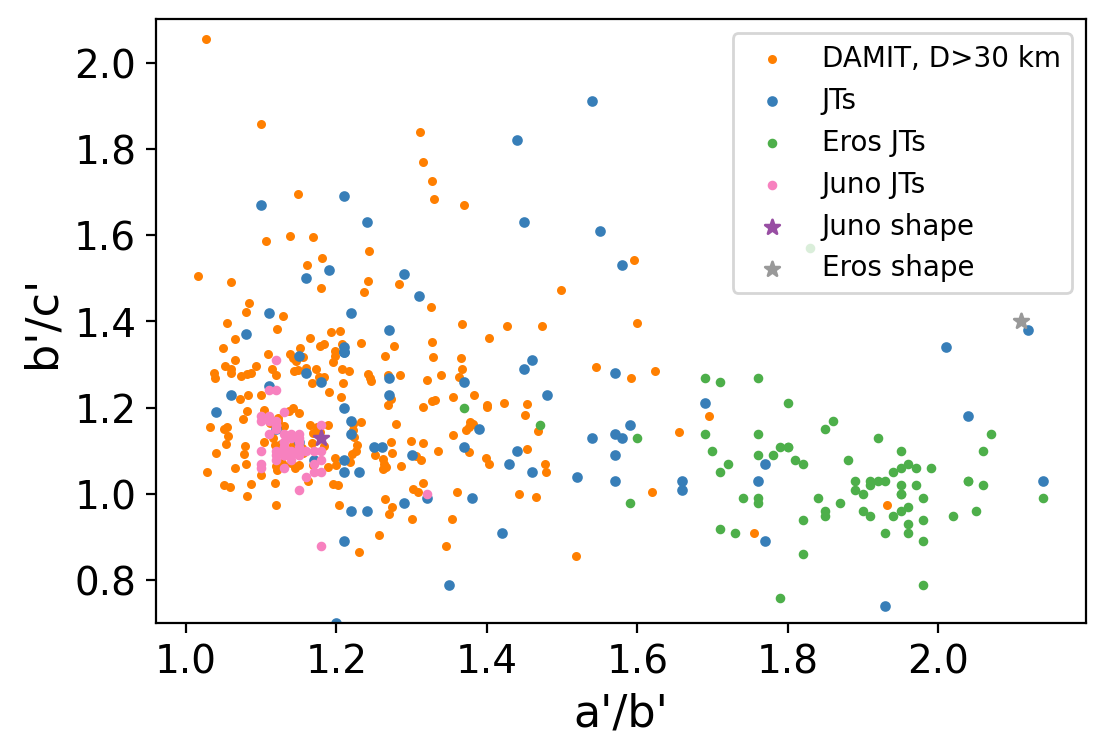}}\\
\end{center}
\caption{\label{fig:ratios_systematics}Assessing the systematic uncertainties by using assumed shapes. Left panel: The cumulative distribution of obliquities for JTs (orange), for synthetic JTs population assuming the shape models of (433)~Eros (green) and (3)~Juno (red), and of planetesimals obtained in our simulations of the streaming instability (SI) model (blue). We also plot the cumulative distributions for randomly oriented spins (brown), and large ($D>100$~km) MBAs (purple, spin states adopted from DAMIT). Right panel: The $a'/b'$ vs. $b'/c'$ axis ratios of JTs, of large MBAs ($D>30$~km), and of synthetic JT populations assuming the shape models of (433)~Eros and (3)~Juno. We also indicate the axis ratios for the original shapes of (433)~Eros and (3)~Juno.}
\end{figure*}

\twocolumn
\begin{figure}[!htb]
\begin{center}
  \includegraphics[width=\columnwidth]{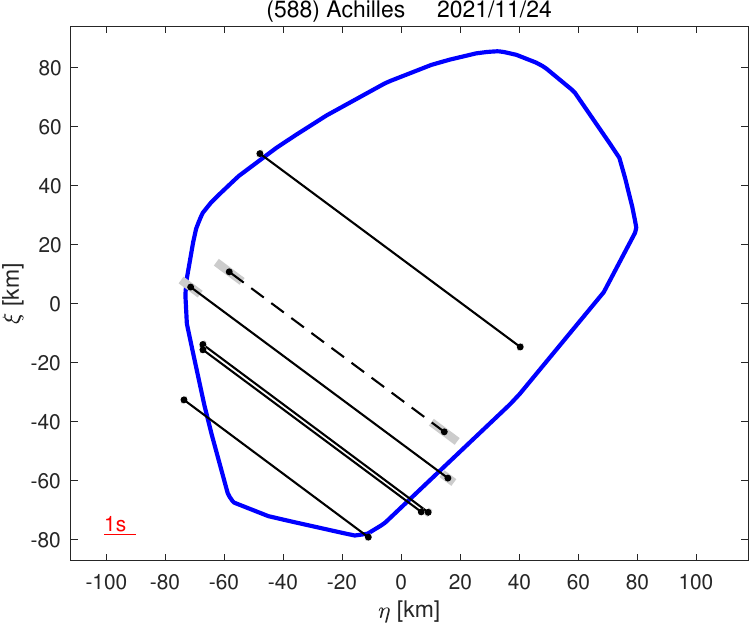}
\end{center}
 \caption{\label{fig:occ_588} Projection of the shape model of (588)~Achilles. The volume equivalent diameter is $131 \pm 8$\,km.}
\end{figure}

% updated
\begin{figure}[!htb]
\begin{center}
  \includegraphics[width=\columnwidth]{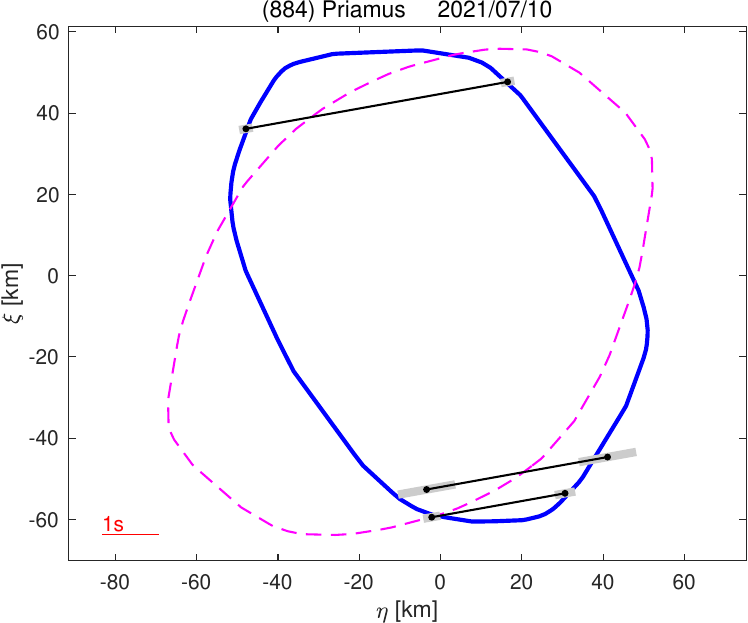}
\end{center}
 \caption{\label{fig:occ_884} Projection of the shape model of (884)~Priamus. Only the model with the pole $(1^\circ, -32^\circ)$ agrees with the observed chords, its equivalent diameter is $105 \pm 4$\,km.}
\end{figure}

% updated
\begin{figure}[!htb]
\begin{center}
  \includegraphics[width=\columnwidth]{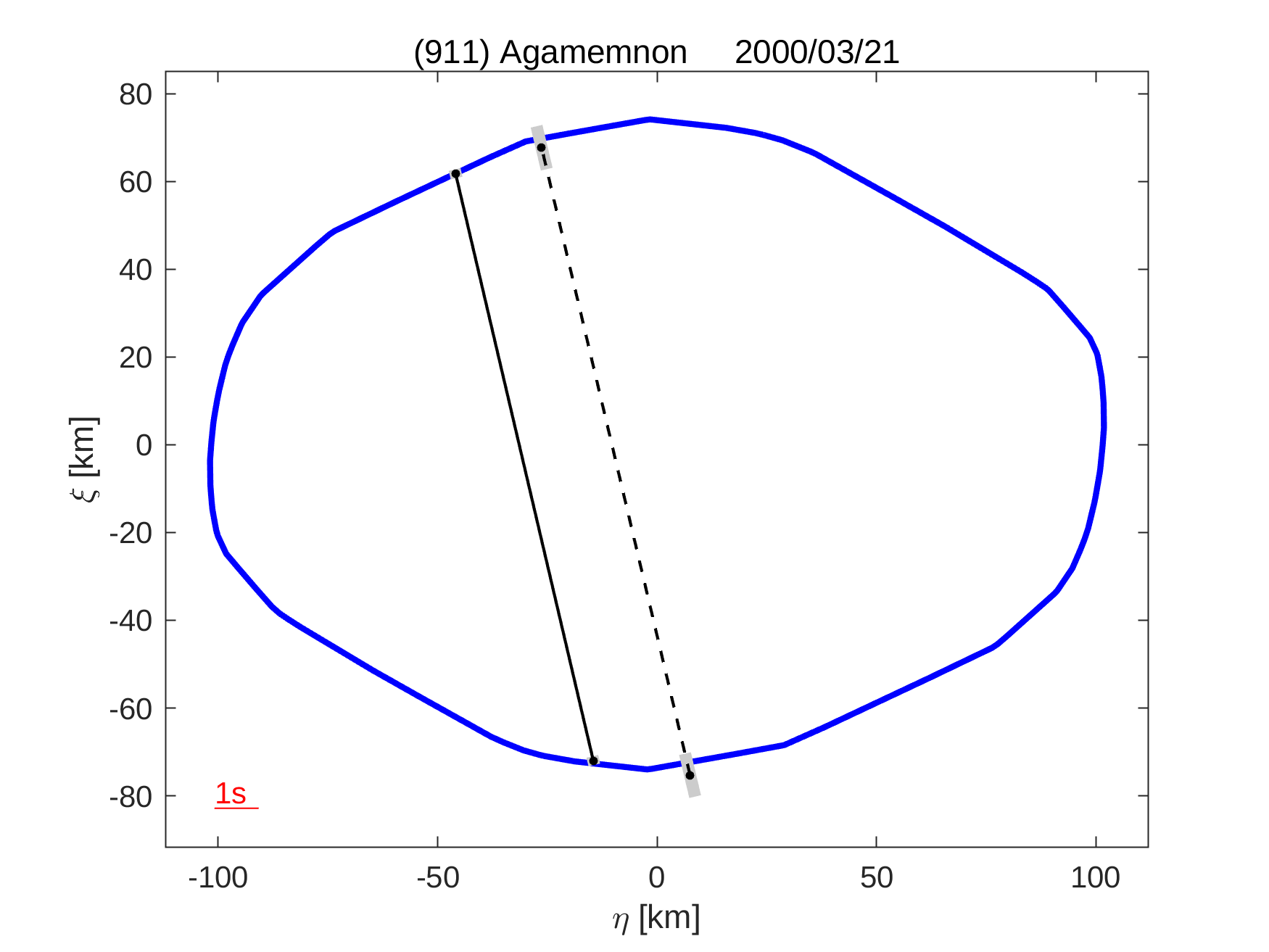}
  \includegraphics[width=\columnwidth]{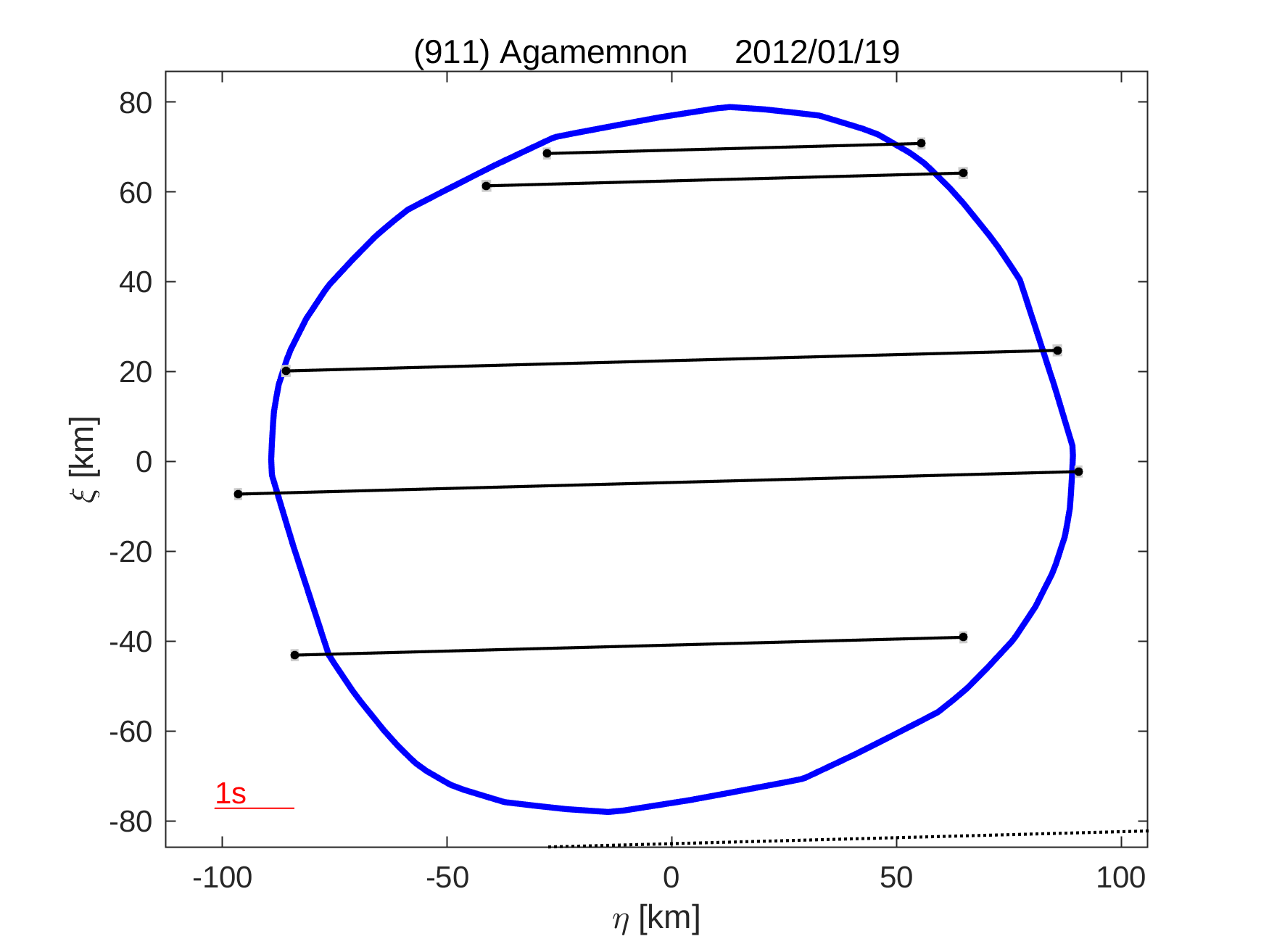}
\end{center}
 \caption{\label{fig:occ_911} Projection of the shape model of (911)~Agamemnon. The model with the pole $(290^\circ, 35^\circ)$ agrees with the observed chords, its equivalent diameter is $153 \pm 6$\,km. %The nonconvex ADAM model (orange, spin axis $(292^\circ, 29^\circ)$ and diameter $154 \pm 5$\,km) provides a better fit.
 }
\end{figure}

% updated
\begin{figure*}[!htb]
\begin{center}
  \includegraphics[width=\columnwidth]{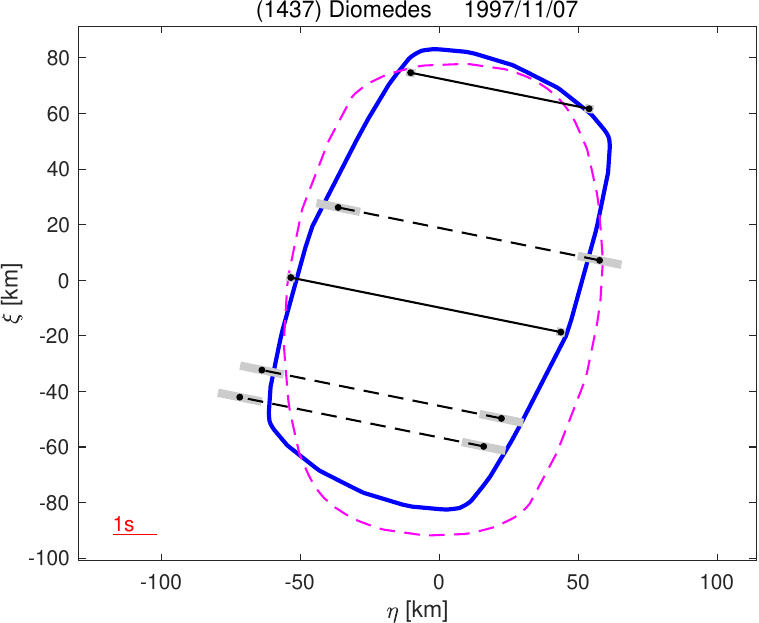}
  \includegraphics[width=\columnwidth]{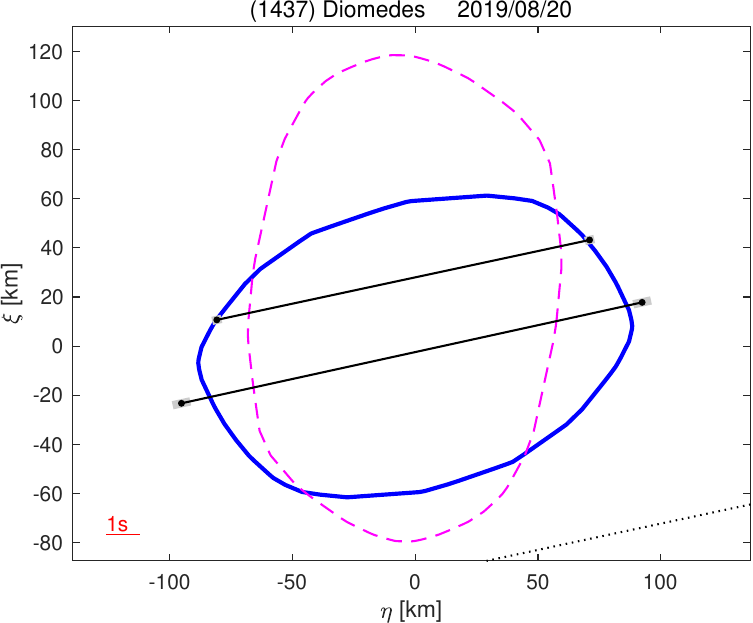}
  \includegraphics[width=\columnwidth]{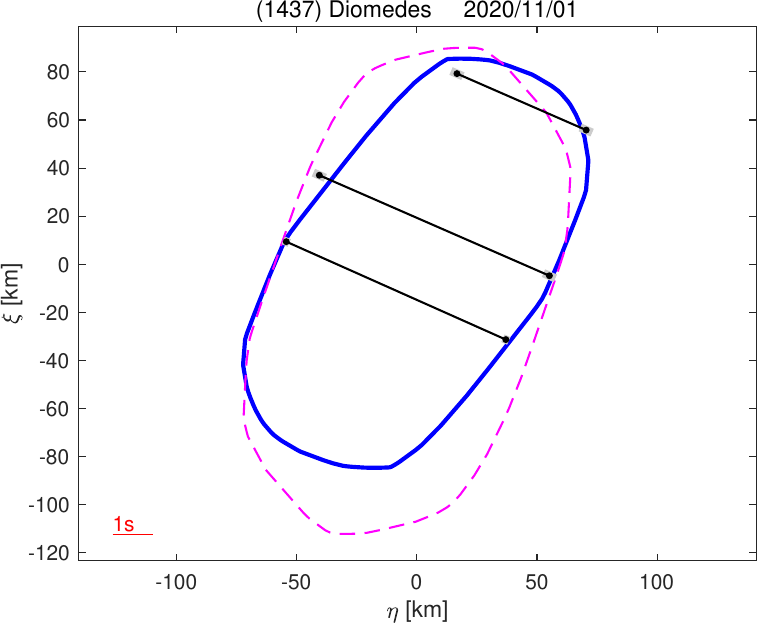}
  \includegraphics[width=\columnwidth]{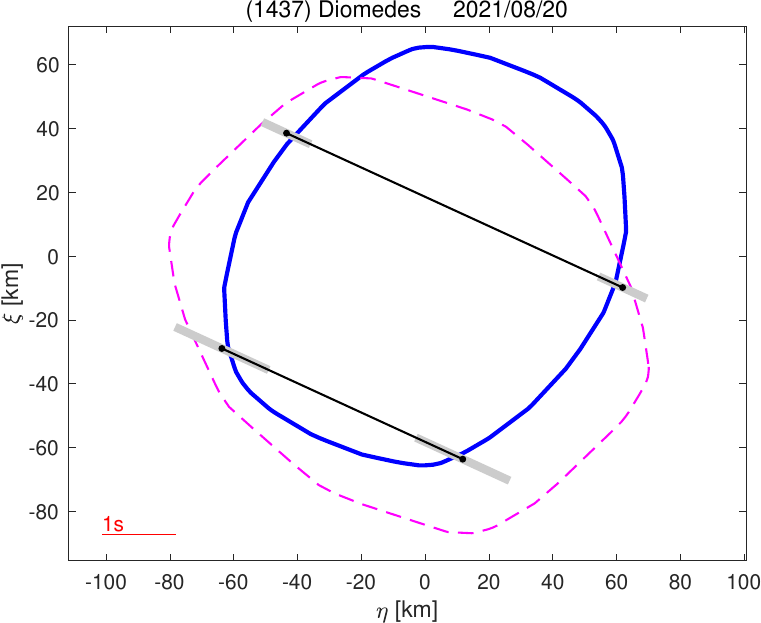}
\end{center}
 \caption{\label{fig:occ_1437} Projection of the shape model of (1437)~Diomedes. Only the model with the pole $(147^\circ, 5^\circ)$ agrees with the observed chords, its equivalent diameter is $133 \pm 5$\,km.}
\end{figure*}

% updated
\begin{figure}[!htb]
\begin{center}
  \includegraphics[width=\columnwidth]{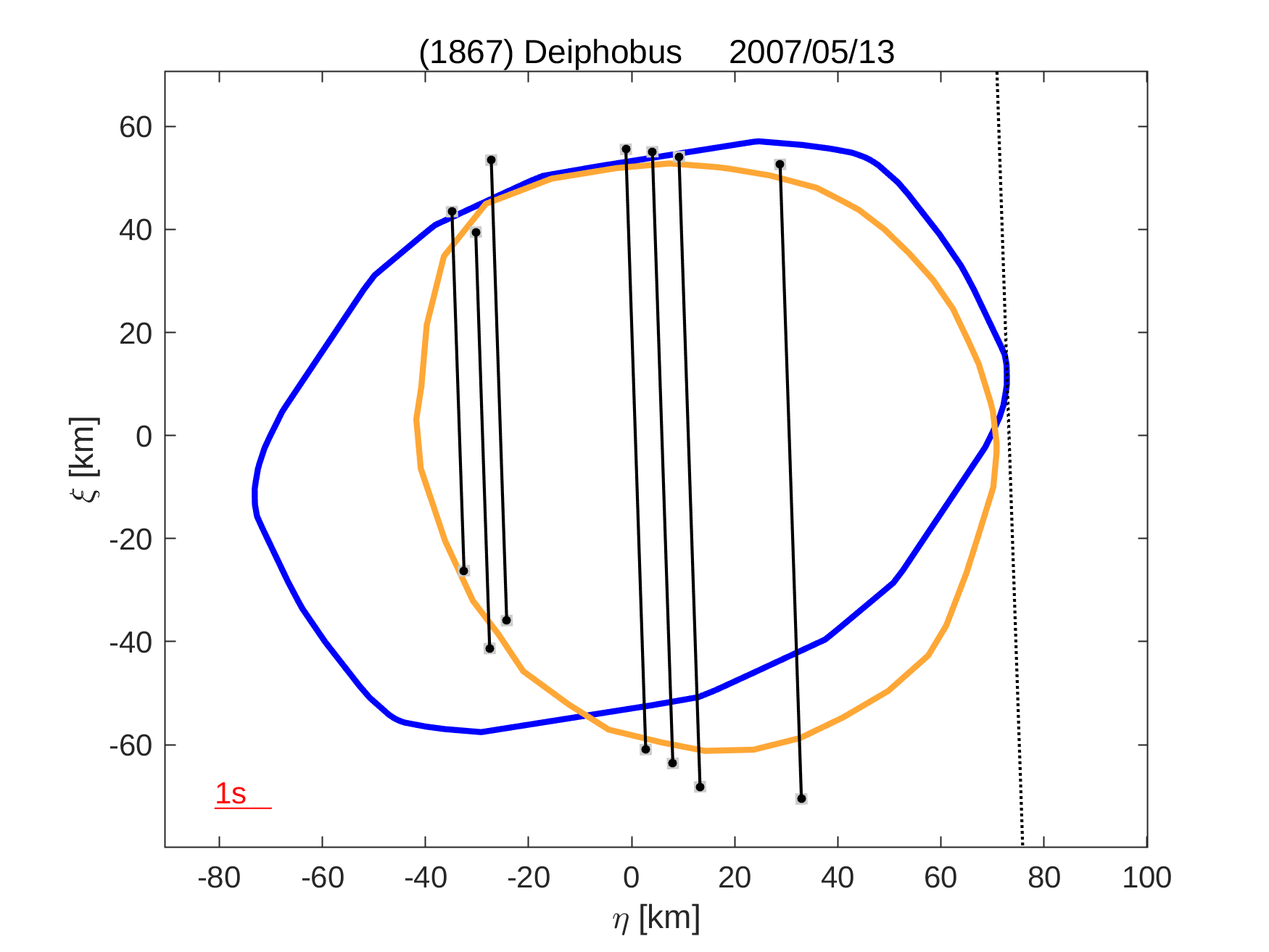}\\
\end{center}
 \caption{\label{fig:occ_1867} Projection of the shape model of (1867)~Deiphobus. The convex shape model (blue) with the pole $(338^\circ, 79^\circ)$ does not agree well with the observed chords. The nonconvex ADAM model (orange) provides a much better fit with the pole direction $(329^\circ, 67^\circ)$ and an equivalent diameter of $108 \pm 4$\,km. The second chord from the left is not consistent with the two other chords next to it, probably due to some systematic error (or the shape is highly concave at this place). Due to the negative observation (miss chord on the right), the shape cannot be larger and its size is significantly smaller than the values derived from Akari ($131 \pm 2$\,km) and IRAS ($123 \pm 4$\,km) data.}
\end{figure}

% updated
\begin{figure}[!htb]
\begin{center}
  \includegraphics[width=\columnwidth]{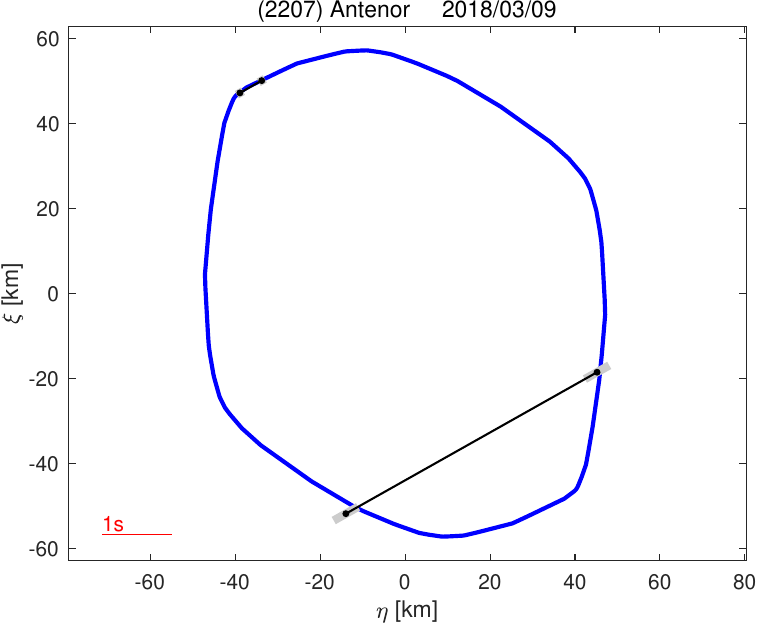}
  \includegraphics[width=\columnwidth]{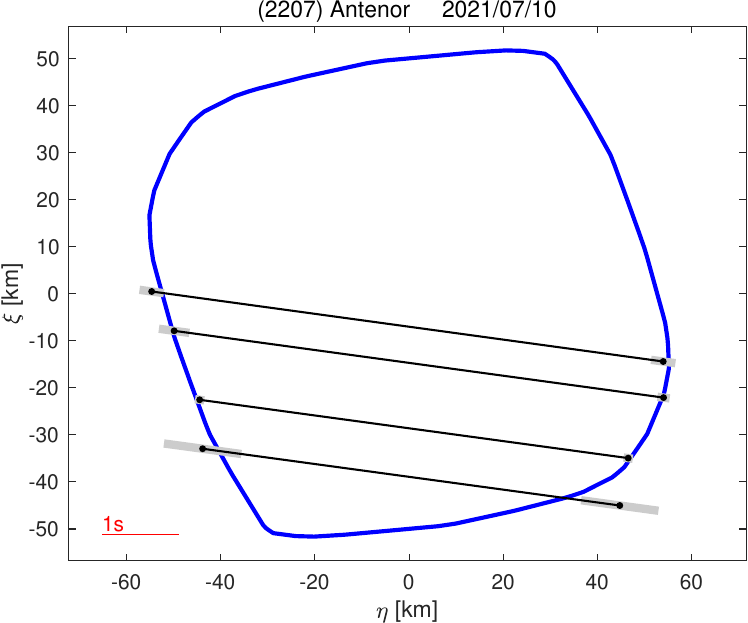}
  \includegraphics[width=\columnwidth]{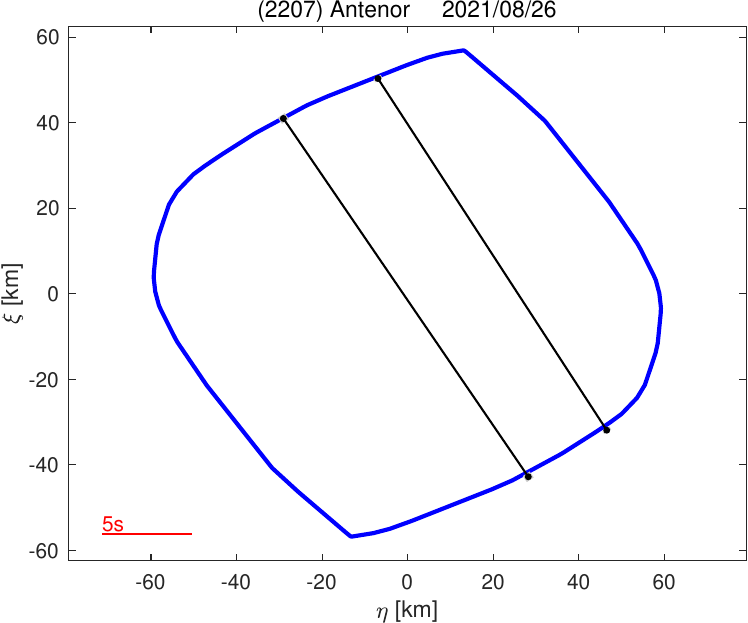}
\end{center}
 \caption{\label{fig:occ_2207} Projection of the shape model of (2207)~Antenor. The model with the pole $(86^\circ, -19^\circ)$ and equivalent diameter of $101 \pm 3$\,km agrees well with the observed chords.}
\end{figure}

% updated
\begin{figure*}[!htb]
\begin{center}
  \includegraphics[width=\columnwidth]{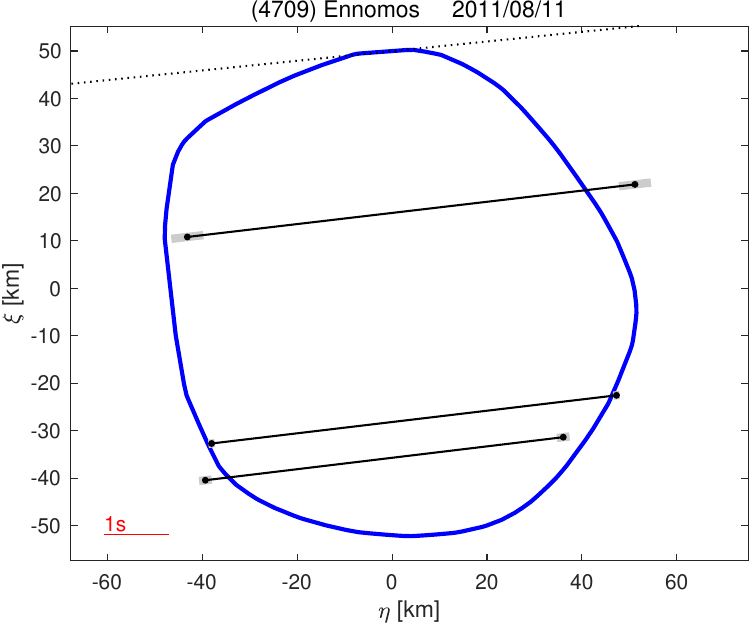}
  \includegraphics[width=\columnwidth]{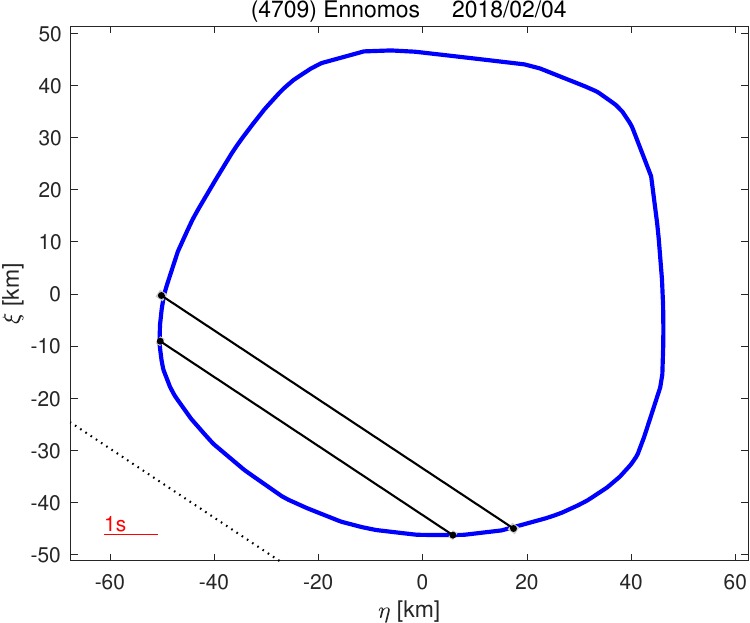}
  \includegraphics[width=\columnwidth]{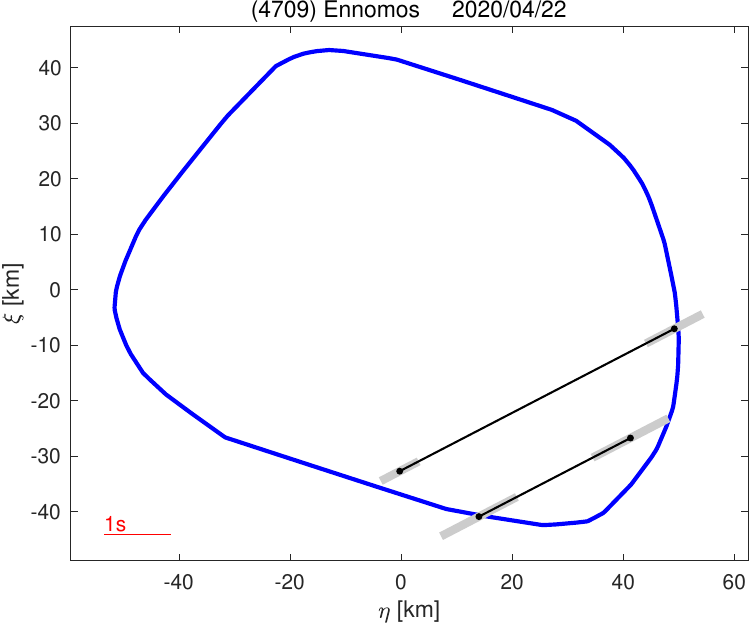}
  \includegraphics[width=\columnwidth]{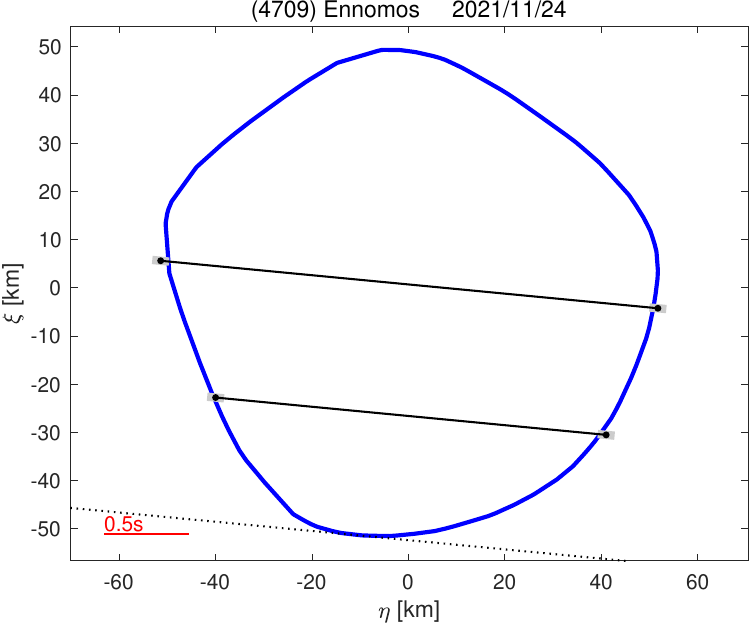}
\end{center}
 \caption{\label{fig:occ_4709} Projection of the shape model of (4709)~Ennomos. The model with the pole $(241^\circ, 75^\circ)$ and equivalent diameter of $91 \pm 4$\,km agrees with the observed chords.}
\end{figure*}

% updated
\begin{figure}[!htb]
\begin{center}
  \includegraphics[width=\columnwidth]{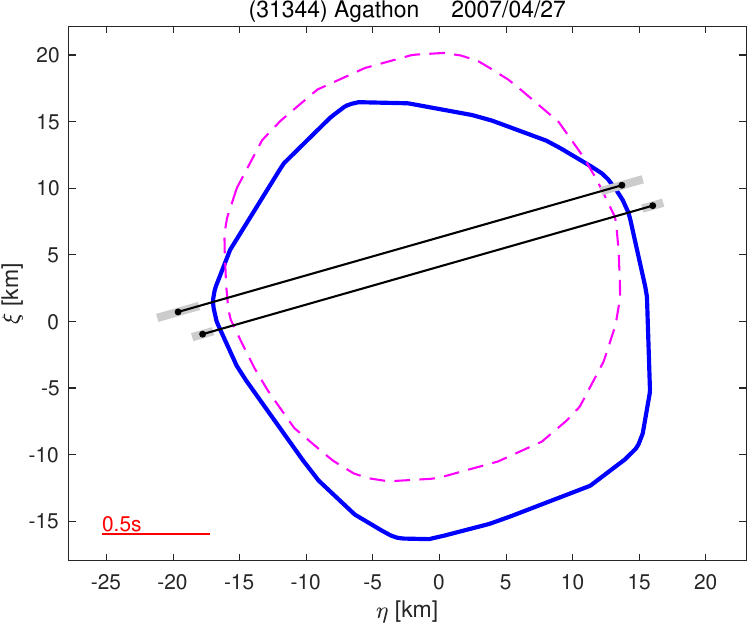}
  \includegraphics[width=\columnwidth]{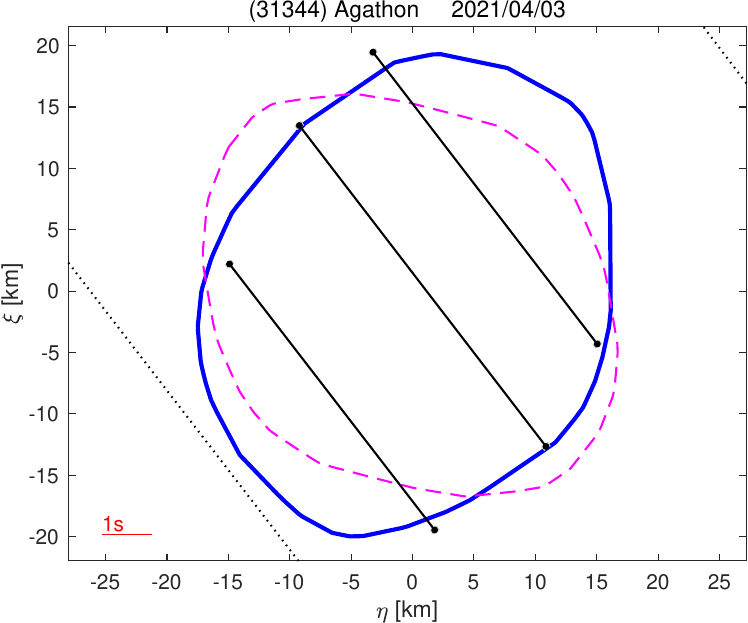}
 \end{center}
 \caption{\label{fig:occ_31344} Projection of the shape model of (31344)~Agathon. The model with the pole $(112^\circ, 52^\circ)$ and equivalent diameter of $38 \pm 1.5$\,km agrees with the observed chords better than the second pole solution $(294^\circ, 36^\circ)$ and equivalent diameter of $35 \pm 3$\,km.}
\end{figure}

\onecolumn
\section{Additional tables}

%%%%%%%%%%%%%%%%%%%%%%%%%%%  Table A.1  %%%%%%%%%%%%%%%%%%%%%%%%%%
\onecolumn
\begin{landscape}
\begin{table*}[!ht]
    \caption{Partial solutions: Rotation state properties and available photometric data.}
    \label{tab:partials}
    \centering
    \begin{tabular}{rl cccc cc cccccccc}
    \hline \hline
    \multicolumn{2}{c}{Asteroid} & $\beta_\mathrm{P}$ & $\delta\beta_\mathrm{P}$ & $P$ & Camp & N$_\mathrm{LC}$ & N$_\mathrm{app}$ & N$_\mathrm{USNO}$ & N$_\mathrm{CSS}$ & N$_\mathrm{GAIA}$ & N$_\mathrm{ASAS}$ & N$_\mathrm{ATLAS}$ & N$_\mathrm{ZTF}$ & N$_\mathrm{PTF}$ & N$_\mathrm{TESS}$\\
    Number & Name  & (deg) & (deg) & (h) &  & & & & & & & & &\\
    \hline
4827    &  Dares        &$-$66  &  17   &  18.9575   &L$_5$&      &      &       &  87   &  28  &  20   &  544  &  52   &  49   &   \\
4946    &  Askalaphus   &  52   &  25   &  22.7129   &L$_4$&      &      &       &  103  &  20  &  20   &  442  &       &       &   \\
5123    &  1989BL       &$-$61  &  15   &  9.89803   &L$_4$&  15  &  2   &       &  54   &  25  &  22   &  512  &       &       &   \\
5476    &  1989TO11     &$-$57  &  14   &  5.77883   &L$_5$&  8   &  1   &       &  63   &  22  &       &  391  &       &       &   \\
5648    &  1990VU1      &$-$57  &  27   &  37.672    &L$_5$&      &      &       &  120  &  30  &  39   &  647  &  150  &  94   &   \\
6997    &  Laomedon     &  39   &  9    &  22.3002   &L$_5$&      &      &       &  71   &  31  &       &  412  &       &       &   \\
11869   &  1989TS2      &  58   &  23   &  14.0224   &L$_5$&      &      &       &  65   &  23  &       &  228  &  87   &  87   &   \\
15398   &  1997UZ23     &$-$65  &  23   &  5.72241   &L$_4$&      &      &       &  116  &  27  &       &  475  &  25   &       &   \\
17419   &  1988RH13     &$-$37  &  9    &  19.1955   &L$_5$&      &      &       &  128  &  33  &       &  380  &  35   &  138  &   \\
17492   &  Hippasos     &  38   &  28   &  17.7413   &L$_5$&  21  &  2   &       &  140  &  30  &       &  683  &  112  &       &   \\
18278   &  Drymas       &$-$63  &  11   &  28.0814   &L$_5$&      &      &       &  95   &  23  &       &  294  &  43   &  37   &   \\
19020   &  2000SC6      &  39   &  19   &  9.80797   &L$_5$&      &      &       &  47   &  23  &       &  464  &  65   &       &   \\
23968   &  1998XA13     &$-$43  &  11   &  11.1282   &L$_4$&      &      &       &  67   &  29  &       &  150  &       &  33   &   \\
24018   &  1999RU134    &  63   &  22   &  6.15487   &L$_5$&      &      &       &  89   &  33  &       &  235  &       &  62   &   \\
29314   &  1994CR18     &  59   &  29   &  15.0371   &L$_5$&      &      &       &  33   &  30  &       &  214  &  22   &       &   \\
30506   &  2000RO85     &$-$51  &  20   &  7.98528   &L$_5$&      &      &       &  106  &  38  &       &  362  &  26   &  52   &  608\\
32370   &  2000QY151    &  63   &  26   &  71.864    &L$_5$&  1   &  1   &       &  40   &  31  &       &       &       &       &   \\
32451   &  2000SP25     &  14   &  18   &  51.507    &L$_5$&      &      &       &  55   &  20  &       &  78   &       &       &   \\
36269   &  1999XB214    &  39   &  21   &  6.77044   &L$_4$&      &      &       &  83   &  38  &       &  199  &  36   &  38   &   \\
41379   &  2000AS105    &$-$55  &  28   &  13.9683   &L$_4$&      &      &       &  70   &  30  &       &  223  &  36   &       &   \\
47957   &  2000QN116    &$-$69  &  15   &  10.9245   &L$_5$&      &      &       &  49   &  24  &       &  140  &  56   &  86   &   \\
48764   &  1997JJ10     &$-$59  &  20   &  6.24927   &L$_5$&      &      &       &  63   &  35  &       &  341  &  49   &       &   \\
51910   &  2001QQ60     &$-$56  &  22   &  302.87    &L$_5$&      &      &       &  95   &  31  &       &  150  &  70   &       &   \\
56976   &  2000SS161    &  45   &  5    &  348.8     &L$_5$&      &      &       &  44   &  43  &       &  253  &  32   &       &   \\
59049   &  1998TC31     &$-$63  &  23   &  6.17235   &L$_4$&      &      &       &  77   &  28  &       &  181  &       &  59   &   \\
65109   &  2002CV36     &$-$51  &  21   &  9.40501   &L$_4$&      &      &       &  79   &  25  &       &  170  &  59   &  73   &   \\
76836   &  2000SB310    &$-$62  &  22   &  10.70035  &L$_5$&      &      &       &  32   &  26  &       &  39   &       &  25   &   \\
84709   &  2002VW120    &  12   &  17   &  533.4     &L$_5$&      &      &       &  44   &  26  &       &  52   &       &       &   \\
88229   &  2001BZ54     &  42   &  20   &  1503.4    &L$_4$&      &      &       &  36   &  24  &       &  148  &       &       &   \\
98362   &  2000SA363    &$-$66  &  16   &  529.2     &L$_5$&      &      &       &  34   &  24  &       &  72   &       &       &   \\
114141  &  2002VX60     &$-$56  &  14   &  8.66685   &L$_5$&      &      &       &  63   &  18  &       &  112  &  79   &  53   &   \\
\hline
    \end{tabular}
\tablefoot{
The table contains the mean value of the ecliptic latitude $\beta_\mathrm{P}$, 1/2 of the range in latitude within the multiple pole solutions $\delta\beta_\mathrm{P}$, the sidereal rotation period $P$, the clan membership (L$_4$ or L$_5$), and the information about the lightcurve dataset -- the number of dense lightcurves N$_\mathrm{LC}$ spanning N$_\mathrm{app}$ apparitions and the number of measurements in each sparse dataset.}
\end{table*}
\end{landscape}

%%%%%%%%%%%%%%%%%%%%%%%%%%%  Table A.2  %%%%%%%%%%%%%%%%%%%%%%%%%%
%%%%%%%%%%%%%%%%%%%%%%%%%%%  Table A.1 %%%%%%%%%%%%%%%%%%%%%%%%%%
\onecolumn
\begin{landscape}
\begin{longtable}{rl ccccc c    cc cccccccc}
    \caption{\label{tab:NewModels}New shape solutions: Rotation state properties and available photometric data.}\\
    \hline \hline
    \multicolumn{2}{c}{Asteroid}& $\lambda_1$ & $\beta_1$ & $\lambda_2$ & $\beta_2$ & $P$ & Camp & N$_\mathrm{LC}$ & N$_\mathrm{app}$ & N$_\mathrm{USNO}$ & N$_\mathrm{CSS}$ & N$_\mathrm{GAIA}$ & N$_\mathrm{ASAS}$ & N$_\mathrm{ATLAS}$ & N$_\mathrm{ZTF}$ & N$_\mathrm{PTF}$ & N$_\mathrm{TESS}$\\
    Number & Name  & (deg) & (deg) & (deg) & (deg) & (h) & & & & & & & & & & &\\
    \hline
    \endfirsthead
    \caption{continued.}\\
    \hline
    \multicolumn{2}{c}{Asteroid}& $\lambda_1$ & $\beta_1$ & $\lambda_2$ & $\beta_2$ & $P$ & Camp & N$_\mathrm{LC}$ & N$_\mathrm{app}$ & N$_\mathrm{USNO}$ & N$_\mathrm{CSS}$ & N$_\mathrm{GAIA}$ & N$_\mathrm{ASAS}$ & N$_\mathrm{ATLAS}$ & N$_\mathrm{ZTF}$ & N$_\mathrm{PTF}$ & N$_\mathrm{TESS}$\\
    Number & Name  & (deg) & (deg) & (deg) & (deg) & (h) & & & & & & & & & & &\\
    \hline\hline
    \endhead
    \hline
    \endfoot
    \hline
624     &  Hektor       &  331  &$-$25  &        &        &  6.920509  &  L$_4$                    &  87  &  16  &  249  &  87   &  38  &  354  &  324  &       &       &  1201\\
624     &  Hektor*      &  333  &$-$31  &        &        &  6.920509  &  \multicolumn{11}{c}{}\\
659     &  Nestor       &  124  &$-$73  &        &        &  15.97896  &  L$_4$                    &  1   &  1   &  101  &  106  &  60  &  258  &  599  &  63   &  94   &   \\
659     &  Nestor*      &  293  &$-$48  &        &        &  15.97894  &  \multicolumn{11}{c}{}\\                                                                          
884     &  Priamus      &  1    &$-$32  &\sout{170}&\sout{$-$46}&6.86133& L$_5$                    &  20  &  4   &  135  &  73   &  21  &  205  &  666  &       &       &  1002\\
884     &  Priamus      &  1    &$-$32  &        &        &  6.861368  &  \multicolumn{11}{c}{\citet{Stephens2017}}\\
911     &  Agamemnon    &  290  &  34   &        &        &  6.58138   &  L$_4$                    &  12  &  4   &  162  &  78   &  27  &  270  &  154  &       &       &     \\
911     &  Agamemnon*   &  126  &   0   &        &        &  6.58179   &  \multicolumn{11}{c}{}\\                                                                          
1143    &  Odysseus     &  62   &$-$54  &  249   &$-$49   &  10.11212  &  L$_4$                    &  4   &  2   &  184  &  69   &  31  &  243  &  575  &       &  44   &   \\
1173    &  Anchises     &  202  &$-$54  &        &        &  11.60963  &  L$_5$                    &  33  &  4   &  127  &  69   &  31  &  159  &  728  &  22   &       &  1070\\
1173    &  Anchises*    &  200  &$-$16  &  20    &$-$4    &  11.60945  &  \multicolumn{11}{c}{}\\                                                                          
1208    &  Troilus      &  225  &$-$29  &  74    &$-$11   &  56.244    &  L$_5$                    &      &      &  105  &  195  &  29  &  179  &  632  &  52   &       &   \\
1437    &  Diomedes     &  146  &  5    &\sout{319}&\sout{13}&  24.4984&  L$_4$                    &  6   &  2   &  159  &  72   &  33  &  192  &  313  &  53   &       &   \\
1437    &  Diomedes*    &  150  &  5    &        &        &  24.4987   &  \multicolumn{11}{c}{}\\                                                                          
1867    &  Deiphobus    &  338  &  79   &        &        &  59.182    &  L$_5$                    &  51  &  4   &  127  &  63   &  42  &  246  &  610  &       &       &   \\
1867    &  Deiphobus**  &  329  &  67   &        &        &  59.182    &  L$_5$                    &  51  &  4   &  127  &  63   &  42  &  246  &  610  &       &       &   \\
1868    &  Thersites    &  93   &  43   &        &        &  10.47526  &  L$_4$                    &  20  &  3   &  102  &  95   &  21  &  74   &  597  &       &       &   \\
1870    &  Glaukos      &  114  &  15   &  298   &  26    &  5.98051   &  L$_5$                    &  5   &  1   &       &  64   &  44  &       &  527  &  37   &  114  &   \\
1873    &  Agenor       &  348  &  36   &  152   &  8     &  20.6334   &  L$_5$                    &      &      &       &  70   &  38  &       &  500  &       &       &   \\
1873    &  Agenor*      &  328  &  28   &        &        &  20.6338   &  \multicolumn{11}{c}{}\\                                                                          
2207    &  Antenor      &  86   &$-$19  &        &        &  7.96476   &  L$_5$                    &  29  &  4   &  36   &  73   &  33  &  284  &  611  &       &  141  &   \\
2241    &  Alcathous    &  108  &  68   &  225   &  50    &  7.68986   &  L$_5$                    &  44  &  6   &  37   &  91   &  30  &  293  &  488  &       &       &  965\\
2363    &  Cebriones    &  107  &$-$38  &        &        &  20.0764   &  L$_5$                    &  23  &  5   &  29   &  84   &  49  &  230  &  657  &       &       &   \\
2797    &  Teucer       &  291  &$-$30  &  120   &$-$2    &  10.14681  &  L$_4$                    &  11  &  4   &  78   &  73   &  31  &  197  &  378  &  20   &       &   \\
2893    &  Peiroos      &  78   &  72   &        &        &  8.94891   &  L$_5$                    &  19  &  4   &       &  83   &  27  &  205  &  651  &  36   &       &   \\
2895    &  Memnon       &  293  &  8    &        &        &  7.51979   &  L$_5$                    &  24  &  6   &       &  108  &  34  &  46   &  635  &  62   &       &   \\
2920    &  Automedon    &  100  &$-$5   &        &        &  10.21237  &  L$_4$                    &  19  &  3   &  36   &  123  &  25  &  239  &  618  &  59   &       &   \\
3240    &  Laocoon      &  342  &  72   &        &        &  11.3122   &  L$_5$                    &      &      &       &  70   &  19  &       &  575  &  33   &       &   \\
3317    &  Paris        &  119  &  36   &  344   &  30    &  7.08099   &  L$_5$                    &  31  &  5   &  59   &  102  &  24  &  194  &  603  &  56   &       &   \\
3391    &  Sinon        &  100  &  89   &        &        &  8.13675   &  L$_4$                    &  5   &  1   &       &  87   &  25  &       &  264  &       &       &   \\
3391    &  Sinon*       &  63   &$-$52  &  251   &$-$71   &  8.13809   &  \multicolumn{11}{c}{}\\                                                                          
3451    &  Mentor       &  85   &  15   &        &        &  7.69657   &  L$_5$                    &  33  &  5   &  77   &  109  &  52  &  282  &  700  &       &  88   &   \\
3451    &  Mentor*      &  237  &  57   &  82    &  7     &  7.69659   &  \multicolumn{11}{c}{}\\                                                                          
3564    &  Talthybius   &  231  &$-$57  &  80    &$-$70   &  40.518    &  L$_4$                    &  17  &  2   &  34   &  94   &  25  &  146  &  487  &       &       &   \\
3709    &  Polypoites   &  156  &  1    &  346   &  24    &  10.03688  &  L$_4$                    &  41  &  6   &  45   &  115  &  39  &  157  &  630  &  48   &       &   \\
4063    &  Euforbo      &  321  &  75   &  54    &  40    &  8.84545   &  L$_4$                    &  15  &  3   &  50   &  94   &  34  &  224  &  604  &       &  214  &   \\
4068    &  Menestheus   &  45   &$-$57  &  243   &$-$23   &  14.3355   &  L$_4$                    &  31  &  5   &       &  126  &  35  &  115  &  632  &  23   &       &   \\
4086    &  Podalirius   &  13   &  16   &  214   &  44    &  10.43312  &  L$_4$                    &  3   &  1   &  20   &  112  &  31  &  68   &  324  &       &  35   &   \\
4086    &  Podalirius*  &  22   &  12   &  207   &  12    &  10.43320  &  \multicolumn{11}{c}{}\\                                                                          
4348    &  Poulydamas   &  85   &  47   &        &        &  9.92021   &  L$_5$                    &  23  &  4   &       &  64   &  33  &  118  &  581  &  65   &  60   &   \\
4489    &  1988AK       &  100  &  28   &  296   &  11    &  12.5828   &  L$_4$                    &  30  &  3   &  24   &  118  &  43  &  193  &  493  &       &       &   \\
4489    &  1988AK*      &  136  &$-$27  &        &        &  12.5842   &  \multicolumn{11}{c}{}\\                                                                          
4543    &  Phoinix      &  132  &$-$55  &  352   &$-$45   &  38.840    &  L$_4$                    &      &      &       &  94   &  37  &  36   &  349  &       &       &   \\
4707    &  Khryses      &  126  &  76   &  291   &  70    &  6.86340   &  L$_5$                    &  12  &  3   &       &  101  &  30  &       &  563  &       &       &  778\\
4709    &  Ennomos      &  241  &  75   &        &        &  12.2687   &  L$_5$                    &  40  &  3   &       &       &  30  &  232  &  686  &       &       &  797\\
4709    &  Ennomos*     &  174  &  25   &        &        &  12.2688   &  \multicolumn{11}{c}{}\\                                                                          
4722    &  Agelaos      &  346  &  7    &  165   &  25    &  18.4579   &  L$_5$                    &  7   &  2   &       &  72   &  30  &  44   &  716  &  121  &  78   &   \\
4792    &  Lykaon       &  334  &  44   &  156   &  67    &  40.070    &  L$_5$                    &      &      &       &  84   &  22  &       &  588  &  55   &       &   \\
4828    &  Misenus      &  99   &  84   &  268   &  54    &  12.8579   &  L$_5$                    &      &      &       &  100  &  34  &       &  649  &  53   &       &  979\\
4834    &  Thoas        &  94   &  33   &        &        &  18.1930   &  L$_4$                    &  21  &  3   &       &  100  &  21  &  75   &  502  &       &       &   \\
4836    &  Medon        &  279  &$-$57  &  131   &$-$23   &  9.84146   &  L$_4$                    &  10  &  11  &       &  105  &  64  &  75   &  452  &       &       &  966\\
5027    &  Androgeos    &  53   &  31   &  287   &  34    &  11.3192   &  L$_4$                    &  24  &  2   &       &  102  &  29  &  31   &  519  &       &       &   \\
5027    &  Androgeos*   &  276  &  36   &  56    &  28    &  11.3196   &  \multicolumn{11}{c}{}\\                                                                          
5144    &  Achates      &  132  &$-$30  &  323   &$-$31   &  5.95401   &  L$_5$                    &  28  &  20  &  38   &  97   &  23  &  88   &  568  &  45   &       &   \\
5209    &  1989CW1      &  119  &  77   &  259   &  66    &  11.6026   &  L$_4$                    &  2   &  8   &       &  91   &  25  &       &  348  &  40   &       &   \\
5244    &  Amphilochos  &  317  &  78   &  130   &  65    &  9.78574   &  L$_4$                    &  12  &  1   &       &  85   &  30  &       &  447  &       &  37   &   \\
5285    &  Krethon      &  280  &$-$74  &        &        &  12.0246   &  L$_4$                    &  13  &  2   &       &  85   &  61  &       &  476  &       &       &   \\
11089   &  1994CS8      &  54   &$-$68  &        &        &  7.72676   &  L$_5$                    &  21  &  4   &       &  79   &  33  &       &  408  &  25   &       &   \\
13229   &  Echion       &  4    &$-$64  &  185   &$-$82   &  8.45144   &  L$_4$                    &  5   &  6   &       &  50   &  23  &       &  158  &       &  131  &   \\
13372   &  1998VU6      &  96   &  43   &  290   &  47    &  20.1715   &  L$_4$                    &      &      &       &  73   &  35  &       &  272  &       &       &   \\
15436   &  1998VU30     &  39   &  1    &        &        &  8.96656   &  L$_4$                    &  18  &  3   &       &  81   &  21  &  173  &  537  &       &       &   \\
15502   &  1999NV27     &  221  &$-$67  &  40    &$-$32   &  15.1177   &  L$_5$                    &  36  &  5   &       &  67   &  22  &       &  524  &       &       &  831\\
15527   &  1999YY2      &  352  &  86   &  82    &  73    &  6.99181   &  L$_4$                    &  5   &  1   &       &  111  &  29  &       &  478  &       &       &   \\
15663   &  Periphas     &  87   &$-$20  &        &        &  9.9218    &  L$_4$                    &      &      &       &  72   &  32  &       &       &       &       &   \\
15663   &  Periphas*    &  296  &  13   &        &        &  9.9200    &  \multicolumn{11}{c}{}\\                                                                          
16428   &  1988RD12     &  63   &  65   &  245   &  57    &  20.3310   &  L$_5$                    &      &      &       &  50   &  30  &       &  170  &  94   &  116  &   \\
16560   &  Daitor       &  61   &  59   &  333   &  71    &  13.7994   &  L$_5$                    &      &      &       &  67   &  29  &       &  512  &  68   &       &   \\
17314   &  Aisakos      &  291  &$-$81  &  108   &$-$57   &  9.72467   &  L$_5$                    &  6   &  1   &       &  110  &  33  &       &  539  &  23   &       &   \\
17365   &  1978VF11     &  60   &  68   &        &        &  12.6726   &  L$_5$                    &  4   &  1   &       &  69   &  31  &       &  470  &  28   &  28   &  430\\
17414   &  1988RN10     &  285  &$-$78  &        &        &  32.7838   &  L$_5$                    &      &      &       &  71   &  29  &       &  151  &       &       &   \\
18062   &  1999XY187    &  286  &  20   &        &        &  9.7728    &  L$_4$                    &  10  &  1   &       &  82   &  13  &       &  372  &       &  42   &   \\
18263   &  Anchialos    &  73   &$-$81  &  358   &$-$49   &  10.3310   &  L$_4$                    &  5   &  7   &       &  63   &  19  &       &  241  &  38   &       &   \\
23135   &  Pheidas      &  15   &  30   &  190   &$-$3    &  8.71422   &  L$_4$                    &  10  &  8   &       &  75   &  29  &       &  438  &  32   &       &   \\
23549   &  Epicles      &  351  &$-$90  &        &        &  279.56    &  L$_5$                    &      &      &       &  51   &  38  &       &  302  &  46   &       &   \\
23694   &  1997KZ3      &  10   &$-$8   &  191   &  5     &  161.25    &  L$_5$                    &      &      &       &  74   &  29  &       &  189  &  32   &       &   \\
24471   &  2000SH313    &  20   &  74   &        &        &  341.3     &  L$_5$                    &      &      &       &  35   &  41  &       &  314  &  74   &  95   &   \\
24485   &  2000YL102    &  128  &$-$77  &  276   &$-$67   &  14.6842   &  L$_4$                    &      &      &       &  84   &  32  &       &  387  &  27   &       &   \\
25911   &  2001BC76     &  338  &  27   &        &        &  8.76058   &  L$_4$                    &      &      &       &  77   &  27  &       &  266  &  21   &       &   \\
30705   &  Idaios       &  247  &  48   &  15    &  54    &  15.7318   &  L$_5$                    &  31  &  3   &       &  125  &  33  &       &  544  &  89   &  70   &   \\
31342   &  1998MU31     &  106  &  46   &  321   &  25    &  9.16914   &  L$_5$                    &      &      &       &  84   &  18  &       &  537  &  66   &       &   \\
31344   &  Agathon      &  112  &  52   &\sout{294}&\sout{36}&  42.428 &  L$_5$                    &      &      &       &  96   &  34  &       &  488  &  106  &  61   &   \\
31819   &  1999RS150    &  79   &  48   &  255   &  24    &  34.7124   &  L$_5$                    &      &      &       &  85   &  30  &       &  325  &       &       &  245\\
32339   &  2000QA88     &  85   &  33   &        &        &  9.77594   &  L$_5$                    &      &      &       &  67   &  32  &       &  276  &  38   &  89   &   \\
32615   &  2001QU277    &       &  42   &  88    &  86    &  6.71702   &  L$_5$                    &  4   &  1   &       &  62   &  37  &       &  305  &  42   &       &   \\
32811   &  Apisaon      &  314  &  20   &  125   &  61    &  19.0108   &  L$_5$                    &      &      &       &  102  &  28  &       &  428  &  71   &       &   \\
34746   &  2001QE91     &  257  &  25   &  61    &  1     &  19.6248   &  L$_5$                    &      &      &       &  90   &  28  &       &  629  &  59   &       &   \\
34835   &  2001SZ249    &  313  &  50   &  144   &  46    &  7.7401    &  L$_5$                    &      &      &       &       &  16  &       &       &       &       &   \\
51364   &  2000SU333    &  287  &  67   &  57    &  52    &  30.4571   &  L$_5$                    &      &      &       &  99   &  27  &       &  474  &  32   &  26   &   \\
51984   &  2001SS115    &  276  &$-$83  &  119   &$-$78   &  692.1     &  L$_5$                    &      &      &       &  50   &  41  &       &  76   &  31   &       &   \\
55474   &  2001TY229    &  216  &  73   &  51    &  40    &  20.5316   &  L$_5$                    &      &      &       &  92   &  20  &       &  232  &  34   &       &   \\
63923   &  2001SV41     &  12   &$-$59  &  214   &$-$40   &  8.75811   &  L$_5$                    &      &      &       &  41   &  37  &       &  222  &  63   &  57   &   \\
76867   &  2000YM5      &  196  &$-$42  &  74    &$-$32   &  9.11947   &  L$_5$                    &  18  &  3   &       &  116  &  33  &       &  610  &  52   &       &   \\
99943   &  2005AS2      &  280  &$-$18  &  108   &$-$13   &  36.888    &  L$_5$                    &      &      &       &  80   &  35  &       &  316  &  95   &       &   \\
124729  &  2001SB173    &  94   &  47   &  275   &  54    &  23.303    &  L$_5$                    &      &      &       &       &  32  &       &       &  54   &       &   \\
\hline
\end{longtable}
\tablefoot{
The table contains ecliptic longitudes and latitudes of the spin axis directions $\lambda$ and $\beta$ for one or two possible solutions, the sidereal rotation period $P$, the clan membership (L$_4$ or L$_5$), and information about the lightcurve dataset -- the number of dense lightcurves N$_\mathrm{LC}$, the number of apparitions N$_\mathrm{app}$ and the number of measurements in each sparse dataset. The uncertainty of the pole direction is usually about 10\degr and of the period of the order of the last decimal digit. We also list the rotation state properties for solutions in the DAMIT database (indicated by an asterisk), if available. The ADAM model of Deiphobus is indicated by a double asterisk. Finally, pole solutions that do not match the stellar occultations (i.e., the rejected ones) are crossed out.}
\end{landscape}

%%%%%%%%%%%%%%%%%%%%%%%%%%%  Table A.3  %%%%%%%%%%%%%%%%%%%%%%%%%%
\onecolumn
\begin{table*}[!ht]
    \centering
    \caption{\label{tab:models_damit}DAMIT shape solutions: Rotation state properties.}
\begin{tabular}{r@{\,\,\,}l rrrr D{.}{.}{6} c }
    \hline 
    \multicolumn{2}{c} {Asteroid} & $\lambda_\mathrm{D1}$ & $\beta_\mathrm{D1}$ & $\lambda_\mathrm{D2}$ & $\beta_\mathrm{D2}$ & \multicolumn{1}{c} {$P_\mathrm{D}$} & Clan\\
    \multicolumn{2}{l} { } & [deg] & [deg] & [deg] & [deg] & \multicolumn{1}{c} {[hours]} & \\
    \hline\hline
588    &  Achilles   &  172  &$-$5   &        &        &  7.30630  &L$_4$ \\
%624    &  Hektor     &  333  &$-$31  &        &        &  6.920509 &L$_4$ \\
%911    &  Agamemnon  &  126  &  0    &        &        &  6.58179  &L$_4$ \\
1404   &  Ajax       &  294  &  42   &        &        &  29.3874  &L$_4$ \\
4715   &$-$          &  335  &  55   &        &        &  8.81418  &L$_5$ \\
5130   &  Ilioneus   &  307  &$-$26  &        &        &  14.7357  &L$_5$ \\
5283   &  Pyrrhus    &  332  &$-$12  &  152   &  22    &  7.31916  &L$_4$ \\
5436   &  Eumelos    &  272  &  22   &  96    &  11    &  21.2689  &L$_4$ \\
9694   &  Lycomedes  &  133  &  54   &  323   &  53    &  18.1094  &L$_4$ \\
11429  &  Demodokus  &  117  &$-$3   &  306   &$-$38   &  50.219   &L$_4$ \\
13060  &$-$          &  339  &$-$28  &        &        &  482.5    &L$_4$ \\
14268  &$-$          &  97   &$-$39  &  304   &$-$30   &  7.51225  &L$_4$ \\
58931  &$-$          &  56   &$-$83  &  218   &$-$52   &  11.3322  &L$_5$ \\
    \hline
    \end{tabular}
\tablefoot{
The table contains ecliptic longitudes and latitudes of the spin axis directions $\lambda_\mathrm{D}$ and $\beta_\mathrm{D}$ for one or two possible solutions, the sidereal rotation period $P_\mathrm{D}$, and the clan membership (L$_4$ or L$_5$).}
\end{table*}

%%%%%%%%%%%%%%%%%%%%%%%%%%%  Table A.4  %%%%%%%%%%%%%%%%%%%%%%%%%%
\onecolumn
\begin{longtable}{r rrrrrrrr}
    \caption{\label{tab:bootstrap}The mean values and standard deviations of pole directions within the bootstrapped solutions.}\\
    \hline \hline
    \multicolumn{1}{c} {Asteroid} & $\lambda_\mathrm{b1}$ & $\Delta\lambda_\mathrm{b1}$ & $\beta_\mathrm{b1}$ & $\Delta\beta_\mathrm{b1}$ & $\lambda_\mathrm{b2}$ & $\Delta\lambda_\mathrm{b2}$ & $\beta_\mathrm{b2}$ & $\Delta\beta_\mathrm{b2}$\\
    \multicolumn{1}{l} { } & [deg] & [deg] & [deg] & [deg] & [deg] & [deg] & [deg] & [deg] \\
    \hline
    \endfirsthead
    \caption{continued.}\\
    \hline
    \multicolumn{1}{c} {Asteroid} & $\lambda_\mathrm{b1}$ & $\Delta\lambda_\mathrm{b1}$ & $\beta_\mathrm{b1}$ & $\Delta\beta_\mathrm{b1}$ & $\lambda_\mathrm{b2}$ & $\Delta\lambda_\mathrm{b2}$ & $\beta_\mathrm{b2}$ & $\Delta\beta_\mathrm{b2}$\\
    \multicolumn{1}{l} { } & [deg] & [deg] & [deg] & [deg] & [deg] & [deg] & [deg] & [deg] \\
    \hline\hline
    \endhead
    \hline
    \endfoot
    \hline
624     &  330.9  &  0.6    & --24.8  &  1.0   &          &         &          &        \\
659     &  124.5  &  2.4    & --71.0  &  1.8   &          &         &          &        \\
884     &  356.1  &  1.8    & --28.8  &  1.6   &  171.9   &  1.4    & --41.3   &  2.6   \\
911     &  293.6  &  2.6    &  27.7   &  5.3   &          &         &          &        \\
1143    &  62.6   &  2.4    & --53.0  &  5.5   &  251.9   &  3.3    & --51.2   &  4.6   \\
1173    &  200.4  &  3.5    & --52.3  &  1.6   &          &         &          &        \\
1208    &  226.5  &  1.7    & --27.6  &  1.9   &  73.8    &  1.5    & --12.0   &  2.0   \\
1437    &  145.3  &  1.8    &  4.2    &  3.3   &          &         &          &        \\
\hline
\end{longtable}
\tablefoot{
The table contains ecliptic longitudes and latitudes of the spin axis directions $\lambda_\mathrm{b}$ and $\beta_\mathrm{b}$ and its uncertainties $\Delta\lambda_\mathrm{b}$ and $\Delta\beta_\mathrm{b}$ for one or two possible solutions based on ten bootstrapped datasets.}

%%%%%%%%%%%%%%%%%%%%%%%%%%%  Table A.5  %%%%%%%%%%%%%%%%%%%%%%%%%%
\onecolumn
%\scriptsize{
\begin{longtable}{lr rrr l l}
\caption{\label{tab:lcs}Optical disk-integrated data used for the shape modeling. }\\
\hline 
\multicolumn{1}{c} {Epoch} & \multicolumn{1}{c} {$N_p$} & \multicolumn{1}{c} {$\Delta$} & \multicolumn{1}{c} {$r$} & \multicolumn{1}{c} {$\varphi$} & \multicolumn{1}{c} {Filter} & Reference \\
 &  & (AU) & (AU) & ($^{\circ}$) &  &  \\
\hline\hline
\endfirsthead
\caption{continued.}\\
\hline
\multicolumn{1}{c} {Epoch} & \multicolumn{1}{c} {$N_p$} & \multicolumn{1}{c} {$\Delta$} & \multicolumn{1}{c} {$r$} & \multicolumn{1}{c} {$\varphi$} & \multicolumn{1}{c} {Filter} & Reference \\
 &  & (AU) & (AU) & ($^{\circ}$) &  &  \\
\hline\hline
\endhead
\hline
\endfoot
\hline

\multicolumn{7}{c}         {(659)    Nestor}\\             
1988-01-25.2               &         10               &  4.70  &  5.55  &  5.6   &  B    &  \citet{Binzel1992}        \\
\multicolumn{7}{c}         {(884)    Priamus}\\            
2010-07-16.3               &         53               &  3.61  &  4.60  &  3.5   &  R    &  \citet{French2011}   \\
2010-08-14.2               &         29               &  3.62  &  4.61  &  3.3   &  R    &  \citet{French2011}   \\
2010-08-15.1               &         5                &  3.63  &  4.61  &  3.6   &  R    &  \citet{French2011}   \\
2010-08-17.2               &         15               &  3.64  &  4.61  &  4.0   &  R    &  \citet{French2011}   \\
2010-08-18.3               &         14               &  3.65  &  4.61  &  4.3   &  R    &  \citet{French2011}   \\
\end{longtable}
\tablefoot{
    For each lightcurve, the table gives the epoch, the number of individual measurements $N_p$, asteroid's distances to the Earth $\Delta$ and the Sun $r$, phase angle $\varphi$, photometric filter and the reference to the data. %For sparse data, we provide the campaign name for clarity, the proper references are: Catalina Sky Survey -- \citet{Larson2003}, Gaia DR2 -- \citet{Spoto2018}, PTF -- \citet{Waszczak2015}, ATLAS -- \citet{Tonry2018}, ZTF -- \citet{bellm2019}, TESS -- \citet{Pal2020}, USNO-Flagstaff -- downloaded from AstDyS.
    }

\end{appendix}

\end{document}